\documentclass[12pt]{article}
\usepackage{lscape}
\usepackage{geometry}
\usepackage[onehalfspacing]{setspace}
\usepackage{amssymb}
\usepackage{amsfonts}
\usepackage{graphicx}
\usepackage{amsmath}
\usepackage{epstopdf}
\usepackage{natbib}
\usepackage{bibunits}
\usepackage{morefloats}
\usepackage{float}
\usepackage{comment}
\usepackage{pdflscape}
\usepackage{adjustbox}
\usepackage{subcaption}
\usepackage{longtable}
\usepackage{apalike}
\usepackage{xfrac}
\usepackage[dvipsnames]{xcolor}
\usepackage{threeparttable}
\usepackage{tabularx}
\usepackage{algorithm,algorithmic}

\usepackage{multicol}
\usepackage{booktabs}
\usepackage{dcolumn}
\newcolumntype{d}[1]{D..{#1}}

\definecolor{DarkerPineGreen}{RGB}{0, 90, 80} % Define a custom darker shade

\usepackage{etoolbox,siunitx}
\robustify\bfseries

\usepackage{mathpazo} % math & rm
\usepackage[scaled]{helvet} % ss
\usepackage{courier} % tt
\normalfont
\usepackage[T1]{fontenc}

\usepackage[pdftex,bookmarks,colorlinks]{hyperref}
\hypersetup{
  colorlinks,
  citecolor=DarkerPineGreen,
  linkcolor=red,
  urlcolor=blue}
\usepackage{cleveref}

\usepackage{etoolbox}

\usepackage{xcolor,colortbl}

\begin{document}
\title{\huge \hspace*{-0.54cm}  Time-Varying Parameters as Ridge Regressions}
\author{Philippe Goulet Coulombe\thanks{%
\footnotesize Département des Sciences Économiques, \href{mailto:p.gouletcoulombe@gmail.com}{{goulet\_coulombe.philippe@uqam.ca}}. For many helpful discussions, I would like to thank Edvard Bakhitov, Julien Champagne, Frank Diebold, Maximilian G\"obel,  Nicolas Harvie,  Magnus Reif, Frank Schorfheide, Dalibor Stevanovic, Luis Uzeda and Boyuan Zhang. For excellent research assistance at various stages of this project, I am grateful to Preston Ching, Isaac Tham and David Wigglesworth. Finally, I want to thank, for useful suggestions and remarks, participants at the Penn Econometrics Lunch Seminar, Symposium of the Society for Nonlinear Dynamics and Econometrics, Vienna High-Dimensional Time Series Workshop, Nordic Econometric Meeting and Bank of Canada.  The R package \texttt{TVPRidge} is available  \href{https://philippegouletcoulombe.com/code}{here}.}} %
\date{\vspace{-0.4cm}
Université du Québec à Montréal \\[2ex]%
\small
\small
First Draft: December 2, 2018 \\
This Draft: July 31, 2024 \\ 
%\href{https://drive.google.com/file/d/1aVqXxCABMlYrk0PoD45L8WOM-s9LapCw/view?usp=sharing}{Latest Draft Here} \\ %\today
\vspace{0.3cm}
\large
%{\color{PineGreen} Preliminary and incomplete: \\
 %please do not circulate without permission.}
 }
\maketitle

\begin{abstract}
%\vspace{-0.7cm}

%\noindent I show that time-varying parameters (TVPs) models, frequently used in economics to model structural change, can be rewritten as ridge regressions. This instantly makes computations, tuning and implementation much easier than in the state-space paradigm. Among other things, solving the equivalent \textit{dual} ridge problem is computationally very fast even in high dimensions and the crucial "amount of time-variation" is tuned by cross-validation. Evolving volatility is dealt with using a two-step ridge regression. I consider extensions that incorporate sparsity (the algorithm selects which parameters vary and which do not) and reduced-rank restrictions (variation is tied to a factor model). To demonstrate the usefulness of the approach, I use it to study the evolution of monetary policy in Canada. The application requires the estimation of more than 4k TVPs, a task well within the reach of the new method.

\noindent Time-varying parameters (TVPs) models are frequently used in economics to capture structural change.  I highlight a rather underutilized fact---that these are actually ridge regressions.  Instantly, this makes computations, tuning, and implementation much easier than in the state-space paradigm. Among other things, solving the equivalent \textit{dual} ridge problem is computationally very fast even in high dimensions, and the crucial "amount of time variation" is tuned by cross-validation. Evolving volatility is dealt with using a two-step ridge regression. I consider extensions that incorporate sparsity (the algorithm selects which parameters vary and which do not) and reduced-rank restrictions (variation is tied to a factor model).  To demonstrate the usefulness of the approach, I use it to study the evolution of monetary policy in Canada using large time-varying local projections and TVP-VARs with demanding lag lengths.  The applications require the estimation of up to 4\,600 TVPs,  a task within the reach of the new method.

%This paper aims at making Time-varying parameters (TVP) models used in Macroeconomics both computationally and statistically efficient. The end goal is to substantially mitigate the tradeoff between unrealistically small time-varying model and bigger constant model. To achieve that, I first show that TVP models can be seen as a standard Ridge Regression problem for which a closed-form solution exists. Solving the numerically equivalent dual Ridge problem makes this computationally very fast, even in high dimensions. Second, I generalize the solution so that it can handle general specifications like TVP-VARs à la \cite{primiceri2005}. Third, I address directly the problem that time-variation is mechanically eradicated in high-dimensional macro models by extending the methodology to obtain sparse and dense TVPs. That is, not all parameters vary and when they do, it is according to a few latent factors. Hence, this paper not only makes large TVP models computationally feasible, but also provides a way to retain and model efficiently time variation in models of considerable size. I apply the tool in a substantive forecasting exercise of US macro variables and provide new evidence on the evolution of Canadian monetary policy using time-varying local projections.

\end{abstract}

\thispagestyle{empty}%EndExpansion

%\noindent \textit{JEL codes: C32, C55, E27}

%\noindent \textit{Keywords: Vector Autoregressions, Lasso, Ridge, Time-varying Parameters, Factor models}

\clearpage
\pagenumbering{arabic} % Start page numbering at 1

\section{Introduction}

Economies change. Intuitively, this should percolate to the parameters of models characterizing them. To econometrically achieve that, a popular approach is Time-Varying Parameters (TVPs), where a linear equation’s coefficients follow stochastic processes --- usually random walks.  {\color{black} There is a wide body of work utilizing TVPs to study structural changes in key macroeconomic equations, such as Phillips curves and Taylor rules \citep{stock1996evidence,boivin2005has,blanchard2015inflation}. In the multivariate setting,  } classic papers consider TVP Vector Autoregressions (TVP-VARs) to examine changing monetary policy \citep{primiceri2005} and evolving inflation dynamics \citep{cogley2001evolving,cogley2005drifting}.  Recently, these ideas have been extended to \cite{jorda2005}'s local projections (LPs) to obtain directly time-varying impulse response functions \citep{ruisi2019time,lusompa2020local}.

%\footnote{Well-known applications where time variation in LPs is obtained by interacting a linear specification with a "state of the economy" variable are \cite{AG2012} and \cite{RZ2018}.} 

In both the VAR and LP cases, important practical obstacles reduce the scope and applicability of TVPs. One is prohibitive computations limiting the size of the model.  Another is the difficulty of tuning the crucial amount of time variation. To address those and other  problems, I leverage the underutilized fact that TVP models are ridge regressions (RR). The connection is useful: 50 years \citep{hoerl1970ridge} of RR widespread use, research and wisdom is readily transferable to TVPs.  {\color{black}Among other things, this provides fast computations via a closed-form \textit{dual} solution only using matrix operations.   Therefore, it avoids potential initialization and convergence issues because it does not rely on MCMC simulations and filtering.  Next, we have that the amount of time variation is automatically tuned by cross-validation (CV).  Adjustments for evolving residuals' volatility and heterogeneous parameter drifting speeds (random walk variances) are implemented via a 2-step ridge regression (henceforth 2SRR),  the flagship model of this paper.   As a result, it is easy to implement and operate, and it integrates seamlessly into existing machine learning-based macroeconomic forecasting pipelines. } Moreover, the setup is directly extendable to deploy additional shrinkage schemes (sparse TVP, reduced rank TVPs) recently proposed in the literature \citep{stevanovic2016common,bitto2018achieving}.  

{\color{black} In sum,  by introducing 2SRR and related methods,  the paper's key contributions are threefold: (i) easier computations,  (ii) easier tuning,  and (iii) a deeper understanding of standard TVP models used in empirical macroeconomics through the lenses of ridge regression.   Since steps have already been taken in the direction of each of these  contributions, I will now review the relevant literature and discuss where the ridge approach sits.}

%For the remainder of this introduction, I review the issues facing current TVP models, survey their related literature, and explain .

%\footnote{In its simplest form, it consists or creating many new regressors out of the original data and using it as a feature matrix in any RR software, which requires 3 lines of code (cross-validation, estimation, prediction).} 

%Finally, credible regions are available since RR is alternatively a plain Bayesian regression. 

\vskip 0.2cm

{\sc \noindent \textbf{Computations}.} Using standard implementations allowing for stochastic volatility (SV) in TVP-VARs, researchers are limited to a few lags  and a small number of variables (not more than 4 or 5) \citep{kilian2017SVAR}. This constraint leaves the reader ever-wondering whether time variation is not merely the byproduct of omitted variables.  A similar problem occurs if one wants to study time-varying local projections of a certain size,  which imply repeated estimation over many horizons.  Consequently, a growing number of contributions attempt to deal with the computational problem.  Within the state-space paradigm, \cite{koop2013large} and \cite{huber2020bayesian} proposed approximations to speed up MCMC simulations.   \cite{giraitis2014inference} and \cite{kapetanios2017large} drop the state-space paradigm altogether in favor of a nonparametric kernel-based estimator. \cite{chen2012testing} consider a similar approach to develop a test for smooth structural change while \cite{petrova2016quasi} develops a Bayesian version particularly apt with large multivariate systems.  Also in the Bayesian arena,  \cite{hauzenberger2022fast} drop the random walk in favor of a hierarchical prior that lends itself well to computations through matrix operations using the singular value decomposition.  

This paper's contribution along the computational angle is to provide quick and off-the-shelf point estimates for the original random-walk based model.  It demonstrates that this can be achieved using reasonably elementary methods without compromising reliability.  However, unlike Bayesian methods, it does not inherently provide inference, necessitating an additional procedure to obtain it.

%Also,  while some of the alternative  frameworks mentioned above allow the estimation of the desired large models, it is sometimes unclear how we can incorporate useful features such as heterogeneous variances for parameters (as in \cite{primiceri2005}). 

%Notable contributions include dynamic shrinkage priors for large time-varying parameter regressions using scalable Markov chain Monte Carlo methods \citep{hauzenberger2024dynamic}, 

%and inducing sparsity and shrinkage in time-varying parameter models \citep{huber2021inducing}.
%\cite{giannone_lenza_primiceri_2015}

\vskip 0.2cm

{\sc \noindent \textbf{Tuning}.}  \cite{AAG2013}, \cite{BaumeisterKilian2014},  \cite{pettenuzzo2017forecasting},  and several others have investigated, with varying angles, the usefulness of time variation to increase prediction accuracy.  A critical choice underlying forecasting successes and failures is the amount of time variation. Notoriously, tuning parameter(s) determining it can largely influence prediction results and estimated coefficients, accounting for much of the transparency and reliability concerns regarding TVP models. \cite{amir2018choosing} and \cite{cadonna2020triple} propose to treat those pivotal hyperparameters as another layer of parameters to be estimated within the Bayesian procedure --- and find this indeed helps.  %Even in traditional BVARs, carefully estimating the overall tightness of priors can make a sizable difference, as plain VARs are often significantly overparameterized \citep{giannone_lenza_primiceri_2015}.

By demonstrating the TVP-RR equivalence, this paper contributes to the literature by clarifying further the fundamental tuning problem for this class of models. TVP models are standard high-dimensional regressions that require regularization.  When rewritten as a ridge regression, the overall level of regularization is determined by the well-known \(\lambda\).  Thus, the proposed solution to the tuning problem is intuitive: \(\lambda\) is determined using standard cross-validation techniques.  This measure is crucial for both predictive accuracy and economic analysis, making it reassuring that it can be tuned using the same methods that have been applied to ridge \(\lambda\) for decades. 
 Additionally, since ridge regression can equivalently be rewritten as a constrained optimization problem where the constraint is a "parameter budget constraint"—here translating to a "time-variation budget constraint"—we gain a new perspective on how the amount of time variation should be set.  That is, it should balance accommodating relevant time-variation in the data while addressing the reality of short estimation samples.

% balancing bias and variance to maximize out-of-sample predictive accuracy.

%

\vskip 0.2cm

{\sc \noindent \textbf{Interpretation Trinity: Ridge Regressions,  Splines,  and State Space Models.}} The TVP regression is closely related to the Hodrick-Prescott (HP) filter. HP filtering can be seen as a special case of the ridge approach to TVP with only an intercept as the regressor, and it penalizes double rather than first differences. In fact, \cite{schlicht2005estimating}, \cite{paige2010ridge}, and \cite{yamada2016hodrick} rewrite the HP filter as a ridge regression. This connection allows for more principled methods to determine \(\lambda\), which dictates the allocation between the cyclical and trend components. From this perspective, my work extends their contributions to a multiple time-series model with evolving volatility.

Furthermore, there appears to be an artificial division in the recent TVP literature between nonparametric and law-of-motion approaches. By demonstrating that random walk TVPs can be represented as a ridge regression, which is a special case of smoothing splines in this context, it becomes evident that random walk TVPs are just as nonparametric as those derived from the "nonparametric approach." Such equivalences are well-known in the context of local-level models \citep{kimeldorf1970, koop2003bayesian, durbin2012time}, which are TVP models with only an intercept.

%The are even more useful here for understanding the behavior of TVP models, especially in high dimensions.

 %Moreover, ridge regression implements a model nearly identical to classical TVP models and maintains the interpretation of the resulting coefficients as latent stochastic states. Preserving the parallel to a law of motion offers advantages, such as providing a clear prediction for tomorrow’s coefficients. 

%By construction, the unique ridge tuning parameter, in this context, \citep{golub1979gcv} mechanically corresponds to a ratio of two variances, that of parameter innovations and that of residuals. 
%\cite{wahba1978}
%\cite{kimeldorf1970}
 % Also, it has the flavor of \cite{hoover1998nonparametric} for time series rather than panel data.  (moothing splines)
  
%  \cite{durbin2012time}

Basis expansion and reparametrization arguments are crucial for establishing the "ridge link",  and have seldom been used in related applications.  For example, \cite{tibshirani2015statistical} evokes reparametrization as a possibility to estimate "fused" Lasso via plain Lasso, and \cite{korobilis2019high} incorporates related ideas as a core component of his "message-passing" algorithm. Additionally, the structure of the expanded matrix in 2SRR mirrors the "indicator-saturation" approach by \cite{castle2015detecting} for detecting location shifts in the intercept using model selection tools. With the growing use of machine learning methods in time series analysis, the basis expansion and reparametrization ideas developed in this paper are particularly relevant.  These concepts can be applied to easily induce random walk parameters in any regularized models, thereby achieving similar benefits beyond the realm of penalized regressions and those with explicitly specified regularization \citep{MDTM,HNN}.

%TVPs via RR first generalize the approach by interacting all individual regressors with a stray of shifting indicators. Then, rather than selecting one or few of them (sparsity), they are all kept in the model by constraining to incremental location shifts only via ridge shrinkage.

% and the HP filter, which can also be viewed as a local-level model.  

%

%\cite{schlicht2005estimating} also utilizes related derivations to provide an estimator for the ratio of trend and cycle variances.  

%\cite{schlicht2005estimating}
%\cite{durbin2012time}

%and \cite{MDTM} use derivations inspired by those below to implement regularized lag polynomials in  Machine Learning models.   

\vskip 0.2cm

{\sc \noindent \textbf{Extensions and Additional Features}.} Extending the two-step RR to a multistep one brings forth two interesting refinements that have been of interest in the literature.  Firstly, I consider \textit{Sparse} TVPs.  This means that some parameters will vary and some will not,  which can provide efficiency and interpretability gains. \cite{belmonte2014hierarchical}, \cite{korobilis2014data},  \cite{bitto2018achieving}, \cite{cadonna2020triple},  \cite{huber2021inducing},  and \cite{hauzenberger2024dynamic} have proposed such extensions to MCMC-based procedures.   In the Ridge paradigm,  similar shrinkage is obtainable by iterating 2SRR (thus,  a multistep RR) and continuously updating heterogeneous  parameter variance  estimates. 

Another natural way to discipline TVPs is to impose a factor structure.  This means that instead of trying to filter, say, 20 independent states, we can span these with a parsimonious set of latent factors.  This extension is considered in \cite{stevanovic2016common}, \cite{dewindgambetti2014} and \cite{chan-ftvp2018}.  Such reduced rank restrictions are brought to this paper’s arsenal using a different multistep RR that alternates between loading and a factor estimation steps.  

The paper also discusses how to tackle efficiently multivariate models,  and in particular,  how the same ridge apparatus can be used to obtain a time-varying covariance matrix,  a necessary input for structural TVP-VAR analysis.  Lastly,  a scalable Weighted Bayesian Bootstrap procedure (inspired from \cite{newton2018weighted} and \cite{ng2022random}) is proposed to quantify parameter uncertainty taking into account that of $\lambda$.

\vskip 0.2cm

{\sc \noindent \textbf{Simulations and Empirical Results}.} I first evaluate the method with simulations. For models of smaller size, where traditional Bayesian procedures can also be used, 2SRR does as well and sometimes better at recovering the true parameter path than the (Bayesian) TVP-VAR. This is true whether SV is involved or not. This is practical given that running \textit{and} tuning 2SRR for such small models (300 observations, 6 TVPs per equation) takes less than 5 seconds to compute.  Since large single equations will be at the heart of the forecasting and empirical application,  I also benchmark with a state-of-the-art single equation Bayesian method \citep{cadonna2020triple} that (i) optimizes the amount of shrinkage and (ii) can work in modestly higher dimensions with an acceptable computing time.  Again,  2SRR provides estimates of comparable quality at a fraction of the computational cost.  Additionally, I evaluate the performance of 3 variants of the RR approach in a substantive forecasting experiment.  2SRR and its iterated extension provide sizable gains for interest rates and inflation, two variables traditionally associated with the need for time variation. 

I complete with an application to estimating large time-varying LPs (more than 4\,600 TVPs) and equally demanding TVP-VARs in a Canadian context using \cite{champagnesekkel2018}'s narrative monetary policy (MP) shocks.  It is found that MP shocks' long-run impact on inflation increased substantially starting from the 1990s (onset of inflation targeting), whereas the effects on real activity indicators (GDP, unemployment) became milder.   To give a sense of the practicality of the method,  estimating the necessary elements to generate a 3D plot of all impulse response functions (IRFs) from a TVP-VAR with 8 variables and 24 (monthly) lags takes 1:34 minutes and requires very limited user input beyond specifying $Y$.   %,   with $8 \times 193$ conditional mean TVPs and 36 additional TVPs in the covariance matrix,  took 1:17 minute for the first step, and 17 seconds for the second.  

%\vskip 0.2cm

%{\sc \noindent \textbf{Intended Use}.} 

%However, as we will see, it is possible and can yield inference results roughly in line with fully-fledged Bayesian procedures.

%In fact, it sometimes increases it with respect to more sophisticated methods. This practicality and scalability make it a handy generic tool in the applied econometrician's arsenal,  for both forecasting and analysis.\footnote{For example, it was used in widely circulated economic analysis notes by Nancy Lazar,  Jake Oubina,  and Dave Wigglesworth at Piper Sandler on price (wage) inflation stickiness and embeddedness in August (September) 2022.}  What this approach does not do as readily as Bayesian methods, by construction, is provide inference. However, as we will see, it is possible and can yield inference results roughly in line with fully-fledged Bayesian procedures. Nonetheless, the paper focuses primarily on point estimates, whether in simulations, forecasting, or TVP-VAR applications. Additionally, it emphasizes various practical aspects of applying the method and the interesting econometric insights it yields. However, the paper does not offer any theory beyond the original link to the state-space representation and high-dimensional linear regression.

\vskip 0.2cm

{\sc \noindent \textbf{Outline}.}  Section  \ref{sec:2SRR} presents the ridge approach, its extensions, and related practical issues. Sections \ref{sec:simulations} and  \ref{sec:forecasting} report simulations and forecasting results, respectively.  Section \ref{sec:app} applies 2SRR to (large) time-varying LPs. Tables, additional graphs and technical details are in the Appendix.

\vskip 0.2cm

{\sc \noindent \textbf{Notation}.} $\beta_{t,k}$ refers  to the coefficient on regressor $X_k$ at time $t$. To make things lighter, $\beta_t \in {\rm I\!R}^K$ or $\beta_0 \in {\rm I\!R}^K$ always refers to {all} coefficients at time $t$ or time zero, respectively. Analogously, $\beta_k$ represents the whole time path for the coefficient on $X_k$. $\boldsymbol{\beta} \in {\rm I\!R}^{KT}$ is stacking all $\beta_k$'s one after the other, for $k=1,...,K$. All this also applies to $u$ and $\theta$.

\section{Time-Varying Parameters and Ridge Regressions}\label{sec:2SRR}

This covers the ridge approach, its extensions, and associated practical considerations.

\subsection{A Useful Observation}\label{sec:homovar}
Consider a generic linear model with random walk time-varying parameters 
\begin{subequations}
\label{eqn:model1}
\begin{align}
y_{t} &=X_{t}{\beta_{t}}+{\epsilon_{t}}, \quad {\epsilon}_{t}\sim{N(0,\sigma_{\epsilon_t}^{2}}) \\
{\beta_{t}} &={\beta_{t-1}}+{u}_{t}, \quad {u}_{t}\sim{N(0,\Omega_u})
\end{align}
\end{subequations}
where $\beta_t \in {\rm I\!R}^K$, $X_t' \in {\rm I\!R}^K$, $u_t \in {\rm I\!R}^p$ and both $y_t$ and $\epsilon_t$ are scalars. This paper first considers a general single equation time series model and then discuss its generalization to the multivariate case in Section \ref{MV}. For clarity, a single equation in a VAR with $M$ variables and $P$ lags has $K=PM+1$ parameters for each equation. For simplicity of exposition, I first impose $\Omega_u = \sigma_{u}^{2} I_K$ and $\sigma_{\epsilon_t}^{2}=\sigma_{\epsilon}^{2} \quad \forall t$. This means all parameters are assumed to vary equally a priori and constant variance of residuals. These assumptions will be relaxed in Section (\ref{gensol}). The textbook way of estimating \eqref{eqn:model1} is maximum likelihood using the Kalman filter for linear Gaussian model \citep{hamilton1994}. The advantages of the newly proposed methods will be more apparent when considering complications typically encountered in macroeconometric modeling (e.g. evolving volatility, heterogeneous variances for the coefficients paths, and a large $X_t$). %For expository purposes, it is nevertheless legitimate to begin with the model in its simplest form. 

A useful observation is that \eqref{eqn:model1}'s parameters can be equivalently obtained from solving the penalized regression problem
\begin{align}\label{eqn:model2}
\min_{\beta_{1}\dots\beta_{T}}\frac{1}{T}{\sum_{t=1}^{T}}\frac{\left(y_{t}-X_{t}{\beta_{t}}\right)^{2}}{\sigma^2_{\epsilon}}+\frac{1}{KT}{\sum_{t=1}^{T}}\frac{\Vert{\beta}_{t}-\beta_{t-1}\Vert^{2}}{{\sigma^2_{u}}}.
\end{align}
This is merely an implication of the well-known fact that $l_2$ regularization is equivalent to opting for a standard normal prior on the penalized quantity (see, for instance Sections 7.5-7.6 in \citealt{murphy2012}). Hence, model (\ref{eqn:model2}) implicitly poses ${\beta_{t}}-{\beta_{t-1}}\sim{N(0,\sigma_{u}^{2}})$, which is exactly what model (\ref{eqn:model1}) also does. 
Defining $\lambda \equiv \frac{\sigma^2_{\epsilon}}{\sigma^2_{u}}\frac{1}{K}$, the problem has the more familiar look of
\begin{align}\label{eqn:model2}
\min_{\beta_{1}\dots\beta_{T}}{\sum_{t=1}^{T}}\left(y_{t}-X_{t}{\beta_{t}}\right)^{2}+\lambda{\sum_{t=1}^{T}}\Vert{\beta}_{t}-\beta_{t-1}\Vert^{2}.
\end{align}
The sole hyperparameter of the model is $\lambda$ and it can be tuned by cross-validation.\footnote{This definition of $\lambda$ does not imply it decreases in $K$ since $\sigma^2_{u}$ will typically decrease with $K$ to avoid overfitting.} This model has a closed-form solution as an application of generalized ridge regression \citep{hastie2015statistical}. In particular, it can be seen as the $l_2$ norm version of the "Fused" Lasso of \cite{fusedlasso2005} and embeds the economic assumption that coefficients evolve slowly. However, as currently stated, solving directly \eqref{eqn:model2} may prove unfeasible even for models of medium size.   Note that the layout of the regularization apparatus implies that \textit{a priori}, \(u_{t}\) and \(\epsilon_{t}\) are serially and mutually uncorrelated.  However, estimation results may, for instance,  exhibit some persistence in \(u_{t}\), as regularization is a soft rather than a hard constraint.

\subsection{Getting a Ridge Regression by Reparametrization}

The goal of this subsection is to rewrite the problem \eqref{eqn:model2} as a ridge regression so it is easy and fast to implement in any software.  Doing so will prove useful at the conceptual level, but also to dramatically alleviate the computational burden.  From now on, it is less tedious to use matrix notation. The fused ridge problem reads as
	\begin{align*}
\min_{\boldsymbol{\beta}}\left(\boldsymbol{y}-\boldsymbol{W\beta}\right)'\left(\boldsymbol{y}-\boldsymbol{W\beta}\right)+\lambda \boldsymbol{\beta' D'D \beta}
\end{align*}
where $\boldsymbol{D}$ is the first difference operator.
 $\boldsymbol{W}=[diag(X_1) \quad \dots \quad diag(X_K)]$ is a $T \times KT$ matrix. To make matters more visual, the simple case of $K=2$ and $T=4$ gives rise to
$$
\boldsymbol{W}=
\begin{bmatrix}
    X_{11} & 0 & 0 & 0  &  X_{21} & 0 & 0 & 0 \\
    0 & X_{12} & 0 & 0  &  0 & X_{22} & 0 & 0 \\
    0 & 0 &  X_{13} & 0  & 0 & 0 & X_{23} & 0 \\
   0 & 0 & 0 &  X_{14}  &  0 & 0 & 0 & X_{24} \\

\end{bmatrix}.
$$
The first step is to reparametrize the problem by using the relationship $\beta_k = C \theta_k$ that we have for all $k$ regressors. $C$ is a lower triangular matrix of ones (for the random walk case) and I define ${\theta_k} = [{u_k} \quad {\beta_{0,k}}]$. For the simple case of one parameter and $T=4$:
\[
\begin{bmatrix}
    \beta_0  \\
    \beta_1 \\
    \beta_2 \\
   \beta_3
\end{bmatrix}
=
\begin{bmatrix}
    1 & 0 & 0 & 0 \\
    1 & 1 & 0 & 0 \\
    1 & 1 & 1 & 0 \\
   1 & 1 & 1 & 1
\end{bmatrix}
\begin{bmatrix}
    \beta_0  \\
    u_1 \\
    u_2 \\
   u_3
\end{bmatrix}
\]
For the general case of $K$ parameters, we have 
$$ \boldsymbol{\beta}= \boldsymbol{C} \boldsymbol{\theta}, \quad  \boldsymbol{C} \equiv I_K \otimes C $$
and $\boldsymbol{\theta}$ is just stacking all the $\theta_k$ into one long vector of length $KT$. Note that the summation matrix $C$ could accommodate easily for a wide range of law of motions just by changing summation weights. Actually, any process that can be rewritten (a priori) in terms of uncorrelated $u$'s could be used. For instance, AR models of arbitrary order and RW with drifts would be straightforward to implement.\footnote{In the latter case, it can be shown that one simply needs to add regressors $t*X_t$ to those implied by the RW without drift, that is the $Z_t$'s to be detailed later.} Furthermore, one could use $C^2$ in the random walk setup and obtain smooth second derivatives, e.i. a local-level model. While it is clear that many more exotic configurations are only a $C$ choice away, there is a clear advantage to random walks-based processes: the corresponding $C$ has no parameter to estimate. If we wanted to consider an AR(1) process with a coefficient $\phi \in (0,1]$, either a 2-step estimation procedure  or cross-validating $\phi$ would be necessary. Thus, the ridge approach is possible, whether $\beta_t$'s are random walks or not.

Using the reparametrization $\boldsymbol{\beta}= \boldsymbol{C} \boldsymbol{\theta}$, the fused ridge problem becomes
\begin{align*}
\min_{\boldsymbol{\theta}}\left(\boldsymbol{y}-\boldsymbol{WC \theta}\right)'\left(\boldsymbol{y}-\boldsymbol{WC\theta}\right)+\lambda \boldsymbol{\theta 'C'D'D C\theta}
\end{align*}
and it is now clear what should be done. Let $\boldsymbol{Z} \equiv \boldsymbol{WC}$ and use the fact that $\boldsymbol{D} = \boldsymbol{C}^{-1}$ to obtain the desired ridge regression problem
\begin{align}\label{eqn:basicridge}
\min_{\boldsymbol{\theta}}\left(\boldsymbol{y}-\boldsymbol{Z\theta}\right)'\left(\boldsymbol{y}-\boldsymbol{Z\theta}\right)+\lambda \boldsymbol{\theta'\theta}.
\end{align}
Again, for concreteness, the matrix $\boldsymbol{Z} = \boldsymbol{WC}$ looks like 
$$
\boldsymbol{Z}=
\begin{bmatrix}
    X_{11} & 0 & 0 & 0  &  X_{21} & 0 & 0 & 0 \\
     X_{12} & X_{12} & 0 & 0  &  X_{22} & X_{22} & 0 & 0 \\
     X_{13} &  X_{13} &  X_{13} & 0  & X_{23} & X_{23} & X_{23} & 0 \\
    X_{14} &  X_{14} &  X_{14} &  X_{14}  &  X_{24} & X_{24} & X_{24} & X_{24} 
\end{bmatrix}
$$
in the $K=2$ and $T=4$ case.  The solution to the original problem is thus
\begin{align}\label{hardbeta}
\boldsymbol{\hat{\beta}}=\boldsymbol{C\hat{\theta}}=\boldsymbol{C}(\boldsymbol{Z}'\boldsymbol{Z}+\lambda I_{KT} )^{-1}\boldsymbol{Z}'\boldsymbol{y}.
\end{align}	
This is really just a standard (very) high-dimensional Ridge regression.\footnote{In this Section, I assumed for simplicity that we wish to penalize equally each member of $\theta$ which is not the case in practice. It is easy to see why starting values $\beta_0$ should have different (smaller) penalty weights and this will be relaxed as a special case of the general solution presented in Section \ref{gensol}.} These derivations are helpful to understand TVPs, which is arguably one of the most popular nonlinearity in modern macroeconometrics. \eqref{hardbeta} is equivalent to that of a first-order (multivariate) smoothing splines estimator.  More generally, the equivalence between Bayesian stochastic process estimation and splines has been known since \cite{kimeldorf1970}. Following along, considering a local-level model for $\beta_t$ would yield second order smoothing splines. Clearly, random walk TVPs and their derivatives can hardly be described as "more parametric" than kernel-based approaches: splines methods are prominent within the nonparametric canon.  Furthermore,  the observation by itself can be leveraged to think about how inducing RW-like structural change in more modern ML tools,  like Random Forests \citep{MRFjae}.  

However,  we are not there yet,  since \eqref{hardbeta} implies inverting a $KT \times KT$ matrix,  a major computational bottleneck that would limit the applicability of the  technique to models of similar size to \cite{primiceri2005}.  Fortunately, there is no need to invert that matrix.

\subsection{Solving the Dual Problem}\label{easycase}

The goal of this subsection is to introduce a computationally tractable way of obtaining the ridge estimator $\boldsymbol{\hat{\beta}}$ in \eqref{hardbeta}. It is well known from the splines literature \citep{wahba1990spline} and later generalized by \cite{scholkopf2001generalized} that for a $\boldsymbol{\hat{\theta}}$ that solves problem (\ref{eqn:basicridge}), there exist a $\boldsymbol{\hat{\alpha}} \in {\rm I\!R}^T$ such that $\boldsymbol{\hat{\theta}}=\boldsymbol{Z'\hat{\alpha}}$.   While this may appear to be a constraint, it actually arises directly from the first-order conditions of the original problem, just expressed differently.  To see this,  note that the primal problem \eqref{eqn:basicridge} can be converted into the constrained minimization problem 
\begin{equation} \label{constrained}
\arg \min_{\boldsymbol{\theta},\boldsymbol{r}} \frac{1}{2} \left( \boldsymbol{r}' \boldsymbol{r} + \lambda \boldsymbol{\theta}' \boldsymbol{\theta} \right) \quad \text{subject to} \quad \boldsymbol{r} = \boldsymbol{Z}\boldsymbol{\theta} - \boldsymbol{y} 
\end{equation}
as observed in \cite{saunders1998ridge} and others.   Its Lagrangian is
\begin{equation} \label{lagrange}
L(\boldsymbol{\theta}, \boldsymbol{r}, \boldsymbol{a}) = \frac{1}{2} \boldsymbol{r}' \boldsymbol{r} + \frac{\lambda}{2} \boldsymbol{\theta}' \boldsymbol{\theta} + \boldsymbol{a}' (\boldsymbol{r} - \boldsymbol{Z}\boldsymbol{\theta} + \boldsymbol{y}) \,
\end{equation}
where $\boldsymbol{a} \in {\rm I\!R}^T$ is a vector of Lagrange multipliers.  Setting derivatives with respect to the primal variables $(\boldsymbol{\theta}, \boldsymbol{r})$ to zero, we obtain from first order conditions that the solution should satisfy $\boldsymbol{\theta} = \frac{1}{\lambda} \boldsymbol{Z}' \boldsymbol{a} \quad \text{and} \quad \boldsymbol{r} = -\boldsymbol{a}$.  Making these substitutions to eliminate \(\boldsymbol{r}\) and \(\boldsymbol{\theta}\) gives the dual problem 
\begin{equation} \label{dual}
\arg \min_{\boldsymbol{a}} -\frac{1}{2} \boldsymbol{a}' \boldsymbol{a} - \frac{1}{2\lambda} (\boldsymbol{Z} \boldsymbol{a})' (\boldsymbol{Z} \boldsymbol{a}) + \boldsymbol{a}' \boldsymbol{y}
\end{equation}
where everything is expressed in terms of \(\boldsymbol{a}\) rather than \(\boldsymbol{\theta}\).  We can also reparametrize, defining \(\boldsymbol{\alpha} = \frac{1}{\lambda} \boldsymbol{a}\), and use directly the knowledge about the dual solution (\( \boldsymbol{\theta}= \boldsymbol{Z}' \boldsymbol{\alpha}\)) in the primal problem \eqref{eqn:basicridge} to obtain
\begin{align*}
\min_{\boldsymbol{\alpha}}\left(\boldsymbol{y}-\boldsymbol{ZZ'\alpha}\right)'\left(\boldsymbol{y}-\boldsymbol{ZZ'\alpha}\right)+\lambda \boldsymbol{\alpha'ZZ'\alpha}.
\end{align*}
The solution to the original problem becomes
\begin{align}\label{dualsolhomo}
\boldsymbol{\hat{\beta}}=\boldsymbol{CZ'\hat{\alpha}}= \boldsymbol{CZ'}(\boldsymbol{ZZ'}+\lambda\boldsymbol{I}_T)^{-1}\boldsymbol{y}.
\end{align}
When the number of observations is smaller than the number of regressors, the \textit{dual} problem allows to obtain numerically identical estimates by inverting a smaller matrix of size $T$. Since sample sizes for macroeconomic applications quite rarely exceed ~700 observations (US monthly data from the 1960s), the need to invert that matrix is not prohibitive. While computational burden does still increase with $K$, it increases much slower since the complexity of matrix multiplication is now $O(KT^3)$ and $O(T^3)$ for matrix inversion. Solving the primal problem, one would be facing $O(K^2T^3)$ and $O(K^3T^3)$ complexities respectively. Concretely, solving the dual problem makes the estimation of high-dimensional TVP models significantly more practical because the computations scale \textit{linearly} with \(K\) for a fixed \(T\).

%Estimating a small TVP-VAR with $T$=300 with 6 lags and 5 variables takes roughly 10 seconds on a standard computer. This includes hyperparameters optimization by cross-validation, but excluded inference.   

%However, the latter provides full Bayesian inference. A VAR(20) with the same configuration takes less than 2 minutes. Section \ref{sec:simulations} reports detailed results on this.

\subsection{Heterogenous Parameter and Residual Variances}\label{gensol}

For pedagogical purposes, previous sections considered the simpler case of $\Omega_u = \sigma_u^2 I_K$ and no evolving volatility of residuals. I now generalize the solution \eqref{dualsolhomo} to allow for heterogeneous $\sigma_{u_k}^2$ (a diagonal $\Omega_u \neq \sigma_u^2 I_K$) and $\sigma_{\epsilon,t}^2$. The end product is 2SRR, this paper's flagship model.

New matrices must be introduced. First, we have the standard matrix of time-varying residuals variance $\Omega_{\epsilon}=diag( [\sigma_{\epsilon_1}^2 \quad \sigma_{\epsilon_2}^2 \quad ... \quad \sigma_{\epsilon_T}^2]).$
I assume in this section that both $\Omega_{\epsilon}$ and $\Omega_{u}$ are given and will provide a data-driven way to obtain them later. Departing from the homogeneous parameter variances assumption implies that  the sole hyperparameter $\lambda$ must now be replaced by an enormous $KT \times KT$ diagonal matrix 
$\boldsymbol{\Omega_u}=\Omega_u \otimes I_T$
which is fortunately only used for mathematical derivations. For convenience, I split $\boldsymbol{Z}$ in two parts so they can be penalized differently. Hence, the original $\boldsymbol{Z} \equiv [X \quad \boldsymbol{Z_{-0}}]$. The new primal problem is
\begin{align}\label{genprobtvp}
\min_{\boldsymbol{u},\beta_0}\left(\boldsymbol{y}-\boldsymbol{X}\beta_0-\boldsymbol{Z_{-0}u}\right)'\Omega_{\epsilon}^{-1} \left(\boldsymbol{y}-\boldsymbol{X}\beta_0-\boldsymbol{Z_{-0}u}\right)+ \boldsymbol{u'} \boldsymbol{\Omega_{u}^{-1}} \boldsymbol{u} + \lambda_0 \beta_0' \beta_0.
\end{align}
For convenience, let the $\boldsymbol{\Omega_\theta}$ matrix that stacks on the diagonal all the parameters prior variances, which allow rewriting the problem in a more compact form 
\begin{align*}
\min_{\boldsymbol{\theta}}\left(\boldsymbol{y}-\boldsymbol{Z\theta}\right)'\Omega_{\epsilon}^{-1} \left(\boldsymbol{y}-\boldsymbol{Z\theta}\right)+ \boldsymbol{\theta'} \boldsymbol{\Omega_{\theta}^{-1}} \boldsymbol{\theta}.
\end{align*}
Using a GLS re-weighting scheme on observations \textbf{and} regressors, we get a "new" standard primal ridge problem
	 	\begin{align*}
\min_{\boldsymbol{\tilde{\theta}}}\left(\boldsymbol{\tilde{y}}-\boldsymbol{\tilde{Z}\tilde{\theta}}\right)' \left(\boldsymbol{\tilde{y}}-\boldsymbol{\tilde{Z}\tilde{\theta}}\right)+ \boldsymbol{\tilde{\theta}'} \boldsymbol{\tilde{\theta}}.
\end{align*}
where $\boldsymbol{\tilde{\theta}}= \boldsymbol{\Omega^{-\frac{1}{2}}_{\theta}}\boldsymbol{{u}}$, $\boldsymbol{\tilde{Z}}=\Omega^{-\frac{1}{2}}_{\epsilon}\boldsymbol{Z} \boldsymbol{\Omega^{\frac{1}{2}}_{\theta}}$ and $\boldsymbol{\tilde{y}}=\Omega^{-\frac{1}{2}}_{\epsilon} \boldsymbol{y}$. Solving this problem by the "dual path" and rewriting it in terms of original matrices gives the general formula
\begin{align}\label{2sdualsol}
\hat{\boldsymbol{\theta}} = \boldsymbol{\Omega_{\theta}} \boldsymbol{Z'} (\boldsymbol{Z} \boldsymbol{\Omega_{\theta}} \boldsymbol{Z'} +\Omega_{\epsilon})^{-1}\boldsymbol{y}.
\end{align}
Equation \eqref{2sdualsol} contains all the relevant information to back out the parameters paths, provided some estimates of matrices $ \boldsymbol{\Omega_{\theta}}$ and $ \Omega_{\epsilon}$.\footnote{This two-step procedure is partly reminiscent of \cite{ito2014GLS} and \cite{ito2017alternative}'s non-Bayesian Generalized Least Squares (GLS) estimator, where a two-step strategy is also proposed for reasons similar to the above. Their approach could have a ridge regression interpretation with certain tuning parameters fixed. However, the absence of tuning leads to overfitting and the GLS view cannot handle bigger models because the implied matrices sizes are even worse than that of the \textit{primal} ridge problem discussed earlier.}

%We do not have these matrices. Hence, we must look into a way to obtain them. 

\subsubsection{Implementation}

The solution \eqref{2sdualsol} takes $ \boldsymbol{\Omega_{\theta}}$ and $ \Omega_{\epsilon}$ as given.  In this section, I provide a simple adaptive algorithm to get the heterogeneous variances model estimates empirically. Multi-step approaches to obtain the obtain analogs of $ \boldsymbol{\Omega_{\theta}}$ and $ \Omega_{\epsilon}$ have been proposed in \cite{ito2017alternative} and \cite{giraitis2014inference}.  Algorithm \ref{2srralgo} follows along and describes a two-step ridge regression (2SRR) which uses a first stage plain RR to gather the necessary hyperparameters.  % in one swift blow. 

\begin{algorithm}
\caption{2SRR \label{2srralgo}}
\begin{algorithmic}[1]
  %\footnotesize
  \small
  \STATE Use the homogeneous variances approximation. That is, get $\boldsymbol{\hat{\theta}_1}$ with (\ref{dualsolhomo}).  $\lambda$ is obtained by CV.
\STATE Obtain ${{\hat{\sigma}_{\epsilon,t}^2}}$ by fitting a volatility model to the residuals from step 1.  Normalize ${{\hat{\sigma}_{\epsilon,t}^2}}$'s mean to 1.
\STATE Obtain ${{\hat{\sigma}_{u,k}^2}}=\frac{1}{T}{\sum_{\tau=1}^{T}}\hat{u}_{k,\tau}^{2}$ for $k=1,...,K$. Normalize the new vector to have its previous mean ($\sfrac{1}{\lambda}$).
\STATE Stack these into matrices $\boldsymbol{\Omega_{u}}$ and $\Omega_{\epsilon}$.  Use solution (\ref{2sdualsol}) to rerun CV and get $\boldsymbol{\hat{\theta}_2}$, the final estimator.
\end{algorithmic}
\end{algorithm}

Here are some necessary details.  In step 2,  the prior variances for "starting values" $\beta_0$ can either be fixed of cross-validated -- which will be important for bigger models.  The prior mean can be zero or the OLS solution as in \cite{primiceri2005}.  Regarding step 3,  I use GARCH(1,1) as the volatility model.  Better (and faster) alternatives are available for multivariate cases where we may also want to model time-varying covariances.  Those are discussed in Section \ref{MV}.  Lastly,  the optimal $\lambda$ may change when $\Omega_{u}$'s entries are heterogeneous, justifying a second CV run in step 4.  %In practice,which has been found to help in practice,  albeit marginally.

2SRR (and eventually MSRR$_{\text{S}}$ in Section  \ref{cosso})  makes use of adaptive (or data-driven) shrinkage. Adaptive prior tuning has a long tradition in Bayesian hierarchical modeling \citep{murphy2012} but the term itself came to be associated with the Adaptive Lasso of \cite{zou2006}. To modulate the penalty's strength in Lasso, the latter suggest weights based on preliminary OLS (or Ridge) estimates. Those, taken as given, may be contaminated with a considerable amount of noise, especially when the regression problem is high-dimensional (like the one considered here).  Adaptive weights in 2SRR have a natural  group structure which substantially improves their accuracy through averaging.

\subsection{Choosing $\lambda$}\label{sec:cv}

Derivations from previous sections rely on a given $\lambda$. This section explains how to obtain the amount of time variation by CV (as alluded to in Algorithm \ref{2srralgo}), and how that new strategy compares to more traditional approaches to the problem.

\vskip 0.2cm

{\sc \noindent \textbf{Cross-Validating $\lambda$}.} I use k-fold CV for convenience, but anything could be used -- conditional on some amount of thinking about how to make it computationally tractable. This is also what standard RR implementations use, like \texttt{glmnet} in \texttt{R}. A concern is that k-fold CV might be overoptimistic with time series data. Fortunately, \cite{bergmeir2018note} show that without residual autocorrelation, k-fold CV is consistent. Assuming models under consideration include enough lags of $y_t$, this condition will be satisfied for one-step ahead forecasts. Moreover, \cite{GCLSS2018} report that macroeconomic forecasting performance can often be improved by using k-fold CV rather than a CV procedure that mimics the recursive pseudo-out-of-sample experiment.

\vskip 0.2cm

{\sc \noindent \textbf{Time-Variation Budget Interpretation}.} Like any ridge regression problem, \eqref{eqn:basicridge} can be written as a minimization problem with a "parameter budget" constraint \citep{hastie2015statistical}.  In our context,  the parameter budget has a specific meaning as "Time-Variation Budget"  given  \(\boldsymbol{\theta}\)'s content.  Precisely,  we have
\begin{align}\label{basicridge2}
\min_{\boldsymbol{\theta}} \left(\boldsymbol{y} - \boldsymbol{Z\theta}\right)' \left(\boldsymbol{y} - \boldsymbol{Z\theta}\right) \quad \text{subject to} \quad \boldsymbol{\theta}'\boldsymbol{\theta} \leq \text{TVB}
\end{align}
where TVB stands for the "Time-Variation Budget",  a constant corresponding to the maximal total magnitude of \(\boldsymbol{\theta}\).  Technically (but less so practically), one could tune TVB instead of $\lambda$ and achieve identical results.  Still, this representation is useful to understand how the amount of time-variation is set in the 2SRR context using cross-validation techniques. It asks the question: given a specific dataset finite sample limitations, how much time-variation can we afford? The CV answer is: as much as out-of-sample performance in predicting $y_t$ does not degrade from overly wiggly parameter path estimates. 

This representation also helps in understanding the hypothesized behavior of \(\lambda\) as a function of \(T\) and \(K\). In a standard ridge context, where the number of regressors is fixed as the sample size increases, \(\lambda \rightarrow 0\) as \(T\) grows, and we return to the OLS solution. This will not happen in the TVP setup since the effective number of regressors is \(KT\), which grows as fast as \(T\). Regarding \(K\), \(\lambda\) will tend to increase with it, simply because a larger model requires more regularization. This is a feature of the model, not the ridge estimation strategy, and the representation in \eqref{basicridge2} clarifies that,  for a fixed TVB,  the total amount of time-variation must now be shared across more predictors as \(K\) increases.  This results in less time-variation allocated to each coefficient individually.

\vskip 0.2cm

{\sc \noindent \textbf{Comparison To Bayesian Approaches}.} In the Bayesian TVP-VAR literature, it is common to use a sparse grid of values for the hyperparameters of the smoothing priors, inspired by \cite{primiceri2005}. Primiceri employed a reversible jump MCMC algorithm where, in an initial step, the model must be estimated conditional on every possible combination of hyperparameters from a grid. However, as noted in \cite{amir2018choosing}, it is unclear whether this grid is appropriate for applications beyond those originally considered by Primiceri, and simply using wider grids introduces significant computational challenges. Consequently, \cite{amir2018choosing} and \cite{cadonna2020triple} propose estimating the hyperparameters in the priors of smoothing parameters through a hierarchical structure within the entire Bayesian procedure, finding that this approach can significantly improve forecasting results. 

At a conceptual level, cross-validating \(\lambda\) in the ridge approach aligns with recent Bayesian advances.  That is,  hyperparameters ultimately influencing the smoothness of coefficients should ideally be obtained in a data-driven manner, and allowing for a reasonably wide range of possible values.  However, my approach estimates \(\lambda\) directly (resulting in a single value), whereas \cite{amir2018choosing} estimate the hyperparameters of the prior for parameters that play an equivalent role to that of  \(\lambda\) in their framework.

%At a very general level,  cross-validating \(\lambda\) in the ridge approach is in agreement with recent Bayesian advances,  in the sense  that hyperparameters that have an impact (directly or indirectly) on the resulting smoothness of coefficients  should preferably be obtained in a  data-driven way, and allowing for reasonably wide ranger of possible values.

\vskip 0.2cm

{\sc \noindent \textbf{Comparison to Classical Approaches}.} Within the older literature where TVPs were obtained via classical methods, estimating the parameters variances ($\sigma_u^2$ and $\sigma_{\epsilon}^2$ in my notation) made plain MLE's life particularly difficult \citep{stock1998median,boivin2005has}.   One of such issues is the "pile-up" problem,  which can occur when a variance parameter is estimated to be 0.    In the Ridge Regression (RR) paradigm, where parameter variances are expressed through \(\lambda\), it becomes clear why some of these issues emerged: directly maximizing the \textit{in-sample} likelihood to determine ridge's \(\lambda\) is not a common practice.  Seeing the TVP problem as a high-dimensional regression,  one avoids the "pile-up" problem inherent to \textit{in-sample} maximum likelihood estimation \citep{grant2017bayesian} by estimating it separately on a hold-out sample via cross-validation.  

It is natural to ask how we should think of \(\lambda\) obtained from cross-validation compared to estimating \(\sigma_u\) and \(\sigma_{\epsilon}\) jointly with \(\boldsymbol{\theta}\) using (well-behaved) maximum likelihood estimation. This approach is standard for classical state-space estimation of more compact models,  such as local-level models.  In that context,  for the same ratio of variances, the smoothing splines (or ridge) version of the problem yields the same result as the filtering approach \citep{durbin2012time}.  In the context of HP filtering, which can be seen as a special case of the ridge approach with only an intercept and penalizing double rather than first differences, \cite{paige2010ridge} also discusses that setting \(\lambda\) to the true ratio of variances \(\frac{\sigma_{\epsilon}^2}{\sigma_{u}^2}\) delivers the best linear \textit{unbiased} predictor.  Possible ways to estimate \(\lambda\) in this context are Generalized CV \citep{golub1979gcv} or Restricted MLE.   However,  for TVP models,  where \(K\) is often much larger than 1, it is evident that CV may select \(\lambda\) 's yielding less time-variation than MLE would command (thus,  not unbiased). This occurs to maximize out-of-sample prediction accuracy and win at the bias-variance trade-off, simply because \(K\) may be large and \(T\) is small.

\subsection{From Univariate to Multivariate}\label{MV}

So far,  derivations have focused on the univariate case.   This section details the modifications necessary for a multivariate 2SRR and covers de facto  the canonical case of a TVP-VAR and also that of estimating  local projections over a large horizons range.   These will be the two applications considered in Section \ref{sec:app}.

\vskip 0.2cm

{\sc \noindent \textbf{Tackling Many Conditional Means Efficiently  }.}  Since both $\Omega_{u}$ and $\Omega_{\epsilon}$ are equation-specific, we must use (\ref{2sdualsol}) for each $\boldsymbol{y}$. However, all estimation procedures proposed in this paper have the homogeneous case of Section \ref{easycase} as a first step. This is usually the longer step since it is where cross-validation is done.  Hence, it is particularly desirable not to have computations of the first step scaling up in $M$, the number of variables in the multivariate system. Thankfully, in the plain ridge case, we can obtain all parameters of the system in one swift blow, by stacking all $\boldsymbol{y}$'s into $\boldsymbol{Y}$ (a $T \times M$ matrix) and computing
\begin{align}
\boldsymbol{\hat{\Theta}}=\boldsymbol{Z'}(\boldsymbol{ZZ'}+\lambda\boldsymbol{I}_T)^{-1}\boldsymbol{Y} %=\boldsymbol{P}^\lambda_Z \boldsymbol{Y} .
\end{align}
This is precisely the approach that will be used as a first step for any multivariate extension. Of course, this works because the multivariate model has the same regressor matrix for each equation (like VARs and LPs).  In this homogeneous variance case,  $\boldsymbol{Z'}(\boldsymbol{ZZ'}+\lambda\boldsymbol{I}_T)^{-1}$ is the same for all equations and cross-validation still implies inverting $(\boldsymbol{ZZ'}+\lambda\boldsymbol{I}_T)$ as many times as we have candidates for $\lambda$. That is, even if we wish to have a different $\lambda$ for each equation in the first step, computing time does not increase in  $M$, except for matrix multiplication operations which are much less demanding.\footnote{Precisely, cross-validation implies calculating \# of folds $\times$ \# of $\lambda$'s  the $\boldsymbol{P}^\lambda_Z$. Then, these matrices can be used for the tuning of every $m$ equation, which is precisely why the computational burden only very mildly increases in $M$.} 

\vskip 0.2cm

{\sc \noindent \textbf{Time-Varying Covariance Matrix}.}  When entering multivariate territory, modeling the residuals covariances -- a necessary input to structural TVP-VAR analysis (but not point forecasting) -- arises as an additional task.  The number of TVPs entering the $\boldsymbol{\Omega_\epsilon}$ matrix grows quickly with $M$ as there are $\frac{M(1+M)}{2}$ of them.  This is the other key part of the challenges facing traditional Bayesian TVP-VARs computations.  So far,  we have seen how the ridge approach can help in computing the various conditional means efficiently.  The following discusses how it can also be applied to the time-varying covariance matrix. 

I  now describe how similar ideas as above -- applying the same smoothing matrix to many series at once -- can also be applied  so that each parameter of the covariance matrix follows a random walk.  First, let us denote $\boldsymbol{\epsilon_t} \in {\rm I\!R}^M$ as the residuals obtained from running multivariate 2SRR with $\boldsymbol{Y}$ as dependent variables.  Let $\tilde{\eta}_t=vech(\boldsymbol{\epsilon_t}\boldsymbol{\epsilon_t}')$ where the  $vech$ operator means that we vectorize but only keep lower-diagonal elements (i.e.,  only the non-redundant entries).   Also,  denote $\bar{\eta}_t$ to be a vector corresponding to the full-sample average of the columns of $\tilde{\eta}_t$,  or, in other words, a "time series" of the lower-diagonal elements of what would be the time-invariant covariance matrix for those residuals.   To be more concrete,  in the $M=2$ case,   we have $\tilde{\eta}_t = [ \epsilon_{1,t}^2 \enskip  \epsilon_{2,t}^2 \enskip \epsilon_{1,t}\epsilon_{2,t}  ]$ and $\bar{\eta}_t = [ \frac{1}{T}\sum_{\tau=1}^{T}\epsilon_{1,\tau}^2 \enskip \frac{1}{T}\sum_{\tau=1}^{T} \epsilon_{2,\tau}^2 \enskip \frac{1}{T}\sum_{\tau=1}^{T} \epsilon_{1,\tau}\epsilon_{2,\tau}  ]$.   Finally,  let $\tilde{\boldsymbol{\eta}}$ be the $T \times \frac{M(1+M)}{2}$ matrix stacking those $\frac{M(1+M)}{2}$ time series of residuals products together and $\bar{\boldsymbol{\eta}}$ doing an analogous stacking for $\bar{\eta}_t $.  Treating those as observed data,  we can get the whole set of paths $\hat{\boldsymbol{\eta}}$ by writing a multivariate TVP-ridge problem as the above,  but with $\frac{M(1+M)}{2}$ targets and a conditional mean model that is only a time-varying intercept.  Precisely,  we can compute
\begin{align}\label{varcov_est}
\hat{{\boldsymbol{\boldsymbol{\eta}}}}=(\boldsymbol{I}_T+\varphi\boldsymbol{D'D})^{-1}(\tilde{\boldsymbol{\eta}}-\bar{\boldsymbol{\eta}})+\bar{\boldsymbol{\eta}}
\end{align}
where $\varphi$ is a smoothness hyperparameter (just like $\lambda$) and $\boldsymbol{D}$ is the matrix difference operator described in Section \ref{sec:2SRR}.  This implies,  again,  that each column of the $\tilde{\boldsymbol{\eta}}$ matrix is assumed to follow a random walk around its mean value with a velocity inversely linked to $\varphi$.  Clearly,  large $\varphi$'s push the solution towards a time-invariant covariance matrix of residuals ($\bar{\boldsymbol{\eta}}$)  and extremely low values will make it overly wiggly.\footnote{I also note that overly small values of $\varphi$ could threaten the positive-definiteness (and thus validity) of the resulting matrix because,  intuitively,  some minimal form of averaging is necessary to obtain covariances.  This problem has not been encountered in practice because results become implausibly wiggly well before positive-definiteness  breaks down.  This also highlights a homogeneous $\varphi$ is necessary here,  elements of the same matrix averaged differently at different points in time could create momentousness incoherency  problems when putting things back together. } Note that without the pre-demeaning of the $\tilde{\eta}_t$ series (which the use of $\bar{\boldsymbol{\eta}}$ is the formal mathematical characterization of) and adding back the mean is only so that shrinkage pushes towards the solution of time-invariant covariance matrix rather than 0.   Computations-wise,  for a fixed $\varphi$, this amounts to one matrix $T \times T$  inversion and then multiplying it with $\tilde{\boldsymbol{\eta}}$, with both operations fast for $T$'s encountered in macroeconomics. 

In applications,  I opt for a user-specified $\varphi$.  Unlike the conditional mean targets,  covariances in such applications are fundamentally unobserved,  and the elements of $\tilde{\boldsymbol{\eta}}$ in themselves cannot be regarded as realized variances or covariances.  Hence,  it is not entirely clear that the target one would be cross-validating for is a proper target at  all.  With some ingenuity (left for future work\footnote{One possibility would be  to cross-validate $\lambda$ and $\varphi$ together and use multivariate out-of-fold log scores as metric.}) on how to create a proper scoring metric in this case,  CV could be conducted very fast because it only implies inverting as many $T \times T$ matrices as one has $\varphi$'s,  which by all reasonable means,  should be below 50. 

The algorithm proposed in this subsection provides a covariance matrix at each $t$.  This sequence of matrices can then be used to conduct structural TVP-VAR analysis like in Section \ref{sec:app} where a Cholesky decomposition is applied at each $t$.  Evidently,  this is simplest thing one can do,  but as long as we have the sequence of covariance matrices, there are quite a few more sophisticated things one could opt for.

Finally,  two important notes for the practical use of the tools above.  First,  it is best practice to add a sufficient number dummy observations duplicating the boundary value at the beginning and end of $\tilde{\boldsymbol{\eta}}$.\footnote{I include 30 such duplicates in the applications.} This is to avoid that the $\boldsymbol{D'D}$ matrix,  in absentia of observations before 1 and after $T$,  overly shrinks $(\tilde{\boldsymbol{\eta}}-\bar{\boldsymbol{\eta}})$ to 0, and thus in our case, mechanically forcing the top and bottom rows of $\hat{{\boldsymbol{\boldsymbol{\eta}}}}$ towards  $\bar{\boldsymbol{\eta}}$.  This remedy is obviously very much related to a similar tactic employed to deal with boundary problems in kernel regressions.  

Second,  and far more subtle,  is the case of the "ridgeless" solution.   For very large $\frac{K}{T}$ ratios,   it has been observed that cross-validation will sometimes pick $\lambda \approx 0$, which is the so-called ridgeless regression of \citep{hastie2019surprises}.  This is a reflection of the double descent phenomenon in extremely high-dimensional models,  which is ubiquitous in modern deep learning \citep{belkin2019reconciling,bartlett2020benign}.  In a nutshell,  the unregularized model might provide a more regularized (and better) solution \textit{out-of-sample} since, with $\lambda  \rightarrow 0$,  the solution converges to the minimum-norm solution because the inverse of the covariance matrix is Moore-Penrose pseudo-inverse (see \cite{kellyvirtue} for a more detailed discussion applied to finance).  However,  this "benign overfitting"  solution (because forecasting is not impacted) provides perfect fit \textit{in-sample}.  In that scenario,  there are no residuals,  no covariance matrix,  and $\beta_t$ certainly overfit,  rendering historical analysis through IRFs unfeasible.  Fortunately,  there is an easy mechanical fix,  which is limiting the range of $\lambda$ to solutions that exists the classical bias-variance regime, rather than the so-called double descent regime.  In many applications,  it is found that there is a solution in the "usable" regime that is almost as good (in terms of cross-validated MSE) as the ridgeless solution \citep{hastie2019surprises}.

\subsection{Quantifying Uncertainty}\label{sec:uq}
Clearly,  2SRR advantage is in getting point estimates rapidly,  which already goes a long way for many of TVP models applications.  Yet,  there are many others where quantifying  $\boldsymbol{\beta}$'s uncertainty  is instrumental.  In principle,  this would be possible for 2SRR by leveraging the link between ridge and a plain Bayesian regression \citep{murphy2012}.  In the general case,  one would need to obtain
	\begin{align}\label{varform}
	V_{\boldsymbol{\beta}} = \boldsymbol{C}(\boldsymbol{Z'}\Omega^{-1}_{\epsilon} \boldsymbol{Z}+ \boldsymbol{\Omega^{-1}_{\theta}})^{-1}\boldsymbol{C'}
	\end{align}
which is precisely the large matrix we were avoiding to invert earlier.   While this is not an issue for smaller models,  it can become one in medium-sized models because it relies on the primal solution.  Also,  note that  in the simple case where $ \Omega_{u}=\sigma_u^2 I_K$ and $ \Omega_{\epsilon}=\sigma_\epsilon^2 \ I_T$, there is a clear Bayesian interpretation allowing the use of the posterior variance formula for linear Bayesian regression.  However, it treats the cross-validated $\lambda$ as known.  In a similar line of thought,  it treats the hyperparameters inherent to 2SRR as given when computing the bands.  While in traditional (modestly dimensional) applications of ridge regression,  this could be regarded as a minor impediment,  it is not the case here because,  in very high-dimensional models,  a small change in regularization structure can lead to rather different coefficients estimates.   Therefore,  it is important to account not only for uncertainty conditional on hyperparameters,  but also that which emanate directly from them.

\vskip 0.2cm

{\sc \noindent \textbf{Weighted Bayesian Bootstrap}.} I provide a way for users to quantify TVP uncertainty that circumvents both of the aforementioned issues,  namely,  it will not need to invert a $KT \times KT$ matrix and will account for regularization uncertainty.  I adapt the Weighted Bayesian Bootstrap (WBB) of \cite{newton2018weighted},  which is an extension of \cite{newton1994approximate}'s ideas to general machine learning models with regularization.\footnote{More closely related to this paper's framework,    \cite{newton2018weighted} consider a "trend filtering" example,   which is a TVP model with only a intercept where the second difference is penalized with a $l_1$ norm. } In short,  randomization via exponential draws is repeatedly applied to observations \textit{and} regularization weights,  thereby accounting for $\lambda$ and overall regularization uncertainty.  This approximate Bayesian inference technique is increasingly popular for scalable inference with machine learning methods or Bayesian models with demanding posterior computations \citep{nie2022bayesian}.   Clearly,  there are many values of $K$ and $T$ for which \eqref{varform} qualifies as such.

Implementation is convenient,   quick,  and is easily parallelizable if need be.   It implies looping over ridge estimation (without CV and only using the \textit{dual} solution \eqref{2sdualsol}), which is rapid and amenable also to extensions discussed in the following section.  As an introductory example,  applying WBB in its plainest form to the simplest TVP formulation found in equation  \eqref{eqn:model2} implies repeatedly estimating 
\begin{align}\label{eqn:model2_wbb}
\min_{\beta_{1}\dots\beta_{T}}{\sum_{t=1}^{T}}\omega_t \left(y_{t}-X_{t}{\beta_{t}}\right)^{2}+\omega_{\lambda} \lambda_{\text{CV}}{\sum_{t=1}^{T}}\Vert{\beta}_{t}-\beta_{t-1}\Vert^{2}
\end{align}
 with $\omega_t \sim \text{Exp}(1)$,  $\omega_{\lambda} \sim \text{Exp}(1)$,  and where $\lambda_{\text{CV}}$ stands for $\lambda$ as estimated from running cross-validation on the unweighted problem.    As discussed in \cite{ng2022random},  it may be worthwhile in high-dimensional models to have a distinct $\omega_{\lambda}$ for the regularization strength of each predictor.   This therefore account not only for overall regularization strength uncertainty,  but that of \textit{relative} regularization,  i.e.,  how heavily certain regressors' coefficients are shrunk versus others.  Clearly,  this is relevant in the 2SRR context because it explicitly contains regularization heterogeneity via $\Omega_{u}$ and $\Omega_{\epsilon}$ estimated with from a homogeneous RR first step.  Therefore,  what we want is something along the lines of a $\omega_{\lambda,k}$.  Accordingly,  the WBB procedure proposed to accompany 2SRR (and tested via simulations in Section \ref{sec:simulations}) is to divide each diagonal element of the two variance matrices (${\Omega_{\theta}}$ and $\Omega_{\epsilon}$ in equation \eqref{genprobtvp})  by a $\text{Exp}(1)$ draw,  and give $\lambda_0$ its own $\omega_0$.\footnote{In practice, some trimming of the draws is necessary to avoid anything close to dividing by 0,  and thus multiplying variances by very large values.} Then, one can use the draws 
\begin{align}\label{2sdualsol}
\hat{\boldsymbol{\theta}}_{\boldsymbol{b}} = \boldsymbol{\Omega_{\theta}^b} \boldsymbol{Z'} (\boldsymbol{Z} \boldsymbol{\Omega_{\theta}^b} \boldsymbol{Z'} +\Omega_{\epsilon}^{\boldsymbol{b}})^{-1}\boldsymbol{y}.
\end{align}
where $\boldsymbol{\Omega_{\theta}^b}$ and $\Omega_{\epsilon}^{\boldsymbol{b}}$ denote the modulated matrices to calculate percentiles and other relevant metrics.  Note that by the virtue of this being an approximate Bayesian method,  it provides no frequentist coverage guarantees (along with most Bayesian methods).  Nonetheless,   to get a sense how credible regions constructed from a WBB approach behave in comparison a full-fleshed Bayesian procedure, I investigate nominal coverage in Section \ref{sec:simulations} and find they are comparable.\footnote{One could further refine the above by considering a blocked Bayesian Bootstrap approach for the elements of $\Omega_{\epsilon}^{\boldsymbol{b}}$ (as considered for other kinds of ML algorithms in \cite{MRFjae} and \cite{HNN}),  but in the conducted experiments (which includes persistent targets and regressors), this did not seem to play an appreciable role.  Such a modification is easily implementable by drawing exponentials in groups. } {Note that in the context of VAR modeling,  we are typically interested in making inference on impulse responses,  which are nonlinear transformation of $\boldsymbol{\beta}$.  In that scenario,  $\hat{\boldsymbol{\theta}}_{\boldsymbol{b}}$ draws can be transformed in IRFs draws for each ${\boldsymbol{b}}$.

 \vskip 0.2cm

{\sc \noindent \textbf{Other Avenues for Density Forecasting}.} Despite not being exploited in this paper,  the ridge and TVP connection also opens the door for the usage of conformal prediction methods that are increasingly popular in the ML literature to obtain model free prediction intervals \textit{with coverage guarantees}.  WBB draws could be used as original inputs and time series adaptation of these methods could be utilized \citep{chernozhukov2018exact}.  %This possibility is left for future work. 

\subsection{Refining the Conditional Mean:  Two Extensions}

This subsection extends 2-step ridge regression to a multistep ridge regression (MSRR) which,  as the name suggests,  by iterating  further can deliver two refinements of the conditional mean function.  The first will deliver Sparse TVPs (MSRR$_{\text{S}}$) and the second Dense TVPs (MSRR$_{\text{D}}$) through a factor structure. 

\subsubsection{Sparse TVPs via Adaptive Ridge}\label{cosso}

I provide a way to iterate it so that not only it fine tunes $\Omega_u$ but also set some of its elements to zero, obtaining \textit{Sparse} TVPs.  That is, some parameters will vary and some will not,  which can provide efficiency gains.   Those have already been proposed in the standard Bayesian MCMC paradigm most notably by \cite{bitto2018achieving} and  \cite{belmonte2014hierarchical},  so it worthwhile to demonstrate how a similarly sparse behavior can be obtained from a relatively minor twist on 2SRR.   The new primal problem is
	 	\begin{align}\label{sparsetvp}
\min_{\boldsymbol{u},  \beta_0 }\left(\boldsymbol{y}-\boldsymbol{X}\beta_0-\boldsymbol{Z_{-0}u}\right)'\Omega_{\epsilon}^{-1} \left(\boldsymbol{y}-\boldsymbol{X}\beta_0-\boldsymbol{Z_{-0}u}\right)+ \boldsymbol{u'} (\Omega_u^{-1} \otimes I_T) \boldsymbol{u} + \xi tr(\Omega_{u}^{\frac{1}{2}}),
\end{align}
which is just adding the penalty $\xi tr(\Omega_{u}^{\frac{1}{2}})$ to (\ref{genprobtvp}).  In the RR paradigm, it is quite straightforward to implement  since making a parameter constant amounts to dropping the group of regressors $\boldsymbol{Z}_k$ corresponding to the basis expansion of $X_k$ making it time-varying.  Thereby,  it corresponds to some blend of  ridge regression and Group Lasso.  This is done by setting this group's $\sigma^2_{u,k}$ to 0.  The proposed implementation, formalized by Algorithm \ref{gglrralgo} (Appendix  \ref{gglrrdetails}), amounts to iterating ridge regressions (hence the name) and constantly updating the elements of  $\Omega_u$ in a particular way.   This estimation approach is inspired by \cite{grandvalet1998}'s proposition of using Adaptive Ridge to compute the Lasso solution. The insight has since been recuperated by  \cite{frommletnuel2016} and \cite{liuli2014} to implement $l_0$ regularization in a way that makes computations tractable.  MSRR$_{\text{S}}$ goes back to implementing the $l_1$ norm by Adaptive Ridge as in \cite{grandvalet1998} but extend it to do Group Lasso and add a Ridge penalty within selected groups.  Derivation details are omitted from the main text and can be found in Appendix \ref{gglrrdetails}.

\subsubsection{Dense TVPs with Reduced Rank Restrictions}\label{shrinkdatTVP}

A frequent empirical observation, dating back to \cite{cogley2005drifting}, is that $\beta_t$'s can be spanned very well by a handful of latent factors.  \cite{dewindgambetti2014}, \cite{stevanovic2016common} and \cite{chaneisenstat2018} exploit this that directly by implementing directly a factor structure within the model. It is clear that dimensionality can be greatly reduced if we only track a few latent states and impose that evolving parameters are linear combinations of those -- \textit{Dense} TVPs.  Precisely, the model under consideration here is \textit{univariate} but reduced rank restrictions will be applied to a matrix $\boldsymbol{U}=vec^{-1}_{K \times T}(\boldsymbol{u})$ where $\boldsymbol{u}$ are the coefficients from an univariate ridge regression the likes of which we have been running all along.\footnote{This can be done because $\boldsymbol{u}$ has an obvious block structure. It has two dimensions, $K$ and $T$, that we can use to create a matrix. }  The primal problem from TVP model with a dense structure for TVPs and sparse loadings can be written as 
\begin{align}\label{densetvp}
\min_{\Lambda, \boldsymbol{F},\beta_0}\left(\boldsymbol{y}-\boldsymbol{X}\beta_0-\boldsymbol{Z}vec(\Lambda \boldsymbol{F})\right)'\Omega_{\epsilon}^{-1} \left(\boldsymbol{y}-\boldsymbol{X}\beta_0-\boldsymbol{Z} vec(\Lambda \boldsymbol{F})\right)+ \boldsymbol{f'f} + \xi \Vert l \Vert_1
\end{align}
where $\boldsymbol{f}=vec(\boldsymbol{F})$,    $l=vec(\Lambda)$,   $\Lambda$ being $K \times r$,   and $F$ being $r \times T$.\footnote{To make the exposition less heavy, I assume throughout this section that $\Omega_{\epsilon}$ is given and that $\beta_0$ are not penalized in any way.} Appendix \ref{MVGGRRRR} goes through the details of how to estimate this model with MSRR$_{\text{D}}$,  which, by using two vectorization identities,  splits the optimization into a ridge step that estimates "factors" (the same way we typically estimate TVPs in this paper, but now with a smaller number of them ($r<K$)) and a Lasso step that updates loadings.  MSRR$_{\text{D}}$ estimates are obtained from the alternation of those two steps.

\section{Simulations}\label{sec:simulations}
%Moreover, computational times will be reported and discussed for various specifications.

The simulation study investigates how accurately the different estimators proposed in this paper can recover the true parameters path. I consider three numbers of observations $T\in\{150,300,600\}$. Most of the attention will be dedicated to $T=300$ since it is roughly the number of US quarterly observations we will have 15 years from now. The size of the original regressor matrix $X$ is $K \in \{6,20,100\}$ and the first regressor in each is the first lag of $y$. Figure \ref{fig:paths} display the 5 types of parameters path $f_i$ that will serve as basic material: cosine, quadratic trend, discrete break, a pure random walk and a linear trend with a break. $f_1$, $f_2$ and $f_4$ "fit" relatively well with the prior that coefficients evolve smoothly whereas $f_3$ and $f_5$ can pose more difficulties. In those latter situations, TVP models are expected to underperform.\footnote{This partly motivates the creation of Generalized TVPs via Random Forest in \cite{MRFjae}.} The design for simulations $S_1$, $S_2$, $S_3$ and $S_4$ can be summarized in a less cryptic fashion as 
\vspace{-0.2cm}
\begin{itemize} \itemsep -0.5em
\item[$\boldsymbol{S_1}$:] $\beta_{k,t}$ follow the red line or is time invariant 
\item[$\boldsymbol{S_2}$:] $\beta_{k,t}$ follow the yellow line, the negative of it or is time invariant
\item[$\boldsymbol{S_3}$:] $\beta_{k,t}$ is either the green line or the red one in equal proportions, otherwise time-invariant.
\item[$\boldsymbol{S_4}$:] $\beta_{k,t}$ is a random mixture (loadings are drawn from a normal distribution) from the red, purple and blue lines. Some coefficients are also time-invariant. 
\end{itemize}
\vspace{-0.2cm}
The considered proportions of TVPs within the $K$ parameters are $\sfrac{K^*}{K} \in \{0.2,0.5,1\}$. Formally, we have 
\begin{align*}
\beta_{k,t}^{S_1} &= (-1)^k I(k<\sfrac{K^*}{K}) f_{1,t}+I(k>\sfrac{K^*}{K})\beta_{k,0} \\
 \beta_{k,t}^{S_2} &= (-1)^k I(k<\sfrac{K^*}{K})f_{2,t} +I(k>\sfrac{K^*}{K})\beta_{k,0} \\
 \beta_{k,t}^{S_3} &= (-1)^k I(k<\sfrac{K^*}{2K})f_{3,t} + (-1)^k I(\sfrac{K^*}{2K}<k<\sfrac{K^*}{K})f_{1,t}+I(k>\sfrac{K^*}{K})\beta_{k,0}. \\
   \beta_{k,t}^{S_4} &= I(k<\sfrac{K^*}{K}) \sum_{j \in \{1,4,5\}} l_{j,k}f_{j,t} \enskip, \quad l_{j,t} \sim N(0,1).
\end{align*}
\noindent The scale of coefficients is manually adjusted to prevent explosive behavior and/or overwhelmingly high $R^2$'s. The most important transformation in that regard is a min-max normalization on the coefficient of $y_{t-1}$ to prevent unit/explosive roots or simply persistence levels that would drive the true $R^2$ above its targeted range. Regarding the latter, I consider four different types of noise process. Three of them are homoscedastic and have a \{Low, Medium, High\} noise level. Those are calibrated so that $R^2$'s are around 0.8, 0.5 and 0.3 for low, medium and high respectively. The last two noise processes are SV, which is the predominant departure from the normality of $\epsilon_t$ in applied macroeconomics. For better comparison with time-invariant volatility cases, those are "manually" forced (by a min-max normalization) to oscillate between a predetermined minimum and maximum. The first SV process is constrained within the Low and Medium noise level bounds. For the second, it is Low and High, making the volatility spread much higher than in the first SV process case.

Four estimators are considered: the standard TVP-BVAR with SV\footnote{For the TVP-BVAR, I use the \texttt{R} package by Fabian Krueger that implements \cite{primiceri2005}'s procedure (with the \cite{delnegroprimiceri2015} correction), available \href{https://sites.google.com/site/fk83research/code}{here}.  Default parameters are used.  The total number of MCMC iterations is 15\,000 with burn-in of 5\,000. }, the two-step Ridge Regression (2SRR), the sparse version (MSRR$_{\text{S}}$) and the dense one (MSRR$_{\text{D}}$).\footnote{The maximal number of factors for MSRR$_{\text{D}}$ is set to 5 and the chosen number of factors is updated adaptively in the EM procedure according to a share of variance criteria.} TVP-BVAR results are only obtained for $K=6$ for obvious computational reasons. Performance is assessed using the mean absolute error (MAE) with respect to the true path. I then take the mean across 100 simulations for each setup. To make this multidimensional notation more compact, let us define the permutation $\mathcal{J}=\{K,\sfrac{K^*}{K},\sigma_{\epsilon},S_i \}$.  I consider simulations $s=1,...,50$ for all $\mathcal{J}$'s. Formally, for model $m$ and setup $\mathcal{J}$, the reported performance metric is $\frac{1}{50}\sum_{s=1}^{50}MAE_\mathcal{J}^{s,m}$ where 
\begin{align}\label{MAE}
 MAE_{\mathcal{J}}^{s,m} = \frac{1}{K}\frac{1}{T}\sum_{k=1}^{K}\sum_{t=1}^{T}|\beta_{k,t}^{\mathcal{J},s}-\hat{\beta}_{k,t}^{\mathcal{J},s,m}|.
\end{align}

\subsection{Results}

The results for $T=300$ are in \Cref{s1_table,s2_table,s3_table,s4_table}. With these simulations, I am interested in verifying many things,  which are covered aspect by aspect in paragraphs that follow.   First, I want to verify that 2SRR's performance is comparable to that of the BVAR for models' size that can be handled by the latter.  Second, I want to demonstrate that additional shrinkage embedded in  MSRR$_{\text{S}}$ and MSRR$_{\text{D}}$ can help under DGPs that more or less fit the prior of reduced-rank and/or sparsity. To make the investigation of these two points visually easier by looking at the tables, the lowest MAE out of BVAR/2SRR for each setup is in blue while that of the best one out of all algorithms is in bold.  Then, follows a series of investigations,  like the effects of changing sample size,  time-varying volatility in DGP,   benchmarking in higher dimensions,  computational time,   and a look at the empirical properties of the WBB procedure for uncertainty quantification.  %Third,  I want to check how reducing or increasing the sample size affects results.  Fourth,  I want to get a sense how bringing in stochastic volatility in the DGP changes 2SRR vs BVAR performance.  Fift
 
\begin{figure}[ht!]
%\centering
\captionsetup[subfigure]{justification=centering}
\setlength{\lineskip}{0.2ex}% increase spacin
  \begin{subfigure}[b]{0.55\linewidth}
\includegraphics[scale=.4]{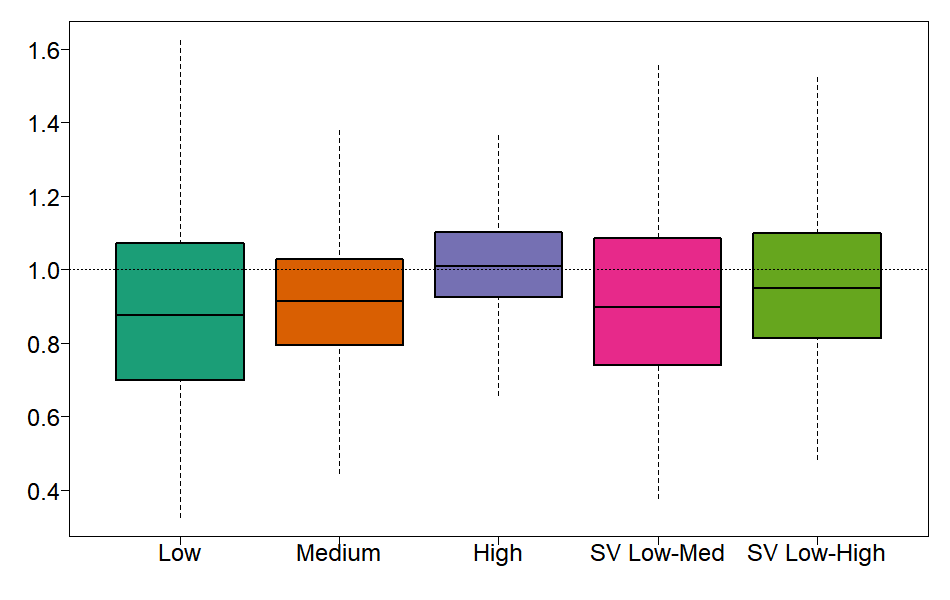}
\caption{By noise process}  
\label{bxp_bynoise}
    \end{subfigure}
      \begin{subfigure}[b]{.45\linewidth}
\hspace*{1cm}\includegraphics[scale=.4]{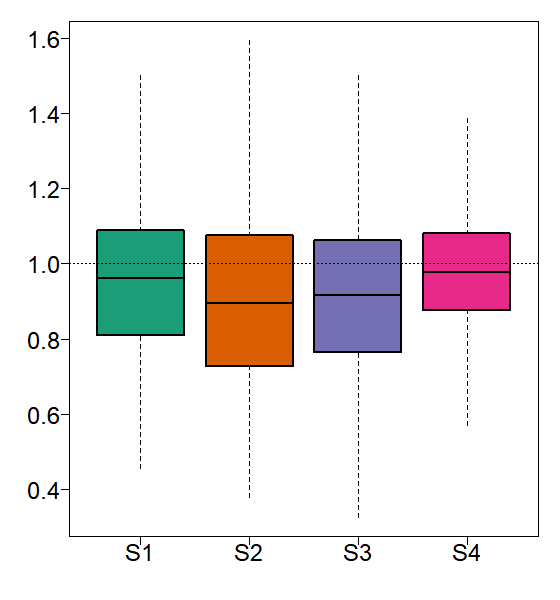}
\caption{By DGP}  
\label{bxp_byDGP}
    \end{subfigure}
      %\vspace{0.5cm}
  \caption{\footnotesize This figure summarizes \cref{s1_table,s2_table,s3_table,s4_table} results comparing 2SRR and the BVAR when $K=6$ and $T=300$. The plotted quantity is the distribution of $\sfrac{MAE_{\mathcal{J}}^{s,2SRR}}{MAE_{\mathcal{J}}^{s,BVAR}}$ for different subsets of interest.}
   \label{bxp_simuls}
\end{figure}

\vskip 0.2cm

{\sc \noindent \textbf{2SRR vs. BVAR}.} Overall, results for 2SRR and the BVAR are very similar and their relative performance depends on the specific setup. These two models are interesting to compare because they share the same prior for TVPs (no additional shrinkage) but address evolving residuals volatility differently. Namely, the BVAR models SV directly within the MCMC procedure whereas 2SRR is a two-step GLS-like approach using a GARCH(1,1) model of the first step's residuals. Figure \ref{bxp_simuls} summarizes results of the 2SRR/BVAR comparison by reporting boxplots showcasing the distribution of relative MAEs (2SRR/BVAR) for different subsets. Overall, 2SRR does marginally better in almost all cases. A lower noise level seems to help its cause. It is plausible that cross-validating $\lambda$ as implemented by 2SRR plays a role (BVAR uses default values).\footnote{Replacing the absolute distance by the squared distance in \eqref{MAE} produces similar looking boxplots as in Figure \ref{bxp_simuls}, with wider dispersion but a near-identical ranking of methods by simulations and noise processes.}

I now dive in specific DGPs.  In Table \ref{s1_table}, where the DGP is the rather friendly cosine-based TVPs, it is observed that the BVAR will usually outperform 2SRR by a thin margin when the level of noise is high. The reverse is observed for low noise environment and results are mixed for the medium one. Results for the SV cases will the subject of its own discussion later.  In Table \ref{s2_table}, where the DGP is a structural break -- at odds with the prior of slow change --  2SRR is clearly performing better than the BVAR, providing a smaller average MAE in 12 out of 15 cases for $K=6$.  In Table \ref{s3_table}, where the DGP is a mix of trending coefficients and cosine ones, Table \ref{s3_table} reports very similar results to that of Simulation 1. 2SRR is better than BVAR except in the high noise setups, where the latter has a minor advantage.  Lastly,  for the simulation in Table \ref{s4_table}, a sophisticated mixture of TVP-friendly and -unfriendly parameters paths, BVAR does a better job than 2SRR for 8 out of 15 cases. The gains are, as before, quantitatively small. When 2SRR does better, gains are also negligible, suggesting that BVAR and 2SRR provide very similar results in this environment. 

\vskip 0.2cm

{\sc \noindent \textbf{MSRR$_{\text{S}}$ and  MSRR$_{\text{D}}$}.} Overall,  the two refinements can help further improving results,  and unsurprisingly,  they tend to do so where the underlying DGP best fits their prior.   In Table \ref{s1_table}, for $K=6$, MSRR$_{\text{S}}$ will marginally improve on 2SRR for most setups, especially those where 2SRR is already better than BVAR. In higher dimensions ($K=20$ or $K=100$), MSRR$_{\text{S}}$ constantly improves on 2SRR (albeit minimally) whereas MSRR$_{\text{D}}$ can provide important gains (see the $\sfrac{K^*}{K}=1$ block for instance) but is more vulnerable in the high noise environment.  Moving to  Table \ref{s2_table},  for the small dimensional case, it is observed that MSRR$_{\text{S}}$ can further reduce the MAE --- albeit by a very small amount --- in many instances. The same is true for MSRR$_{\text{D}}$ when all parameters vary ($\sfrac{K^*}{K}=1$). For MSRR$_{\text{D}}$, this observation additionally extends to $K=20$, an environment where it is expected to thrive.  In Table \ref{s3_table},  MSRR$_{\text{S}}$ often marginally improves upon 2SRR whereas MSRR$_{\text{D}}$'s edge is more visible in low-noise and high-dimensional environments --- factors being more precisely estimated with a large cross-section.  Lastly,  in Table \ref{s4_table},  when it comes to higher-dimensional setups ($K=20$ or $K=100$), MSRR$_{\text{S}}$ emerges as the better option with (now familiar) marginal improvements with respect to 2SRR. This recurrent observation is potentially due to the iterative process producing a more precise $\hat{\Omega}_u$ when $\sigma_{u_k}^2$'s are heterogeneous whether sparsity is involved or not.

\vskip 0.2cm

{\sc \noindent \textbf{Changing Sample Size}.} The results for $T=150$ and $T=600$ are in \Cref{s1_table_150,s2_table_150,s3_table_150,s4_table_150,s1_table_600,s2_table_600,s3_table_600,s4_table_600}. When $T$ is reduced from 300 to 150, the performance of 2SRR relative to that of BVAR remains largely unchanged: both report very similar results. When bumping $T$ to 600, overall performance of all estimators improves, but not by a gigantic leap. This is, of course, due to the fact that increasing $T$ also brings up the number of effective regressors. BVAR has a small edge on Simulation 1 in Table \ref{s1_table_600} whereas 2SRR wins marginally for the more complicated Simulation 4 (Table \ref{s4_table_600}). 

What is perhaps most noticeable from  simulations with a larger $T$ is how much more frequently MSRR$_{\text{S}}$ and especially MSRR$_{\text{D}}$ are preferred, especially in the medium- and high-dimensional cases. For instance, for the Cosine DGP ($S_1$) with $K\in\{20,100\}$, MSRR$_{\text{D}}$ almost always deliver the lowest MAE, and sometimes by good margins (e.g., $\{S_1,\sfrac{K^*}{K}=1,\sigma_{\epsilon}=\text{Low} \}$ for both $K$'s).  Similar behavior is observed for $S_3$ in almost all cases of $K=20$. This noteworthy amelioration of MSRR$_{\text{D}}$ is intuitively attributable to factor loadings being more precisely estimated with a growing $T$. Thus, unlike 2SRR whose performance is largely invariant to $T$ by model design, algorithms incorporating more sophisticated shrinkage schemes may benefit from larger samples.

\vskip 0.2cm

{\sc \noindent \textbf{Time-Varying Volatility}.} A pattern emerges across the four simulations: when SV is built in the DGP ($\sigma_{\epsilon,t}$ in tables), 2SRR either performs better or deliver roughly equivalent results to that of the BVAR. Indeed, with $T=300$, for 17 out of 24 setups with SV-infused DGPs, 2SRR supplants BVAR. The wedge is sometimes small ($\{S_1,\sfrac{K^*}{K}=0.2,\text{SV Low-Med} \}$,$\{S_4,\sfrac{K^*}{K}=1,\text{both SV} \}$), sometimes large ($\{S_1,\sfrac{K^*}{K}=1, \text{SV Low-High} \}$,$\{S_2,\sfrac{K^*}{K}=0.5, \text{both SV}\}$). However, it is fair to say that small gaps between 2SRR and BVAR performances are the norm rather than the exception. Nonetheless, these results suggest that 2SRR is not merely a suboptimal approximation to the BVAR in the wake of computational adversity.  It is a viable statistical alternative with the additional benefit of being easy to compute and to tune.

%\vskip 0.2cm

%{\sc \noindent \textbf{Computational Time}.} Table \ref{time_table} reports how computational time varies in $K$ and $T$ on a cluster,  and how 2SRR compares to a standard BVAR implementation. When $T$ is 300 and $K$ is 6, 2SRR takes one second while BVAR takes 513 seconds. When $T$ increases to 600, BVAR takes over 1\,000 seconds whereas 2SRR takes less than 5. $T=300$ with $K=300$ can be seen as a typical high-dimensional case. It takes 13 seconds to compute (and tune) 2SRRR. When $T$ is reduced to 150, high-dimensional 2SRR runs in less than 3 seconds on average. Only when both $T$ and $K$ gets very large (by traditional macro data sets standards) do things become harder with 2SRR taking 69.5 seconds on average. By construction, MSRR$_{\text{S}}$ takes marginally longer than 2SRR. Finally, by relying on an EM algorithm, MSRR$_{\text{D}}$ inevitably takes longer, yet remains highly manageable for very large models with many observations -- taking a bit more than 100 seconds.

\begin{table}[t!]
	\footnotesize
	\centering
	%\rowcolors{2}{white}{gray!15}
\caption{\normalsize {Comparison in Higher Dimensions for Mixture DGP and $T=300$ } \vspace*{-0.3cm}} \label{tab:shrinktvp}

\begin{threeparttable}
		\setlength{\tabcolsep}{0.40em}%{0.61em}
		  \setlength\extrarowheight{2.5pt}
 \begin{tabular}{l| ccccccc | ccccccc} 
\toprule \toprule
\addlinespace[2pt]
& & \multicolumn{5}{c}{$K=20$} & & & \multicolumn{5}{c}{$K=50$} \\
\cmidrule(lr){3-7} \cmidrule(lr){10-14} \addlinespace[2pt]
& &  2SRR &  &   \texttt{ShTVP}-R &  &\texttt{ShTVP}-3G &  & &
	  2SRR  &  & \texttt{ShTVP}-R &  &\texttt{ShTVP}-3G & \\
\midrule
\addlinespace[5pt] 
\rowcolor{gray!15} 
 \multicolumn{14}{l}{\textbf{$\mathbf{\sfrac{K^*}{K}=0.2}$}} &\cellcolor{gray!15} \\ \addlinespace[2pt] 
$\sigma_{\epsilon} =\text{Low}$ & & 0.067 & & 0.074  & &  0.063  && &  0.103  & & 0.121 & & 0.101   &  \\ \addlinespace[2pt]  
$\sigma_{\epsilon} =\text{Medium}$ & & 0.113 & & 0.130  & &  0.115  && &  0.178  & & 0. 198 & & 0. 166  &  \\ \addlinespace[2pt]  
$\sigma_{\epsilon} =\text{High}$ & & 0.233 & & 0.239  & &  0.202  && &  0. 387 & & 0. 348 & & 0. 241  &  \\ \addlinespace[2pt]  
$\sigma_{\epsilon,t} = \text{SV Low-Med}$ & & 0.087 & & 0.095  & &  0.081  && &  0. 127 & & 0. 148 & & 0. 130 &  \\ \addlinespace[2pt]  
$\sigma_{\epsilon,t} = \text{SV Low-High}$ & & 0.128 & & 0.14  & &  0.121  && &  0. 216 & & 0. 207 & & 0. 169  &  \\ \addlinespace[2pt]

\midrule
\addlinespace[5pt] 
\rowcolor{gray!15} 
 \multicolumn{14}{l}{\textbf{$\mathbf{\sfrac{K^*}{K}=0.5}$}} &\cellcolor{gray!15} \\ \addlinespace[2pt] 
$\sigma_{\epsilon} =\text{Low}$ & & 0.088 & & 0.085  & &  0.081  && &  0. 121 & & 0. 132 & & 0. 117  &  \\ \addlinespace[2pt]  
$\sigma_{\epsilon} =\text{Medium}$ & & 0.130 & & 0.139  & &  0.130  && &  0. 186 & & 0. 205 & & 0. 181  &  \\ \addlinespace[2pt]  
$\sigma_{\epsilon} =\text{High}$ & & 0.239  & & 0.245  & &  0.207  && &  0. 394 & & 0. 353 & & 0. 254  &  \\ \addlinespace[2pt]  
$\sigma_{\epsilon,t} = \text{SV Low-Med}$ & & 0.102 & & 0.105  & &  0.095  && &  0. 142  & & 0. 156 & & 0. 141  &  \\ \addlinespace[2pt]  
$\sigma_{\epsilon,t} = \text{SV Low-High}$ & & 0.137 & & 0.141 & &  0.13  && &  0. 220 & & 0. 215 & & 0. 183  &  \\ \addlinespace[2pt]

\midrule
\addlinespace[5pt] 
\rowcolor{gray!15} 
 \multicolumn{14}{l}{\textbf{$\mathbf{\sfrac{K^*}{K}=1}$}} &\cellcolor{gray!15} \\ \addlinespace[2pt] 
$\sigma_{\epsilon} =\text{Low}$ & & 0.111  & & 0.104  & &  0.109  && &  0. 151 & & 0. 151& & 0. 144  &  \\ \addlinespace[2pt]  
$\sigma_{\epsilon} =\text{Medium}$ & & 0.151 & & 0.155  & &  0.155  && &  0. 213 & & 0. 222 & & 0. 202  &  \\ \addlinespace[2pt]  
$\sigma_{\epsilon} =\text{High}$ & & 0.268 & & 0.258  & &  0.230  && &  0. 415 & & 0. 364 & & 0. 278  &  \\ \addlinespace[2pt]  
$\sigma_{\epsilon,t} = \text{SV Low-Med}$ & & 0.127 & & 0.125  & &  0.125  && &  0. 164 & & 0. 173 & & 0. 159  &  \\ \addlinespace[2pt]  
$\sigma_{\epsilon,t} = \text{SV Low-High}$ & & 0.162 & & 0.156  & &  0.151  && &  0. 229  & & 0. 227 & & 0. 200  &  \\ \addlinespace[2pt]

\midrule
\addlinespace[5pt] 
\rowcolor{gray!15} 
 \multicolumn{14}{l}{\textbf{Running Times}} &\cellcolor{gray!15} \\ \addlinespace[2pt] 
Seconds & & 5$^*$ & & 70  & & 70  && &  13$^*$  & & 360 & &  362   &  \\ \addlinespace[2pt]  

\bottomrule \bottomrule
	\end{tabular}
		\begin{tablenotes}[para,flushleft]
	\scriptsize %\item[] \hspace*{-0.5cm}
		\textit{Notes}: This table reports the average MAE of estimated $\beta_t$'s for various models.  Seconds are from running such models on a 2020 M1 Macbook Air.  MAE are averaged over 20 simulations.  ($*$) It is important to note that the reported times for 2SRR only include \textit{point estimates} and do not account for the additional features that \texttt{ShTVP}-R  and \texttt{ShTVP}-3G offer,  such as inference and density predictions. %The number in bold is the best statistic (lowest MAE or computing time) of all models for a given setup. 
	\end{tablenotes}
\end{threeparttable}
\end{table}

\vskip 0.2cm

{\sc \noindent \textbf{Benchmarking in Higher Dimensions}.} In the years following the original draft of this paper,  new Bayesian procedures have been proposed and packaged in statistical software.  This allows for an easier evaluation of 2SRR in higher dimensions,  both statistically and computationally.  One such package is \texttt{shrinkTVP} from \cite{knaus2021shrinkage}.  It conveniently features C++ based computations,  a very wide set of potential prior specifications,  focus on single equation modeling,  and features a data-driven way of choosing shrinkage intensity.  Two default specifications are used,  one with a ridge prior (\texttt{ShTVP}-R) on coefficients and another with the triple gamma prior (\texttt{ShTVP}-3G) of \cite{cadonna2020triple}).  The former is the closest relative to 2SRR within the suite of available configurations.   The latter should  perform well in the sparse TVP environments for which it has been designed.  The high-dimensional case of $K=100$ has been reduced to $K=50$ since it took \texttt{shrinkTVP} 38 minutes to estimate a single model in the original large $K$ case (whereas 2SRR delivers point estimates in under 30 seconds).

Results in Table \ref{tab:shrinktvp} further establish the relevance of 2SRR for the estimation of models with many TVPs.  Its statistical performance is mostly on par with \texttt{ShTVP}-R, with differences always being marginal in favor of one or another.  \texttt{ShTVP}-3G can provide some less trivial improvements over both 2SRR and\texttt{ShTVP}-R  under a sparse DGP or in the hostile high noise and $K=50$ environment.  Otherwise,  it does not meaningfully outperform 2SRR.   This is evidently the prior's doing,  and it is an empirical question as to which setup in Table  \ref{tab:shrinktvp} most closely resembles real data.

\vskip 0.2cm

{\sc \noindent \textbf{Computational Time}.} In Table \ref{tab:shrinktvp},  we see that the Bayesian computations are manageable for $K=20$,  and becomes rather burdensome for $K=50$ (about 10 minutes vs. less than 15 seconds).  Even at $K=20$, the fact that 2SRR gets similarly valid estimates in 5 seconds rather than in over a minute makes it more convenient in applied work where the trial and error of specifications (and sometimes even grid search of them) is widespread.   Of course,  full-blown Bayesian estimation directly provides many additional quantities that 2SRR does not procure readily,  and its computational time could be brought down to a certain extent by cutting MCMC iterations.  Nonetheless,  the point is that, depending on the intended use,  the ridge-based methods  have non-trivial comparative advantages in many ways that are relevant for applied work.  

It is apparent that for large models, such as those used in rolling/expanding window forecasting evaluations (Section \ref{sec:forecasting}) and large local projections (Section \ref{sec:app}), 2SRR remains the only practical approach.  As discussed in Section \ref{easycase}, we expect computations for 2SRR (for a single equation) to scale (approximately) linearly with \( K \) for a fixed \( T \). Table \ref{tab:2SRR_computational_time} (Appendix) provides additional results, verifying this by showing how 2SRR computational time varies with \( K \) and \( T \) for the Mixture DGP.  I also find, as expected,  that an increasing sample size has a nonlinear effect on computing time. For instance, the most demanding combination (\( K = 100 \), \( T = 600 \)) takes about 153 seconds to compute, whereas the same high-dimensional \( K = 100 \) results in 3 seconds when \( T = 150 \) and 22 seconds when \( T = 300 \).  

The table also reports running times for estimating \(K\) equations, rather than just one, to illustrate how the techniques discussed in Section \ref{MV} help in preventing the running time from being multiplied by \(K\) for large systems. Estimating 50 equations, each with 50 time-varying parameters (TVPs), resulting in a total of 2,500 TVPs,  takes one minute with \(T = 300\).  This is about six times longer than the estimation time for a single equation (\(K = 50\), \(T = 300\)). The most extreme case tested involves estimating 100 equations, each with 100 parameters (totaling 10,000 TVPs) over 600 time periods, which takes 30 minutes to run.

%Table \ref{time_table} reports how computational time varies in $K$ and $T$ on a cluster,  and how 2SRR compares to a standard BVAR implementation. When $T$ is 300 and $K$ is 6, 2SRR takes one second while BVAR takes 513 seconds. When $T$ increases to 600, BVAR takes over 1\,000 seconds whereas 2SRR takes less than 5. $T=300$ with $K=300$ can be seen as a typical high-dimensional case. It takes 13 seconds to compute (and tune) 2SRRR. When $T$ is reduced to 150, high-dimensional 2SRR runs in less than 3 seconds on average. Only when both $T$ and $K$ gets very large (by traditional macro data sets standards) do things become harder with 2SRR taking 69.5 seconds on average. By construction, MSRR$_{\text{S}}$ takes marginally longer than 2SRR. Finally, by relying on an EM algorithm, MSRR$_{\text{D}}$ inevitably takes longer, yet remains highly manageable for very large models with many observations -- taking a bit more than 100 seconds.

\vskip 0.2cm

{\sc \noindent \textbf{Some Evaluation of Uncertainty Quantification for 2SRR}.} How reliable are credible regions obtained from the WBB approach proposed in Section \ref{sec:uq} versus that of fully Bayesian methods deployed above? I provide some answers in Table \ref{tab:uq} (Appendix) where I report nominal coverage for 2SRR (obtained from 250 WBB draws) and those from \texttt{ShTVP}-R and \texttt{ShTVP}-3G.  As mentioned earlier,  Bayesian techniques (approximate or not)  are not formally expected to hit nominal coverage targets,  an inherently frequentist object.  Nonetheless,  looking at such metrics gives an idea of how those methods will behave when using them on real data,  e.g.,  how conservative or not they may be.  For instance,  it could be worrisome if 2SRR is always much higher or lower than either \texttt{ShTVP}-R or \texttt{ShTVP}-3G.  

In all cases,  I use the relevant percentiles from the draws coming out of the procedure and calculate the frequency (over $t$ and 20 simulations) at which the true $\beta_{t,k}$ falls in the interval.  Also,  to give a sense how methods adapt the local width of bands along  with  heteroscedasticity,  in the two DGPs where errors follow a SV process,  I calculate the average correlation between the empirical 12\% and 84\%  interquantile range (IQR) of $\beta_{t,k}$  (averaging over $k$) and the true SV process.  Since we expect heteroscedasticity to drive time-varying width of bands,  those two quantities should be correlated in a non-trivial fashion.

Again,  in Table  \ref{tab:uq},  I focus on the $T=300$ case,  and look at a subset of simulations for this particular aspect.  Namely,  I center the attention on the mixture DGP of Table \ref{tab:shrinktvp} above and look at $K=6$ and $K=20$ in the middle ground case of  ${\sfrac{K^*}{K}=0.5}$.  Except in a few instances,  2SRR provides coverage that sits somewhere between \texttt{ShTVP}-R and \texttt{ShTVP}-3G.  \texttt{ShTVP}-R is conservative,  and so is for the most part 2SRR.  \texttt{ShTVP}-3G is almost always under the other two,  and sometimes by a lot (like for the 68\% high noise case).  2SRR often remains in the middle,  but more on the side of  \texttt{ShTVP}-R,   especially for $K=20$.   The discrepancies are less  substantial for the 95\% target at $K=6$,   with 2SRR oscillating around 95\%,   \texttt{ShTVP}-R being above by 2-3\%,  and \texttt{ShTVP}-3G remaining below.  While  2SRR and \texttt{ShTVP}-R coverage is mostly unchanged moving from $K=6$ to $K=20$ and remains above nominal targets,  \texttt{ShTVP}-3G  often provides sharper posteriors than what frequentist coverage would commend.  

Perhaps the most intriguing observation from Table \ref{sec:uq} is how the relative width of credible regions around TVPs correlates with the true SV process.  In all cases,  2SRR provides bands which IQR is most correlated with the true time-varying volatility.  Its correspondence with it is up to twice as high vis-à-vis \texttt{ShTVP}-R and even more so when compared to \texttt{ShTVP}-3G.  Thus,  their ability  to reflect changing levels of uncertainty in the width of bands, is at the very least, within the simulation setup under study,  on par with two Bayesian methods that explicitly feature a SV component.   Some additional intuition on this -- beyond the fact that 2SRR models volatility in its own way between the first and the second step --  can be obtained from noting that in the frequentist domain,  the analogous nonparametric  "pairs" bootstrap is valid under heteroscedasticity \citep{mackinnon2006bootstrap}.  In fact,  from a Bayesian point of view, \cite{lancaster2003} show that the obtained variance for a linear regression model from using such a bootstrap is asymptotically equivalent to what one would get from White's  sandwich formula.

\section{Forecasting}\label{sec:forecasting}

Machine Learning models' merits are typically evaluated out-of-sample before taking to interpret them.  2SRR lends itself nicely to forecasting because everything is automatically tuned and more importantly,  its quick computations makes it amenable to recursive forecasting backtests where the models needs to be re-estimated at every step.  This section describes forecasting results and discusses several practical aspects (like variations on CV) that may help improve forecasting performance on real data and further inform choices for macroeconomic applications in Section \ref{sec:app}.

I present results for a pseudo-out-of-sample forecasting experiment at the quarterly frequency using the dataset FRED-QD \citep{mccracken2020fred}. The latter is publicly available at the Federal Reserve of St-Louis's 
web site and contains 248 US macroeconomic and financial aggregates observed from 1960Q1. The forecasting targets are real GDP, Unemployment Rate (UR), CPI Inflation (INF), 1-Year Treasury Constant Maturity Rate (IR) and the difference between 10-year Treasury Constant Maturity rate and Federal funds rate (SPREAD). These series are representative macroeconomic indicators of the US economy which is based on \cite{GCLSS2018}'s exercise for many ML models, itself based on \cite{KLS2019} and a whole literature of extensive horse races in the spirit of \cite{SW1998comparison}. The series transformations to induce stationarity for predictors are indicated in \cite{mccracken2020fred}. For forecasting targets, GDP, CPI and UR are considered $I(1)$ and are first-differenced. For the first two, the natural logarithm is applied before differencing. IR and SPREAD are kept in "levels".  Forecasting horizons are 1, 2, and 4 quarters. For variables in first differences (GDP, UR and CPI), average growth rates are targeted for horizons 2 and 4.

The pseudo-out-of-sample period starts in 2003Q1 and ends 2014Q4.  I use expanding window estimation from 1961Q3. Models are estimated \textit{and} tuned at each step. I use direct forecasts, meaning that $\hat{y}_{t+h}$ is obtained by fitting the model directly to $y_{t+h}$ rather than iterating one-step ahead forecasts. Following standard practice in the literature, I evaluate the quality of point forecasts using the root Mean Square Prediction Error (MSPE). For the out-of-sample (OOS) forecasted values at time $t$ of variable $v$ made $h$ steps ahead, I compute 
$$RMSPE_{v,h,m}= \sqrt{ \frac{1}{\#\text{OOS}}\sum_{t \in \text{OOS}} (y_{t}^v-\hat{y}_{t-h}^{v,h,m})^2}.$$
The standard \cite{dieboldmariano} (DM) test procedure is used to compare the predictive accuracy of each model against the reference AR(2) model. $RMSPE$ is the most natural loss function given that all models are trained to minimize the squared loss in-sample.
%feel like nothing, dust.

Three types of TVPs will be implemented: 2SRR (Section \ref{gensol}), MSRR$_{\text{S}}$ (Section \ref{cosso}), MSRR$_{\text{D}}$ (Section \ref{shrinkdatTVP}). I consider augmenting four standard models with different methodologies proposed in this paper. The first will be an \textbf{AR} with 2 lags. The second is the well-known \cite{sw2002} \textbf{ARDI} (Autoregressive Diffusion Index) with 2 factors and 2 lags for both the dependent variable and the factors. The third is a \textbf{VAR(5)} with 2 lags and the system is composed of the 5 forecasted series. Finally, I consider as a fourth model a \textbf{VAR(20)} with 2 lags in the spirit of \cite{banbura2010large}'s medium VAR. Thus, there is a total of $4 \times 4 = 16$ models considered in the exercise. The BVAR used in Section \ref{sec:simulations} is left out for computational reasons  --- models must be re-estimated every quarter for each target. Moreover, the focus of this section is rather single equation \textit{direct} (as opposed to iterated) forecasting.  

The first three constant coefficients models are estimated by OLS, which is standard practice.  Since the constant parameters VAR(20) has 41 coefficients and around 200 observations, it is estimated with a ridge regression.  Potential outliers are dealt with as in \cite{GCLSS2018} for Machine Learning models. If the forecasted values are outside of $[\boldsymbol{\bar{y}}+2*min(\boldsymbol{y}-\boldsymbol{\bar{y}}),\boldsymbol{\bar{y}}+2*max(\boldsymbol{y}-\boldsymbol{\bar{y}})]$, the forecast is discarded in favor of the constant parameters forecast.

\subsection{Results}

I report three sets of results. Table \ref{tab_fcst_main} corresponds exactly to what has been described beforehand.  Table \ref{tab_fcst_modavg} gathers results where TVPs have been additionally shrunk to their constant parameters counterparts by means of model averaging with equal weights. The virtues of this \textit{Half \& Half} strategy are two-fold. First, k-fold CV can be over-optimistic for horizons $h$>1 because of imminent serial correlation. Second, k-fold CV ranks potential $\lambda$'s using the whole sample, whereas in the case of "forecasting", prediction always occurs at the boundary of the implicit kernel. In that region, the variance is mechanically higher and ensuing predictions could benefit from extra shrinkage. Shrinking to OLS in this crude and transparent fashion is a natural way to attempt getting even better forecasts.  Finally, Table \ref{tab_fcst_main_bb} reports results where k-fold CV has been replaced by Blocked k-fold CV (BCV) with blocks of 8 quarters as in, e.g.,  \citep{MRFjae}.  This is a more principled strategy to curb the downward bias of $\lambda$ induced by serial dependence.  Among other things,  it does not rely on blending two models using intuitive but admittedly arbitrary weights. 

\begin{comment}
\begin{figure}[h!]
%\centering
\captionsetup[subfigure]{justification=centering}
\setlength{\lineskip}{1.5ex}% increase spacin
  \begin{subfigure}[b]{0.5\linewidth}
\includegraphics[trim={1cm 1.1cm 0.25cm 1.1cm},clip,scale=.36]{{graphs/BP_REVERSE_vh31}}
\caption{Inflation ($h=1$)}  
\label{vh31}
  \end{subfigure}
  \begin{subfigure}[b]{0.5\linewidth}
\includegraphics[trim={1cm 1.1cm 0.25cm 1.1cm},clip,scale=.36]{{graphs/BP_vh33}}
\caption{Inflation ($h=4$)}  
\label{vh33}
    \end{subfigure}
  \begin{subfigure}[b]{0.5\linewidth}
\includegraphics[trim={1cm 1.1cm 0.25cm 1.1cm},clip,scale=.36]{{graphs/BP_vh41}}
\caption{Interest Rate ($h=1$)}  
\label{vh41}
     \end{subfigure}
             \hfill 
  \begin{subfigure}[b]{0.5\linewidth}
\includegraphics[trim={1cm 1.1cm 0.25cm 1.1cm},clip,scale=.36]{{graphs/BP_REVERSE_vh43}}
\caption{Interest Rate ($h=4$)}  
\label{vh43}
      \end{subfigure}
      \vspace{-0.6cm}
  \caption{\footnotesize A subset of $RMSPE_{v,h,m}/RMSPE_{v,h,\text{Plain AR(2)}}$'s (from Tables \ref{tab_fcst_main} and \ref{tab_fcst_modavg}) for forecasting targets usually associated with the need for time variation. Blue is the benchmark AR with constant coefficients. Darker green means that the competing forecast rejects the null of a Diebold-Mariano test at least at the 10\% level (with respect to the benchmark).}
   \label{results_fig}
\end{figure}
\end{comment}

\vskip 0.2cm

{\sc \noindent \textbf{Results with Basic CV and  Half \& Half}.} Overall, results are in line with evidence previously reported in the TVP literature: very limited improvements are observed for real activity variables (GDP, UR) whereas substantial gains are reported for INF and IR.  For the latter, allowing for time variation in either AR or a compact factor model (ARDI) generate very competitive forecasts. For instance, ARDI-2SRR is the best model for IR with a reduction of 36\% in RMPSE over the AR(2) benchmark which is strongly statistically significant. Still for IR, at horizon 2 quarters, iterating 2SRR to obtain MSRR$_{\text{S}}$ generate sizable improvements for both AR and ARDIs. VAR(20) is largely inferior to alternatives in any of its forms. Two exceptions are IR forecasts at a one-year horizon where combining VAR(20) with MSRR$_{\text{D}}$ yields the best forecast by a wide margin with improvement of 19\% in RMSPE. VAR(20)-MSRR$_{\text{D}}$ also provide a very competitive forecast for IR at an horizon of one quarter. Finally, at horizon 1 quarter, any form of time variation (2SRR, MSRR$_{\text{S}}$, MSRR$_{\text{D}}$) at least increases SPREAD's forecasting accuracy for for all models but the VAR(20). Precisely, it is a 16\% reduction in RMSPE for AR, about 5\% for ARDI and up to 14\% in the VAR(5) case. For the latter, its combination with 2SRR provides the best forecast with a statistically significant improvement of 18\% with respect to the AR(2) benchmark. 

A notable absence from the relatively cheerful discussion above is inflation, which is the first (or second) variable one would think should benefit from time variation. It is clear that, in Table \ref{tab_fcst_main}, any AR at horizon 1 profits rather timidly from it. A similar finding for Half \& Half is reported in Table \ref{tab_fcst_modavg}. What differs, however, are longer horizons results for INF. Indeed, mixing in additional shrinkage to OLS strongly helps results for those targets: every form of time variation now improves performance by a good margin. For instance, any time-varying ARs improves upon the constant benchmark by around 15\%. It is now widely documented that inflation prior to the Pandemic was better predicted by past values of itself and not much else -- besides maybe for recessionary episodes \citep{KLS2019}.  Results of Table \ref{tab_fcst_modavg} comfortably stand within this paradigm except for the noticeable efforts from MSRR$_{\text{S}}$ versions of both ARDI and VAR(20). While those are the best models, they are closely matched in performance by their AR counterparts. Nevertheless, it is noteworthy that this surge in performance mostly occurs for their sparse TVP versions, suggesting time variation is likely crucial for more sophisticated inflation forecasts not to be off the charts. Finally, additional shrinkage marginally improves GDP forecasting at the two longer horizons, with the Half \& Half ARDI-MSRR$_{\text{S}}$ providing the best forecasts.

\begin{figure}[t!]
%\centering
\captionsetup[subfigure]{justification=centering}
\setlength{\lineskip}{1.5ex}% increase spacin
  \begin{subfigure}[b]{0.5\linewidth}
\includegraphics[trim={1cm 1.1cm 0.25cm 1.1cm},clip,scale=.26]{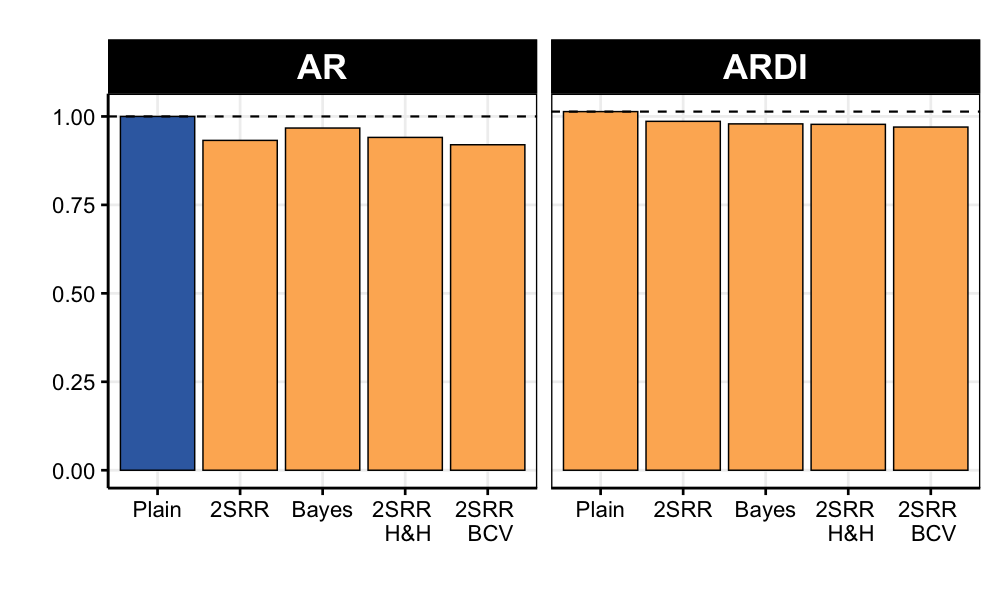}
\caption{Inflation ($h=1$)}  
\label{vh31}
  \end{subfigure}
  \begin{subfigure}[b]{0.5\linewidth}
\includegraphics[trim={1cm 1.1cm 0.25cm 1.1cm},clip,scale=.26]{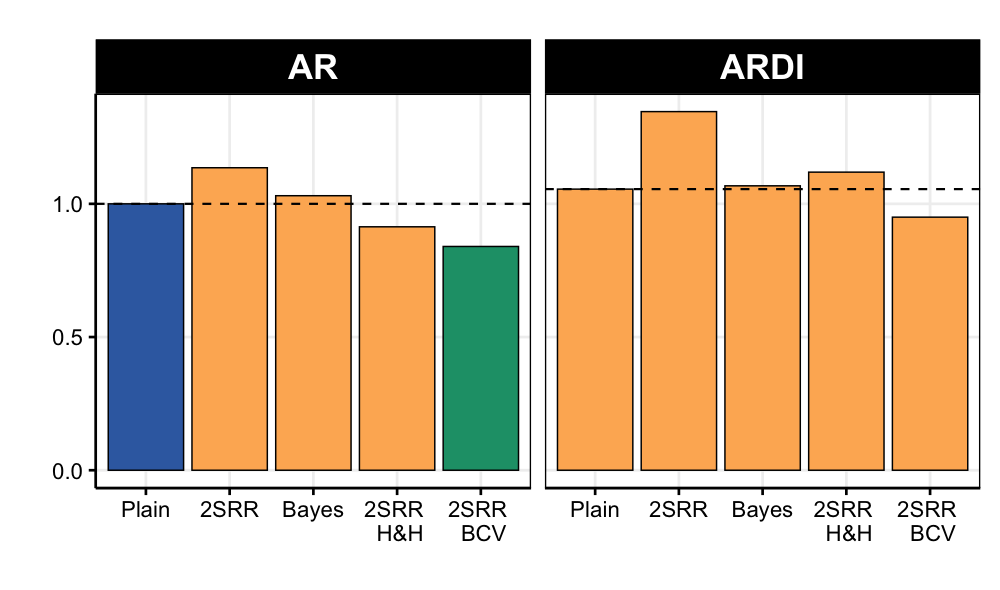}
\caption{Inflation ($h=2$)}  
\label{vh33}
    \end{subfigure}
  \begin{subfigure}[b]{0.5\linewidth}
\includegraphics[trim={1cm 1.1cm 0.25cm 1.1cm},clip,scale=.26]{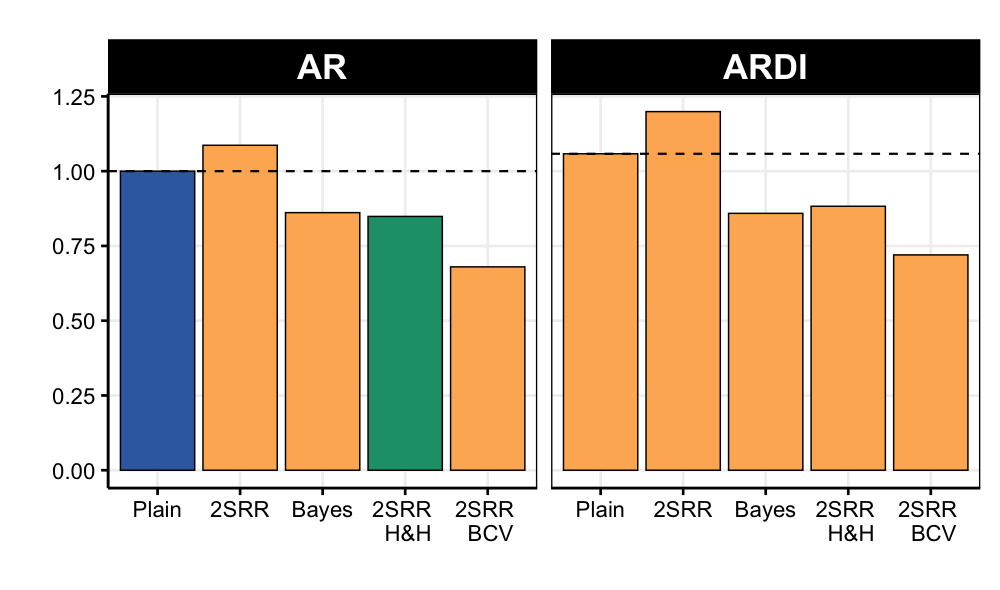}
\caption{Inflation ($h=4$)}  
\label{vh41}
     \end{subfigure}
             \hfill 
  \begin{subfigure}[b]{0.5\linewidth}
\includegraphics[trim={1cm 1.1cm 0.25cm 1.1cm},clip,scale=.26]{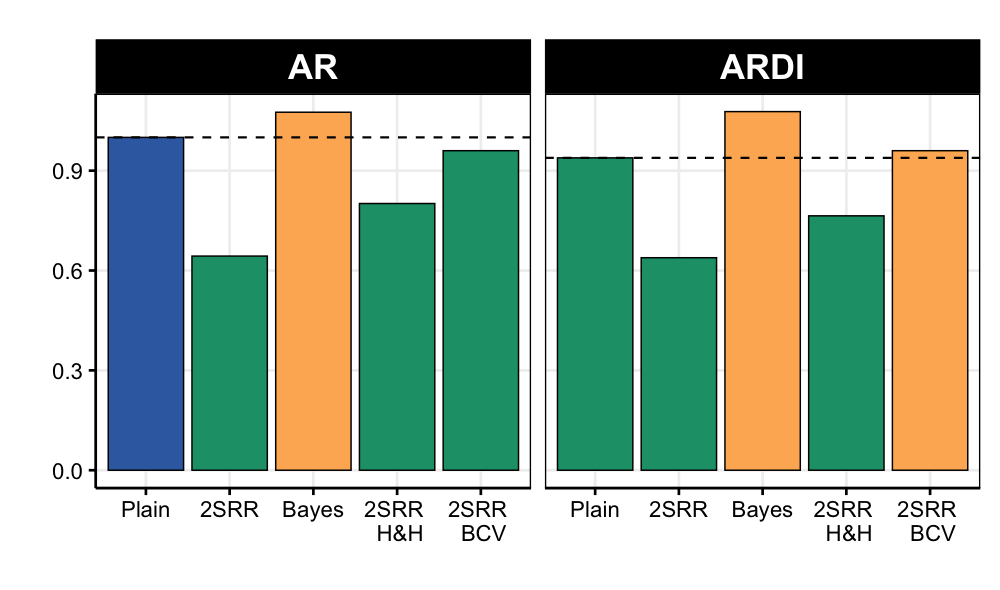}
\caption{Interest Rate ($h=1$)}  
\label{vh43}
      \end{subfigure}
      \vspace{-0.6cm}
  \caption{\footnotesize A subset of $RMSPE_{v,h,m}/RMSPE_{v,h,\text{Plain AR(2)}}$'s (from Tables \ref{tab_fcst_main} and \ref{tab_fcst_modavg}) for forecasting targets usually associated with the need for time variation. Blue is the benchmark AR with constant coefficients.  Dark green means that the competing forecast rejects the null of a Diebold-Mariano test at least at the 10\% level (with respect to the benchmark). Orange means that it does not.  "Bayes" corresponds to results from Bayesian approach,  precisely the Triple Gamma prior model as available from the \texttt{ShrinkTVP} package.}
   \label{results_fig}
\end{figure}

%\textbf{\color{red} {RECHECK}} 
\vskip 0.2cm

{\sc \noindent \textbf{Blocked CV and an Overall Assessment}.} Coming to Table \ref{tab_fcst_main_bb}, we see that switching from CV to Blocked CV offers comprehensive gains.  Large losses to constant parameters models for real activity variables have been substantially mitigated and sometimes turned into marginal gains (for instance,  ARDI-2SRR,  unemployment at $h=1$).   Overall,  at $h=1$, gains from BCV are located where those of  Half \& Half are,  which is intuitive.   Things are more heterogeneous at longer horizons.  The most notable amelioration is that of inflation,  where 2SRR (and variants) now brings important gains across the board.  BCV being such a game changer for INF requires little creativity in explaining it.  Inflation is highly persistent for a great part of the training sample,  which may have misled plain CV in choosing too low of a $\lambda$.  Note that, interestingly,  BCV does not fare as well for IR as other strategies did.  One way to rationalize this is that  "arbitrarily"  opting for a lower $\lambda$ through non-blocked CV may have helped with the massive and rapid structural change that occurred when the zero-lower bound was hit in the test sample.   We can posit that this event with no historical precedent likely induced desirable time variation in  $\lambda$ itself.  Thus,  for a panoply of unforeseen reasons,  it is absolutely possible that $\lambda^\text{CV}$ is closer to the ex-post optimal $\lambda$ than $\lambda^\text{BCV}$,  even if BCV dominates CV in principle for dependent data.

%\textbf{\color{red} {RECHECK}}  

To a large extent, forecasting results suggest that the three main algorithms presented in Section \ref{sec:2SRR} can procure important gains for forecasting targets that are frequently associated with the need for time variation. This subset is put on the spotlight by Figure \ref{results_fig}. The gains for IR at $h=1$ and INF at the two quarters and one-year horizon are particularly visible. For those kinds of targets, it is observed that any form of time-variation will usually ameliorate the constant parameters benchmark, especially in the Half \& Half and BCV cases.  We also see that those perform comparably well or better than a state-of-the-art Bayesian approach,  precisely the Triple Gamma prior model from \cite{cadonna2020triple} using default hyperparameters.\footnote{Marginally less competitive results were obtained using the \texttt{ShTVP}-R configuration also used in simulations,  so I only report\texttt{ShTVP}-3G for parsimony. }  Of course, both the ridge and Bayesian approach have distinct merits,  but it is interesting to note that \textit{ShrinkTVP} marginal advantage in simulations seems to be reversed in this real data application.  %This is convenient given how easy and fast ridge-based forecasts are to generate.  

%, in stark contrast to the typical Bayesian machinery. 

%you always made
 
\section{Time-Varying Effects of Monetary Policy in Canada}\label{sec:app}

%\subsection{Time-Varying Local Projections}
This section shows how to conduct both a TVP-VAR analysis and TVP-local projections (henceforth TVP-LP) on a data set where currently available methods would struggle.   In particular, I study the changing effects of monetary policy (MP) in Canada using the MP shocks series of \cite{champagnesekkel2018}.  The computational difficulties come from mainly two reasons.  First, it is monthly data for 40 years,  which amounts to 480 (generally noisy) observations.  Many traditional methods have computations  which scale badly in $T$,  and as such,  fare better on quarterly data.  Aggregating to quarterly is not an option because it would complicate identification,  by needlessly aggregating into a simultaneous unit things that happened in different months.  A far more challenging obstacle is that 24 lags need to be included in the model for VAR-based impulse response functions to have the "right" sign.  This is true for OLS but also for TVP specifications tried here.  Accordingly,  the TVP-VAR generalization of the "correct" OLS specification will have 97 parameters in each equation for the 4-variate case, and 193 in the 8-variate case.\footnote{Thus, computational struggle mainly comes from conditional means rather the covariance matrix,  but as we have seen computations for the latter scales rather gently in $M$ when using the tools provided in Section \ref{MV}.} Similarly daunting number of lags are utilized in their LP specifications.  

I use the same monthly Canada data set as in \cite{champagnesekkel2018} and the analysis spans from 1976 to 2015. The target variables are unemployment, CPI Inflation and GDP.\footnote{For reference, the three time series being modeled and the shock series can be visualized in Figure \ref{can_ts}. Notably, we can see that the conquest of Canadian inflation (prior to Covid) was done in two steps: reducing the mean from roughly 8\% to 5\% in the 1980s and from 5\% to 2\% in the early 1990s.} The small open economy went  through important structural change over the last 30-40 years. Most importantly, from a monetary policy standpoint, it became increasingly open (especially following NAFTA) and an inflation targeting regime (IT) was implemented in 1991 -- a specific and publicly known date.  Both are credible sources of structural change in the transmission of monetary policy.

\vskip 0.2cm

{\sc \noindent \textbf{Specification \& Implementation : TVP-VAR}.}  \cite{champagnesekkel2018} estimate a parsimonious VARs (4 variables) over two non-overlapping subsamples to check visually whether a break occurred in 1992 following the onset of IT. The reported evidence for a break is rather weak with GDP's response increasing slightly while that of inflation decreasing marginally. While the sample-splitting approach has many merits such as transparency and simplicity, there is arguably a lot it can miss.  

I consider a TVP-VAR(4) with their four original predictors (GDP,  inflation,  commodity prices,  MP shocks),  and a TVP-VAR(8) including additionally unemployment,  USD-CAD exchange rate,  exports,  and imports.   In such dimensions where the time-invariant model itself can overfit,  it is preferable to set $\lambda_0$ in equation \eqref{genprobtvp} (regularization for $\beta_0$) to a value greater than 0.  I set it to the value obtained by cross-validating a plain (thus time-invariant) ridge regression with $\boldsymbol{X}$ as predictors.  Evidently,  this is an educated guess for what one could obtain cross-validating the whole thing over a grid of $\lambda$ and $\lambda_0$.    Also,  the "prior mean" for $\beta_0$ is re-centered at the values corresponding to the coefficients of the time-invariant ridge regression.   As discussed in the forecasting experiments,  opting for a Blocked CV that takes into account the dependence between data points can help in avoiding overfitting. The block size is set to 24 months.  $\varphi$, the sole tuning parameter of the covariance matrix,  is set to 1000,  which made the variances processes move at about the same speed as we typically see in stochastic volatility models.  Increasing it in 2000-3000 range was found to eventually eradicate time variation and decreasing it below 500 made appear unrealistic higher frequency movements (especially in the covariances) that random walks are not built to capture efficiently.   The ordering in the TVP-VAR(4) is the same as in \cite{champagnesekkel2018} (with the MP shocks entering last) and the additional 4 variables in TVP-VAR(8) are also added before the MP shock.\footnote{The exact ordering for the TVP-VAR(4) is GDP,  inflation,  commodity price index,  MP shock.  In TVP-VAR(8),  we have GDP,  inflation,  commodity price index,   unemployment,  USD-CAD exchange rate,  exports, imports,  MP shock. } Lastly,   relative IRFs (dividing by the contemporaneous  effect of MP shock on itself) are reported as they are directly comparable to TVP-LPs \citep{plagborg2021local} and that each IRFs desirably corresponds to a unit shock at each $t$.  

\begin{comment}
usa : Paymens,
WPUSI012011_CCH	W875RX1_PC1	DPCERA3M086SBEA_PC1	INDPRO_PC1	TCU_CHG
HOUST_CH1	WPSFD49207_PCA	PCEPI_PCA	CES0500000030_CCH
M2SL_PCA 
M1SL_PCA	WILL5000INDFC_CCH	GS10_CHG	RNUSBIS_PCH
6 lags TVP.Ridge (3:07,07), IRF (0:28,57)
12 lags TVP.Ridge (5:15,46), IRF (0:46,23)
\end{comment}

In terms of computational results,  the TVP-VAR(4) with a total $4 \times 97$ conditional mean TVPs took 25 seconds to estimate and tune.  It took an additional  6 seconds to calculate 10 (co)variances TVPs \textit{and} calculate/generate IRFs for 48 horizons and 4 variables.  The TVP-VAR(8),  with $8 \times 193$ conditional mean TVPs and 36 additional TVPs in the covariance matrix,  took 1:17 minute for the first step, and 17 seconds for the second.  Hence,  2SRR lives up to the claim that it can estimate  demanding models very quickly with little user interaction.

\begin{figure}[htp!]
%  \begin{subfigure}[b]{0.32\textwidth}
  % \includegraphics[width=\textwidth]{graphs/variables4/CPI}
    %\caption{Unemp  -- TVP-VAR(4)  \label{lp_ur}}
  %\end{subfigure}
  \hspace{0.32\textwidth}
  \begin{subfigure}[b]{0.32\textwidth}
   \includegraphics[width=\textwidth]{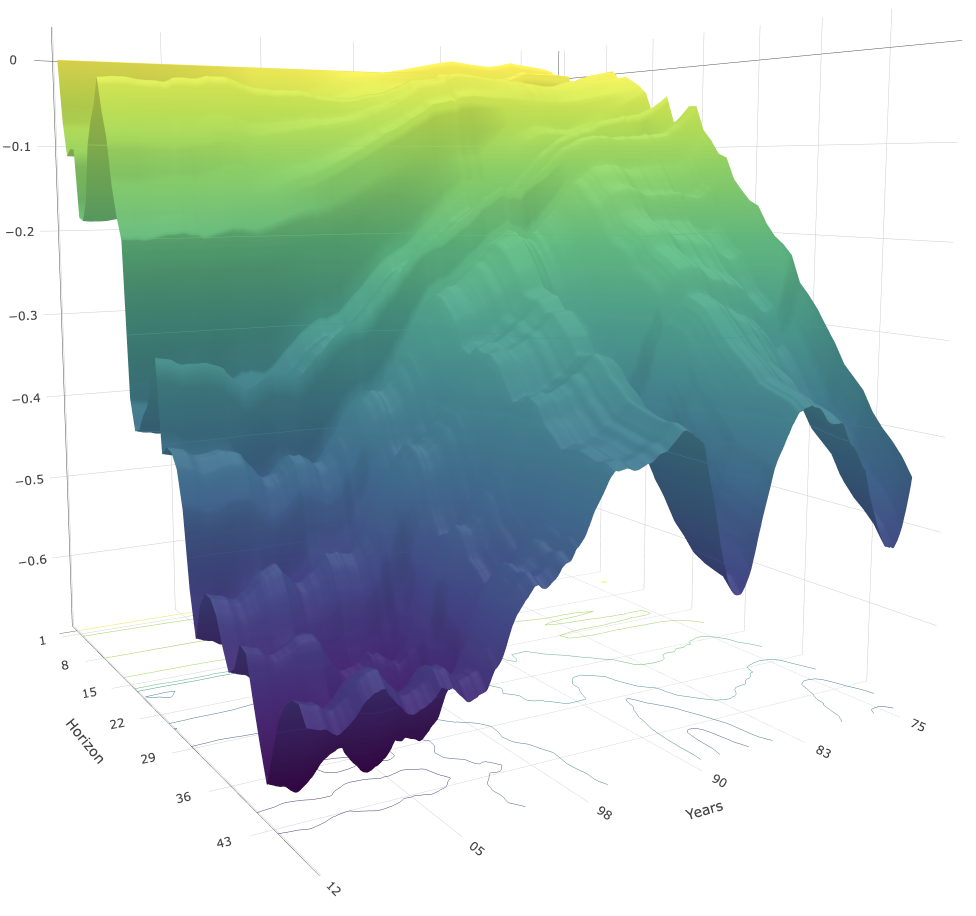}
    \caption{Inflation  -- TVP-VAR(4)   \label{lp_inf}}
  \end{subfigure}
  \centering 
    \begin{subfigure}[b]{0.32\textwidth}
   \includegraphics[width=\textwidth]{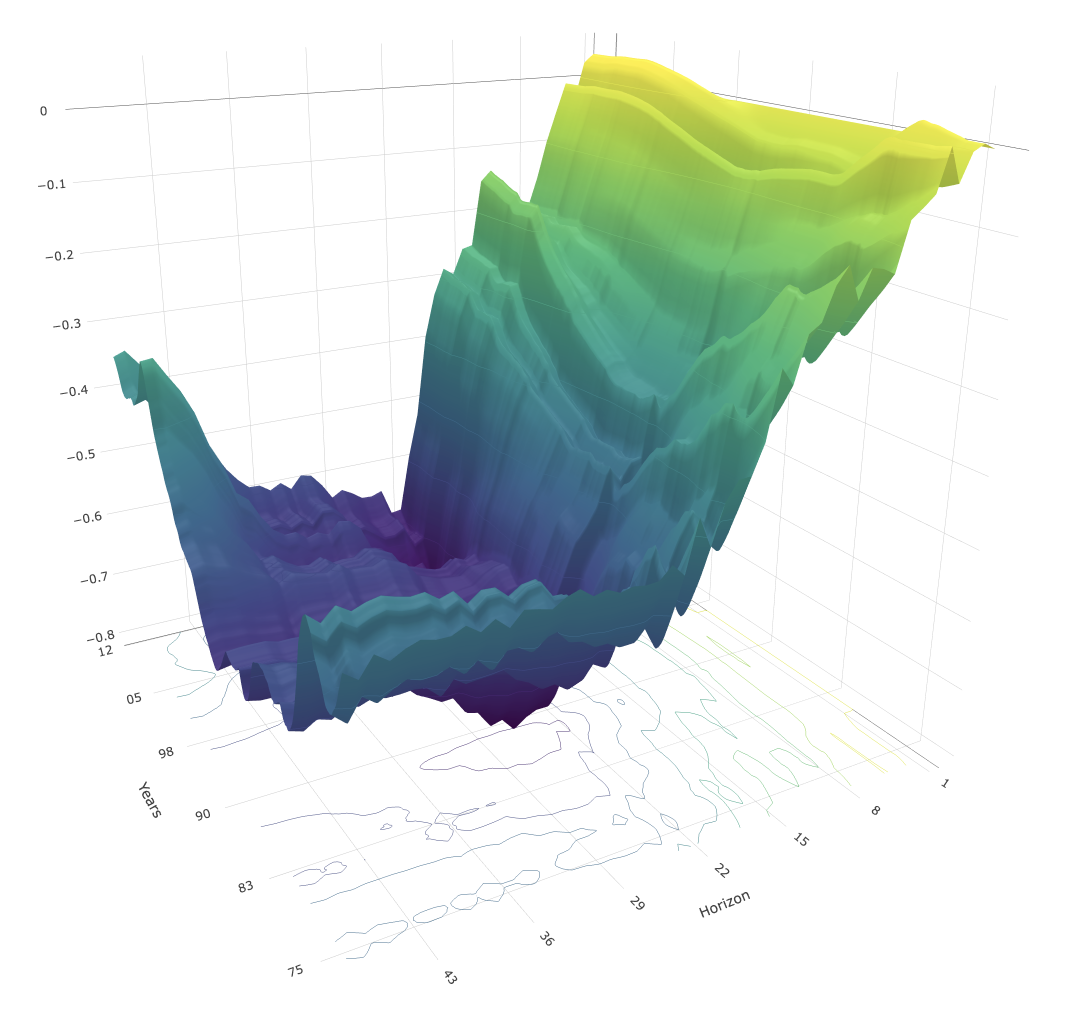}
    \caption{GDP -- TVP-VAR(4)  \label{lp_gdp}}
  \end{subfigure}
    \begin{subfigure}[b]{0.32\textwidth}
   \includegraphics[width=\textwidth]{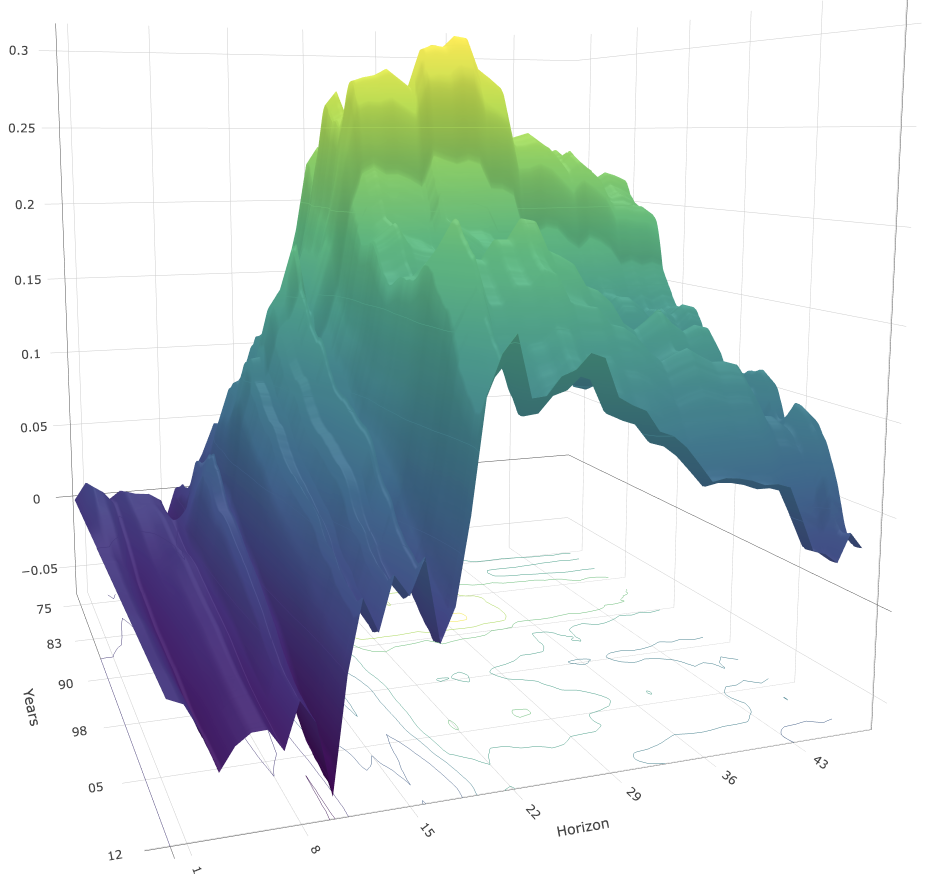}
    \caption{Unemp -- TVP-VAR(8) \label{lp_ur}}
  \end{subfigure}
  %\hspace{1em}
  \begin{subfigure}[b]{0.32\textwidth}
   \includegraphics[width=\textwidth]{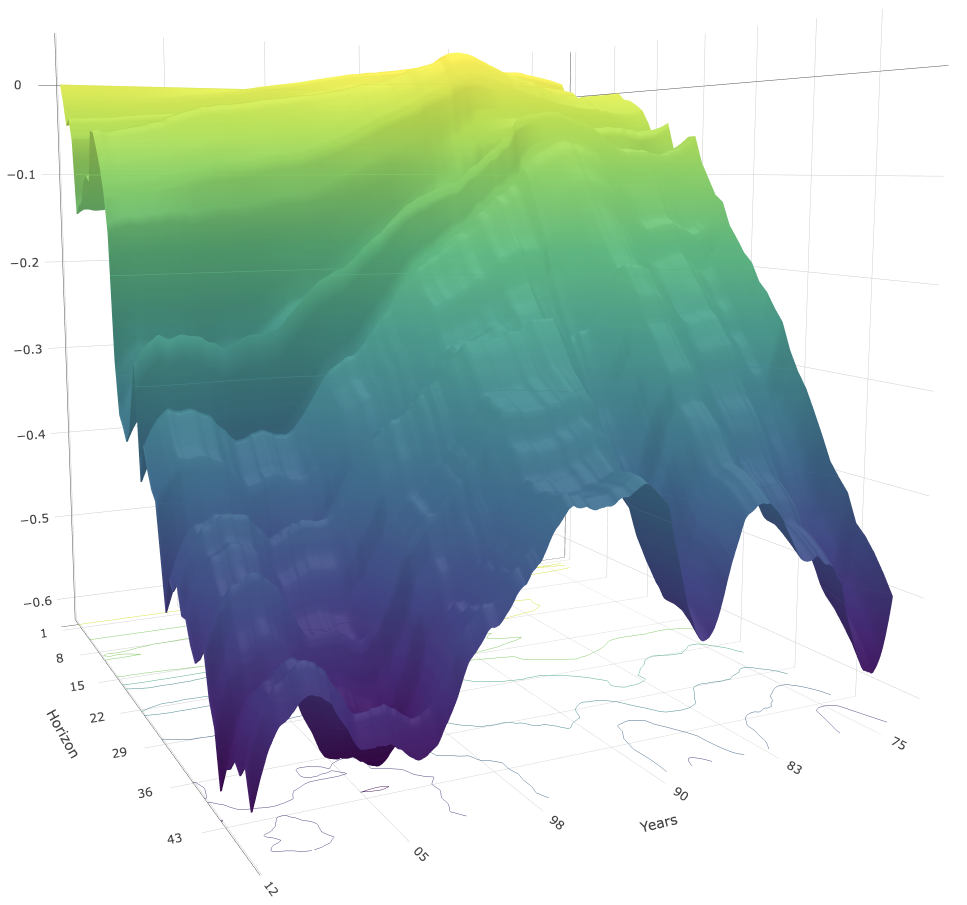}
    \caption{Inflation -- TVP-VAR(8)  \label{lp_inf}}
  \end{subfigure}
  \centering 
    \begin{subfigure}[b]{0.32\textwidth}
   \includegraphics[width=\textwidth]{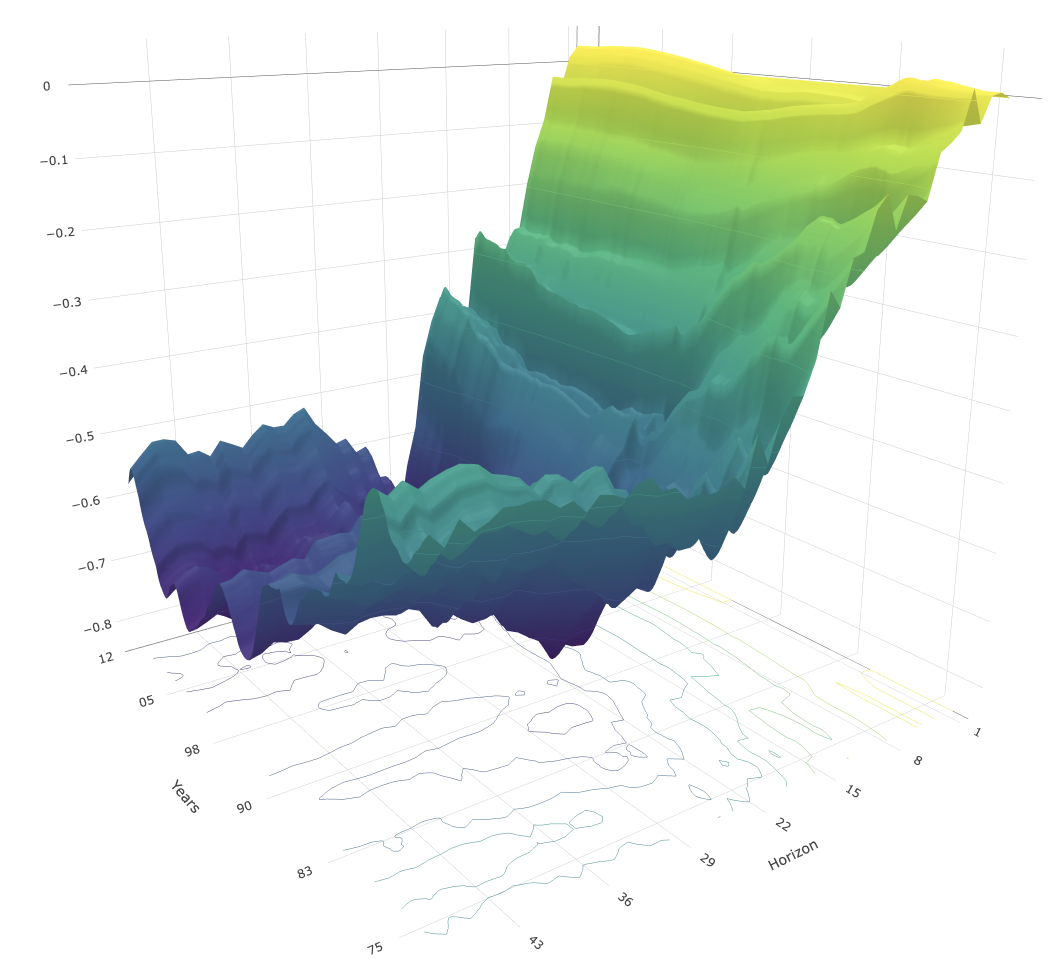}
    \caption{GDP -- TVP-VAR(8)  \label{lp_gdp}}
  \end{subfigure}
    \begin{subfigure}[b]{0.32\textwidth}
   \includegraphics[width=\textwidth]{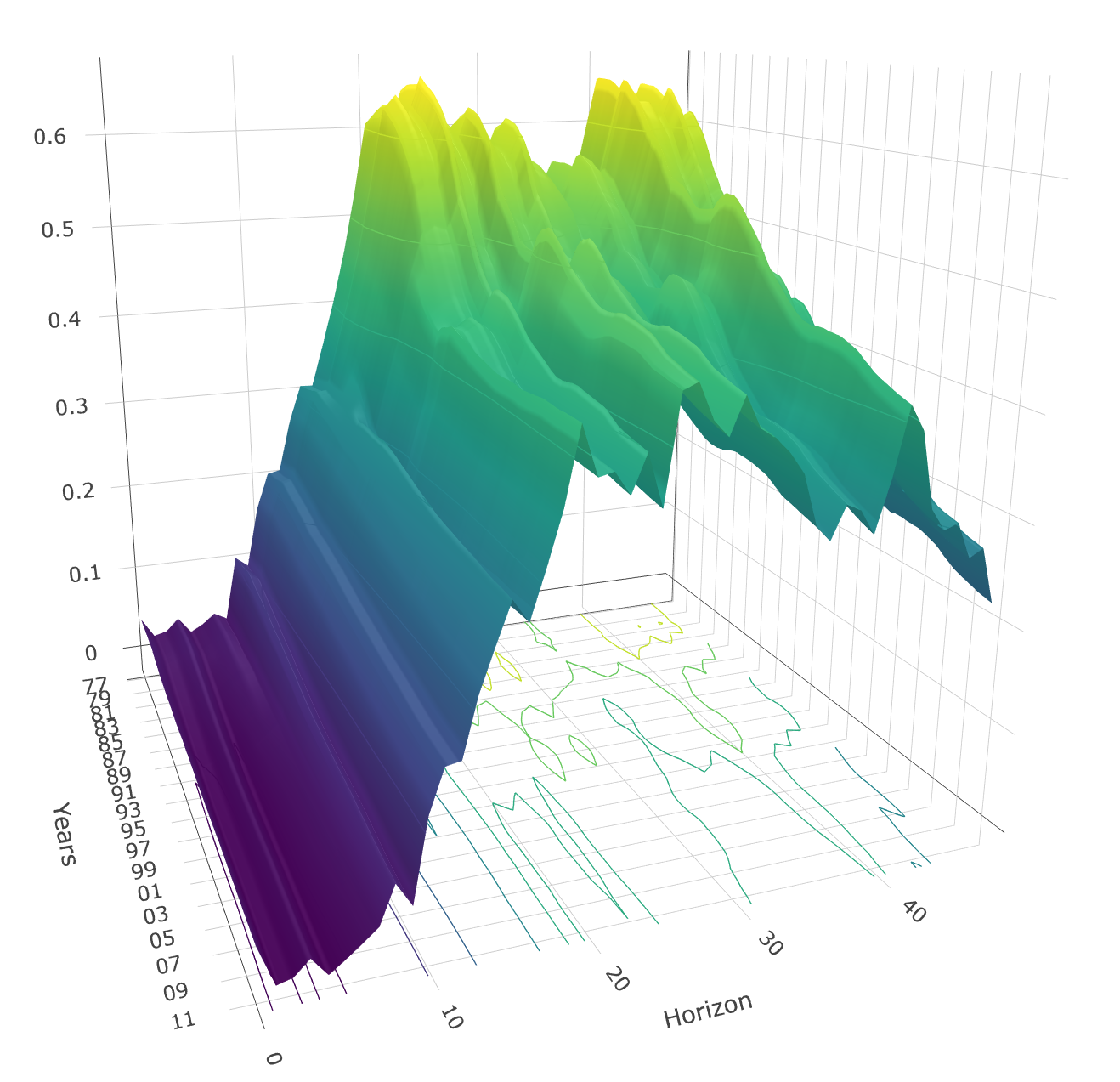}
    \caption{Unemp -- TVP-LP \label{lp_ur}}
  \end{subfigure}
  \hspace{1em}
  \begin{subfigure}[b]{0.32\textwidth}
   \includegraphics[width=\textwidth]{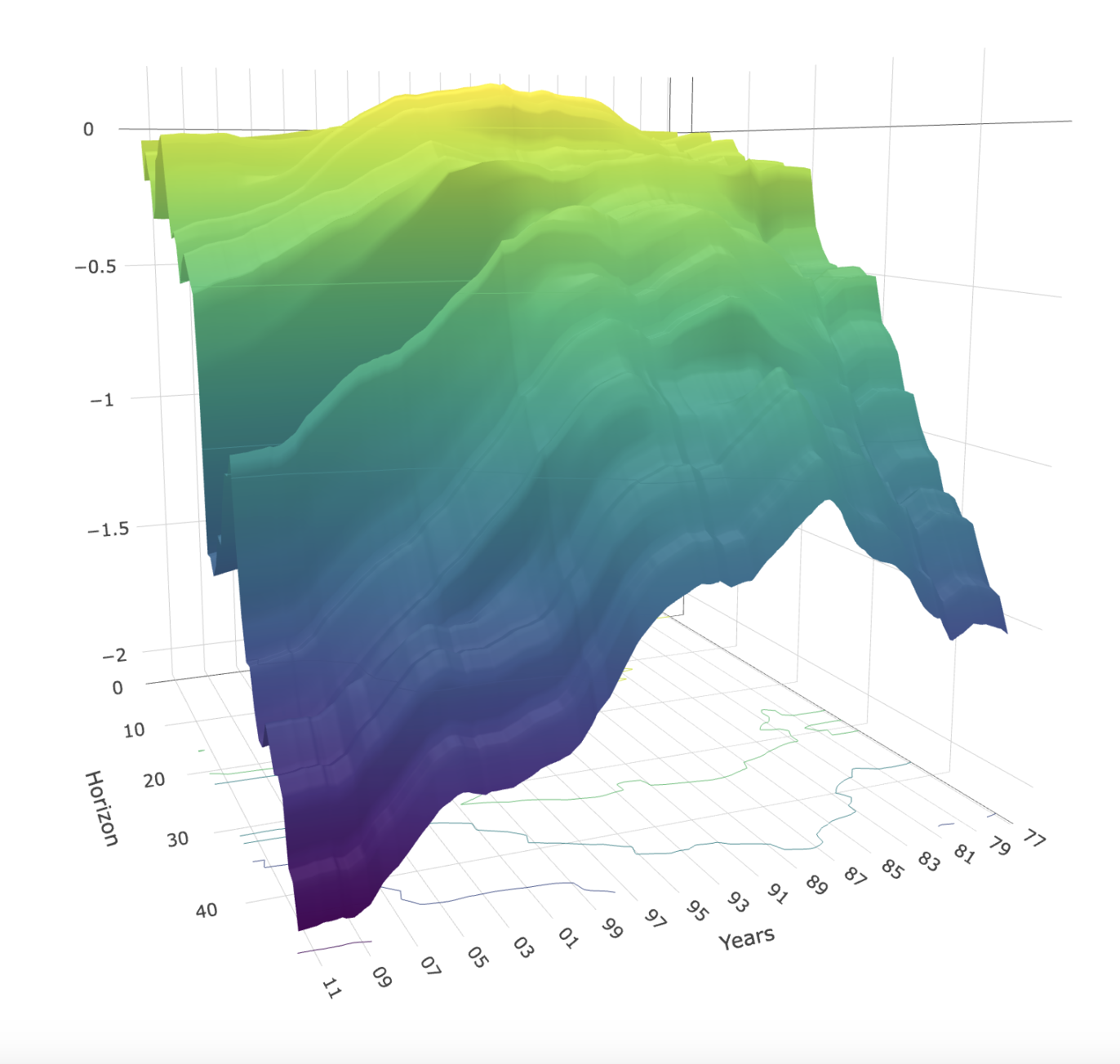}
    \caption{Inflation -- TVP-LP  \label{lp_inf}}
  \end{subfigure}
  \centering 
    \begin{subfigure}[b]{0.32\textwidth}
   \includegraphics[width=\textwidth]{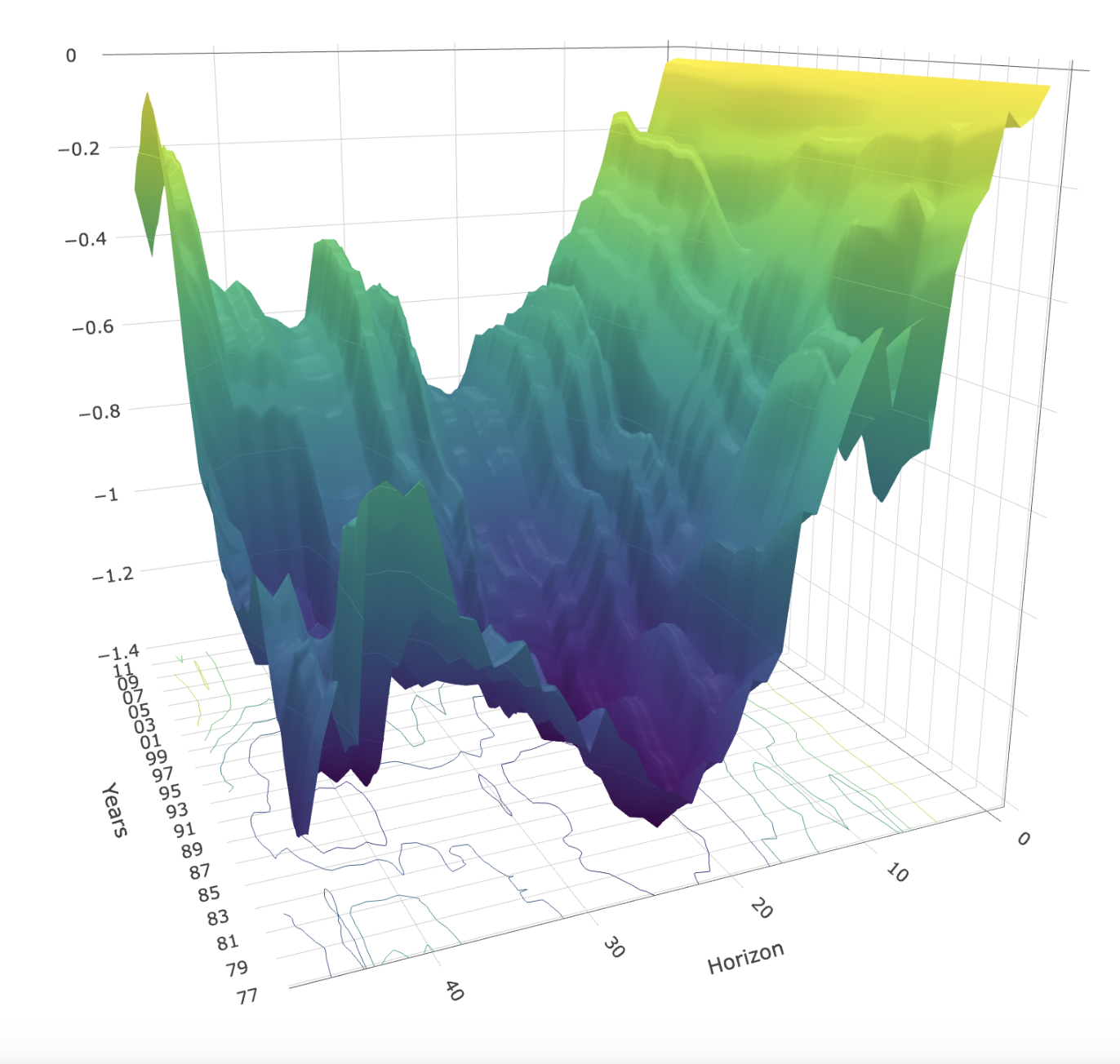}
    \caption{GDP -- TVP-LP  \label{lp_gdp}}
  \end{subfigure}
  \caption{Cumulative Time-Varying Effect of Monetary Policy Shocks from TVP-VAR(4) (which excludes unemployment),  TVP-VAR(8),  and Local Projections.  Rotations of 3D plots are hand-picked to highlight most salient features of each time-varying IRF. Interactive plots where the reader can manually explore different rotations are available \href{https://david-wigglesworth.shinyapps.io/for_pgc/}{here}.}
    \label{lp_main}
\end{figure}

\vskip 0.2cm

{\sc \noindent \textbf{Specification \& Implementation : TVP-Local Projections}.}  VARs do not have a monopoly on the proliferation of parameters. \cite{jorda2005} local projections' -- by running a separate regression for each horizon -- are also densely parametrized. For that reason, constructing a large LP-based time-varying IRF via a MCMC procedure would either be burdensome or unfeasible.  The use of 2SRR for estimation of LPs constitutes a very useful methodological development given how popular local projections have become over recent years.  It is now known that IRFs from LPs and VARs (coming from linear models) are intimately linked \citep{plagborg2019local}.  However,  here,  since we are modeling time variation in a different space than the TVP-VAR (i.e.,  directly in IRF-space rather than in reduced-form VAR coefficients and then the covariance matrix),  it is natural to expect that results can differ for reasons other than finite sample estimation variability.  Moreover,  cross-validation is done for direct forecasts at each  horizon in LPs whereas it is done for each variable and one-step ahead forecasts in the TVP-VAR case.  Hence,  from a methodological standpoint,  it is interesting to compare TVP-VAR and TVP-LP results on a terrain where full agreement is not necessarily the expected outcome. 

\cite{champagnesekkel2018} original LP specification includes 48 lags of the narrative monetary policy (MP) shock series which is constructed in the spirit of \cite{romerromer2004} and carefully adapted to the Canadian context.\footnote{For details regarding the construction of the crucial series --- especially on how to account for the 1991 shift to IT, see \cite{champagnesekkel2018}. Note that a positive shock means (unexpected) MP tightening.}  Furthermore, their regression comprises 4 lags for the controls which are first differences of the log GDP, log inflation and log commodity prices. To certify that time-variation will not be found as a result of omitted variables, I increase the lag order from 4 to 6 months and augment the model with the USD/CAD exchange rate, exports, imports and CPI excluding Mortgage Interest Cost (MIC). In terms of TVP accounting, $\boldsymbol{X}$ contains 97 regressors (including the constant) and $\boldsymbol{Y}$ is $48 \time 419$. Thus, a single TV-LP is assembled from a staggering total of $97 \times 48=4\,656$ TVPs.  The computing time one IRF is 26 minutes.  Thus,  an evident advantage of the VAR approach is that by estimating a single generative model,  one can get  results considerably faster.

\vskip 0.2cm

{\sc \noindent \textbf{Results}.}  Current evidence on the effects of IT is mixed.  \cite{ramey2016hb} and \cite{barakchian2013monetary} obtain counterintuitive results (a price puzzle and MP tightening increases GDP) for post-1988 US data.  \cite{cloyne2016macroeconomic} and \cite{champagnesekkel2018} report more intuitive findings for the UK and Canada: GDP response increases marginally after IT and that of inflation shrinks.   Figure \ref{lp_main} suggests new answers to the evolving effect of monetary policy on the Canadian economy based on 3 TVP models.   Results mostly follow a qualitative common thread.   There are quantitative differences,  but this is expected that as even OLS versions of LPs vs. VARs give quantitatively different answers in    \cite{champagnesekkel2018}.  Yet,   in a pairwise fashion,  TVP-LPs signs and magnitudes are consistent with those reported in \cite{champagnesekkel2018},  and TVP-VARs similarly concur with time-invariant VARs.   Thus,  when comparing IRFs,  it is good to keep in mind that there are two potential sources of dissimilarity: those that emanate from the linear specification itself,  and those that will be attributable to how time variation is captured.

Generally,  short- and medium-run effects ($h<24$ months) have been much more stable than longer-run ones.  This is especially true of unemployment which exhibits a very homogeneous response (through time) for the first year and half after the shock.  GDP's response roughly follows a similar pattern.  When it comes to inflation, its usual long response lag has mildly shortened in the 2000s,  a phenomenon that is visible in all 3 specifications.   For instance,  in the TVP-LP case,  at a horizon of 24 months, the effect of a positive one standard deviation shock was -0.3\% from 1976 to the late 1990s, then slowly increased (in absolute terms) to nearly double at -0.6\%.  A similar relative decrease is observed for TVP-VARs,  but their overall absolute size throughout the sample is smaller.  This is in line with \cite{champagnesekkel2018}'s time-invariant LPs vs. VAR where the overall  IRF magnitudes of the latter are always smaller than the former.  Nonetheless,  all results for horizons up to 18 months suggest that the ability of the central bank to "rapidly" impact inflation has increased.  Despite these insights coming from a sample ending in the mid-2010s,  they appear roughly in accord with the post-Pandemic Canadian inflation experience,  where the Bank of Canada's rate hikes arguably had a material impact (among a myriad of confounding factors) within 1.5 years.

Given the long lags of monetary policy, most of the relevant action from an economic standpoint is also where most time variation is found: from 1.5 to 4 years after the shock. For the two models that include it,  the cumulative long-run effects of MP shocks on unemployment have substantially shrunk over the sample period.  For TVP-LP , the decrease from a 0.6 to 0.4 unemployment percentage points mostly occurred throughout the 1980s, and stabilized at 0.4 thereafter.  In TVP-VAR(8),  the overall effect (even in the OLS case) is more muted,  but the reduction of long-term effects is also visible and of similar relative size.  Additionally,  both models showcase the same attenuation pattern for mid-range horizons,  that is,  the peak effect shrinking to about $\sfrac{2}{3}$ of its early 1980s value.    All inflation IRFs show a declining pattern starting from the early 1990s,  but that of TVP-LP is more visible given its overall smoother behavior.  Pre-1990,  we see TVP-VAR based IRFs displaying mild business-cycle frequency movement,  but those disappear post-1990.

Finally,  GDP IRFs all display a U-shape pattern,  with a trend reversal in the strength of long-run effects happening in the 1990s.  However,  there is discord between TVP-VAR-based estimates as to whether the effects were milder at the beginning or the end of the sample.  TVP-VAR(4) and TVP-LPs agree on it being at the end,  in line with results for unemployment,  but TVP-VAR(8) sees this happening in the late 1970s instead.   Obviously,  decisively choosing between one model would require further work,  and in the machine learning spirit,  the model (here,  direct vs. iterative TVP-VAR forecasts) that predicts best for the contentious horizons should be preferred.  Nonetheless,  key findings are homogeneous across specifications,  that is,  there is (i) a trend reversal in the strength long-run effects of MP shocks on inflation and GDP (getting stronger for the former and weaker for the latter) and (ii) the effects on unemployment became milder starting from the 1990s for all $h$'s greater than 18 months.

\vskip 0.2cm

{\sc \noindent \textbf{Zooming on 1992}.} An important question is what happens to $\beta_t$ around 1992,  after the onset of IT.   Neither TVP-VAR nor TVP-LPs  suggest the occurrence of a structural break in 1992, which is in line with most of the international evidence on IT implementation.  For inflation,  TVP-LPs' results point to a change in coefficients' trending behavior, a subtle phenomenon which would effectively stay under the radar of simpler approaches.  For illustrative purposes,  Figure \ref{TVLPwrtOLS} (Appendix) reports $\beta_{t}^{\text{TVP-LP}}-\beta^{\text{OLS}}$ for the dynamic effect of MP shocks on the three variables.   The response of inflation (in absolute terms) is much larger at the end of the sample than what constant coefficients would suggest.  This is especially true at the 24 months horizon. It is quite clear that, for all horizons, the effect of MP shocks on inflation starts increasing in the years following the implementation of IT.  The downward trend in MP shocks' impact on unemployment seems to have started at least since the 1970s and slowed down in recent years. 

More generally, it is interesting to note that those results are consistent with some flattening of the unemployment- or GDP-based Phillips' curve (but maybe not a more general one,  \citealt{HNN}) which was a well-established observation,  at least,  before the COVID-19 pandemic \cite{blanchard2015inflation,del2020s}.  The combined evidence displayed in Figure \ref{lp_main}  is that,   while the cumulative effect of MP shocks either stood still or became more muted starting from the 1990s for real activity variables, it has increased for inflation. This suggests that stabilizing Canadian inflation is  less costly (in terms of unemployment/GDP variability) within the IT framework.  Current evidence from the post-Pandemic rates hikes cycle seems to partly validate this view.

\vskip 0.2cm

{\sc \noindent \textbf{Testing Limits}.} It is not because we can compute something that we necessarily want to.  To test the limits of the ridge approach,  I also computed results from a TVP-VAR(23), again including the necessary 24 lags, which makes it for a model that, even with time-invariant coefficients, has more parameters than observations ($23 \times 553 + 276$ in total).  The first 8 variables and that of TVP-VAR(8) and the remaining 15 are the US variables included in UK database constructed in \cite{GCMS} and can be seen as a reasonably sparse characterization  of economic conditions facing Canada's main commercial partner.  In such an environment,  it is difficult to avoid the ridgeless solution mentioned earlier,  and all IRFs are much smaller in magnitude falling under the weight of heavy shrinkage.   Nonetheless, it can be computed.  It takes 10 minutes to estimate and less than 3 minutes to run the code evaluating the covariance matrix, calculating IRFs,  and generating 3D plots.  Hence,  it takes less than 15 minutes to obtain results.  However,  in such dimensions, things are heavily compressed and time-variation patterns obtained do not seem as economically plausible.   Reducing the number of lags to 12 and 6,   IRFs become less compressed and ridgeless issues go away,  but as discussed in the intro of this section,  puzzles emerge.  Computations are also much less demanding, with the whole operation taking 3:30 minutes.  Thus,  in this environment where an abundant number of lags appear to be necessary,  the number of variables in the system cannot be too large.

\section{Conclusion}

I provide a new framework to estimate TVP models with potentially evolving volatility of shocks. It is conceptually enlightening and computationally very fast. Moreover, seeing such models as ridge regressions suggest a simple way to tune the amount of time variation, a consequential quantity. The approach is easily extendable to have additional shrinkage schemes like sparse TVPs or reduced-rank restrictions. The proposed variants of the methodology are very competitive against the standard Bayesian TVP-VAR in simulations. Furthermore, they improve forecasts against standard forecasting benchmarks for variables usually associated with the need for time variation (US inflation and interest rates). Finally, I apply the tool to estimate time-varying IRFs via VAR and local projections. The large specifications necessary to adequately characterize  the evolution of monetary policy in Canada rendered this application nearly unfeasible without the newly developed tools.  I report that monetary policy shocks long-run impact on the price level increased substantially starting from the early 1990s (onset of inflation targeting), whereas the effects on unemployment became milder.  %This finding is consistent with the hypothesized flattening of the Phillips' curve in advanced economies.

\clearpage
%\singlespacing

\setlength\bibsep{5pt}
		
\bibliographystyle{apalike}

\setstretch{0.3}
\bibliography{/Users/UQAM/Dropbox/pgc_bib/ref_pgc_v181204}

\begin{thebibliography}{}

\bibitem[Amir-Ahmadi et~al., 2018]{amir2018choosing}
Amir-Ahmadi, P., Matthes, C., and Wang, M.-C. (2018).
\newblock Choosing prior hyperparameters: with applications to time-varying
  parameter models.
\newblock {\em Journal of Business \& Economic Statistics}, pages 1--13.

\bibitem[Bai and Ng, 2017]{bai2017principal}
Bai, J. and Ng, S. (2017).
\newblock Principal components and regularized estimation of factor models.
\newblock {\em arXiv preprint arXiv:1708.08137}.

\bibitem[Ba{\'n}bura et~al., 2010]{banbura2010large}
Ba{\'n}bura, M., Giannone, D., and Reichlin, L. (2010).
\newblock Large bayesian vector auto regressions.
\newblock {\em Journal of Applied Econometrics}, 25(1):71--92.

\bibitem[Barakchian and Crowe, 2013]{barakchian2013monetary}
Barakchian, S.~M. and Crowe, C. (2013).
\newblock Monetary policy matters: Evidence from new shocks data.
\newblock {\em Journal of Monetary Economics}, 60(8):950--966.

\bibitem[Bartlett et~al., 2020]{bartlett2020benign}
Bartlett, P.~L., Long, P.~M., Lugosi, G., and Tsigler, A. (2020).
\newblock Benign overfitting in linear regression.
\newblock {\em Proceedings of the National Academy of Sciences}.

\bibitem[Baumeister and Kilian, 2014]{BaumeisterKilian2014}
Baumeister, C. and Kilian, L. (2014).
\newblock What central bankers need to know about forecasting oil prices.
\newblock {\em International Economic Review}, 55:869--889.

\bibitem[Belkin et~al., 2019]{belkin2019reconciling}
Belkin, M., Hsu, D., Ma, S., and Mandal, S. (2019).
\newblock Reconciling modern machine-learning practice and the classical
  bias--variance trade-off.
\newblock {\em Proceedings of the National Academy of Sciences},
  116(32):15849--15854.

\bibitem[Belmonte et~al., 2014]{belmonte2014hierarchical}
Belmonte, M.~A., Koop, G., and Korobilis, D. (2014).
\newblock Hierarchical shrinkage in time-varying parameter models.
\newblock {\em Journal of Forecasting}, 33(1):80--94.

\bibitem[Bergmeir et~al., 2018]{bergmeir2018note}
Bergmeir, C., Hyndman, R.~J., and Koo, B. (2018).
\newblock A note on the validity of cross-validation for evaluating
  autoregressive time series prediction.
\newblock {\em Computational Statistics \& Data Analysis}, 120:70--83.

\bibitem[Bitto and Fr{\"u}hwirth-Schnatter, 2018]{bitto2018achieving}
Bitto, A. and Fr{\"u}hwirth-Schnatter, S. (2018).
\newblock Achieving shrinkage in a time-varying parameter model framework.
\newblock {\em Journal of Econometrics}.

\bibitem[Blanchard et~al., 2015]{blanchard2015inflation}
Blanchard, O., Cerutti, E., and Summers, L. (2015).
\newblock Inflation and activity--two explorations and their monetary policy
  implications.
\newblock Technical report, National Bureau of Economic Research.

\bibitem[Boivin, 2005]{boivin2005has}
Boivin, J. (2005).
\newblock Has us monetary policy changed? evidence from drifting coefficients
  and real-time data.
\newblock Technical report, National Bureau of Economic Research.

\bibitem[Cadonna et~al., 2020]{cadonna2020triple}
Cadonna, A., Fr{\"u}hwirth-Schnatter, S., and Knaus, P. (2020).
\newblock Triple the gamma—a unifying shrinkage prior for variance and
  variable selection in sparse state space and tvp models.
\newblock {\em Econometrics}, 8(2):20.

\bibitem[Carriero et~al., 2011]{CKM2011}
Carriero, A., Kapetanios, G., and Marcellino, M. (2011).
\newblock Forecasting large datasets with bayesian reduced rank multivariate
  models.
\newblock {\em Journal of Applied Econometrics}, 26(5):735--761.

\bibitem[Castle et~al., 2015]{castle2015detecting}
Castle, J.~L., Doornik, J.~A., Hendry, D.~F., and Pretis, F. (2015).
\newblock Detecting location shifts during model selection by step-indicator
  saturation.
\newblock {\em Econometrics}, 3(2):240--264.

\bibitem[Champagne and Sekkel, 2018]{champagnesekkel2018}
Champagne, J. and Sekkel, R. (2018).
\newblock Changes in monetary regimes and the identification of monetary policy
  shocks: Narrative evidence from canada.
\newblock {\em Journal of Monetary Economics}, 99:72--87.

\bibitem[Chan and Eisenstat, 2018]{chaneisenstat2018}
Chan, J.~C. and Eisenstat, E. (2018).
\newblock Comparing hybrid time-varying parameter vars.
\newblock {\em Economics Letters}, 171:1--5.

\bibitem[Chan et~al., 2018]{chan-ftvp2018}
Chan, J.~C., Eisenstat, E., and Strachan, R.~W. (2018).
\newblock {Reducing dimensions in a large TVP-VAR}.
\newblock CAMA Working Papers 2018-49, Centre for Applied Macroeconomic
  Analysis, Crawford School of Public Policy, The Australian National
  University.

\bibitem[Chen and Hong, 2012]{chen2012testing}
Chen, B. and Hong, Y. (2012).
\newblock Testing for smooth structural changes in time series models via
  nonparametric regression.
\newblock {\em Econometrica}, 80(3):1157--1183.

\bibitem[Chernozhukov et~al., 2018]{chernozhukov2018exact}
Chernozhukov, V., W{\"u}thrich, K., and Yinchu, Z. (2018).
\newblock Exact and robust conformal inference methods for predictive machine
  learning with dependent data.
\newblock In {\em Conference On learning theory}, pages 732--749. PMLR.

\bibitem[Cloyne and H{\"u}rtgen, 2016]{cloyne2016macroeconomic}
Cloyne, J. and H{\"u}rtgen, P. (2016).
\newblock The macroeconomic effects of monetary policy: a new measure for the
  united kingdom.
\newblock {\em American Economic Journal: Macroeconomics}, 8(4):75--102.

\bibitem[Cogley and Sargent, 2001]{cogley2001evolving}
Cogley, T. and Sargent, T.~J. (2001).
\newblock Evolving post-world war ii us inflation dynamics.
\newblock {\em NBER macroeconomics annual}, 16:331--373.

\bibitem[Cogley and Sargent, 2005]{cogley2005drifting}
Cogley, T. and Sargent, T.~J. (2005).
\newblock Drifts and volatilities: monetary policies and outcomes in the post
  wwii us.
\newblock {\em Review of Economic dynamics}, 8(2):262--302.

\bibitem[D'Agostino et~al., 2013]{AAG2013}
D'Agostino, A., Gambetti, L., and Giannone, D. (2013).
\newblock Macroeconomic forecasting and structural change.
\newblock {\em Journal of Applied Econometrics}, 28(1):82--101.

\bibitem[de~Wind and Gambetti, 2014]{dewindgambetti2014}
de~Wind, J. and Gambetti, L. (2014).
\newblock {Reduced-rank time-varying vector autoregressions}.
\newblock CPB Discussion Paper 270, CPB Netherlands Bureau for Economic Policy
  Analysis.

\bibitem[Del~Negro et~al., 2020]{del2020s}
Del~Negro, M., Lenza, M., Primiceri, G.~E., and Tambalotti, A. (2020).
\newblock What’s up with the phillips curve?
\newblock Technical report, National Bureau of Economic Research.

\bibitem[Del~Negro and Primiceri, 2015]{delnegroprimiceri2015}
Del~Negro, M. and Primiceri, G.~E. (2015).
\newblock Time varying structural vector autoregressions and monetary policy: a
  corrigendum.
\newblock {\em The review of economic studies}, 82(4):1342--1345.

\bibitem[Diebold and Mariano, 2002]{dieboldmariano}
Diebold, F.~X. and Mariano, R.~S. (2002).
\newblock Comparing predictive accuracy.
\newblock {\em Journal of Business \& economic statistics}, 20(1):134--144.

\bibitem[Durbin and Koopman, 2012]{durbin2012time}
Durbin, J. and Koopman, S.~J. (2012).
\newblock {\em Time series analysis by state space methods}.
\newblock Oxford university press.

\bibitem[Frommlet and Nuel, 2016]{frommletnuel2016}
Frommlet, F. and Nuel, G. (2016).
\newblock An adaptive ridge procedure for l0 regularization.
\newblock {\em PloS one}, 11(2):e0148620.

\bibitem[Giraitis et~al., 2014]{giraitis2014inference}
Giraitis, L., Kapetanios, G., and Yates, T. (2014).
\newblock Inference on stochastic time-varying coefficient models.
\newblock {\em Journal of Econometrics}, 179(1):46--65.

\bibitem[Golub et~al., 1979]{golub1979gcv}
Golub, G.~H., Heath, M., and Wahba, G. (1979).
\newblock Generalized cross-validation as a method for choosing a good ridge
  parameter.
\newblock {\em Technometrics}, 21(2):215--223.

\bibitem[Goulet~Coulombe, 2022]{HNN}
Goulet~Coulombe, P. (2022).
\newblock A neural phillips curve and a deep output gap.
\newblock {\em Available at SSRN}.

\bibitem[Goulet~Coulombe, 2024]{MRFjae}
Goulet~Coulombe, P. (2024).
\newblock The macroeconomy as a random forest.
\newblock {\em Journal of Applied Econometrics}.

\bibitem[Goulet~Coulombe et~al., 2021a]{MDTM}
Goulet~Coulombe, P., Leroux, M., Stevanovic, D., and Surprenant, S. (2021a).
\newblock Macroeconomic data transformations matter.
\newblock {\em International Journal of Forecasting}, 37(4):1338--1354.

\bibitem[Goulet~Coulombe et~al., 2022]{GCLSS2018}
Goulet~Coulombe, P., Leroux, M., Stevanovic, D., and Surprenant, S. (2022).
\newblock How is machine learning useful for macroeconomic forecasting?
\newblock {\em Journal of Applied Econometrics}, 37(5):920--964.

\bibitem[Goulet~Coulombe et~al., 2021b]{GCMS}
Goulet~Coulombe, P., Marcellino, M., and Stevanovic, D. (2021b).
\newblock Can machine learning catch the covid-19 recession?
\newblock {\em CEPR Discussion Paper No. DP15867}.

\bibitem[Grandvalet, 1998]{grandvalet1998}
Grandvalet, Y. (1998).
\newblock Least absolute shrinkage is equivalent to quadratic penalization.
\newblock In {\em ICANN 98}, pages 201--206. Springer.

\bibitem[Grant and Chan, 2017]{grant2017bayesian}
Grant, A.~L. and Chan, J.~C. (2017).
\newblock A bayesian model comparison for trend-cycle decompositions of output.
\newblock {\em Journal of Money, Credit and Banking}, 49(2-3):525--552.

\bibitem[Hamilton, 1994]{hamilton1994}
Hamilton, J.~D. (1994).
\newblock {\em Time series analysis}, volume~2.
\newblock Princeton university press Princeton, NJ.

\bibitem[Hastie et~al., 2019]{hastie2019surprises}
Hastie, T., Montanari, A., Rosset, S., and Tibshirani, R.~J. (2019).
\newblock Surprises in high-dimensional ridgeless least squares interpolation.
\newblock {\em arXiv preprint arXiv:1903.08560}.

\bibitem[Hastie et~al., 2015]{hastie2015statistical}
Hastie, T., Tibshirani, R., and Wainwright, M. (2015).
\newblock {\em Statistical learning with sparsity: the lasso and
  generalizations}.
\newblock CRC press.

\bibitem[Hauzenberger et~al., 2024]{hauzenberger2024dynamic}
Hauzenberger, N., Huber, F., and Koop, G. (2024).
\newblock Dynamic shrinkage priors for large time-varying parameter regressions
  using scalable markov chain monte carlo methods.
\newblock {\em Studies in Nonlinear Dynamics \& Econometrics}, 28(2):201--225.

\bibitem[Hauzenberger et~al., 2022]{hauzenberger2022fast}
Hauzenberger, N., Huber, F., Koop, G., and Onorante, L. (2022).
\newblock Fast and flexible bayesian inference in time-varying parameter
  regression models.
\newblock {\em Journal of Business \& Economic Statistics}, 40(4):1904--1918.

\bibitem[Hoerl and Kennard, 1970]{hoerl1970ridge}
Hoerl, A.~E. and Kennard, R.~W. (1970).
\newblock Ridge regression: Biased estimation for nonorthogonal problems.
\newblock {\em Technometrics}, 12(1):55--67.

\bibitem[Huber et~al., 2021]{huber2021inducing}
Huber, F., Koop, G., and Onorante, L. (2021).
\newblock Inducing sparsity and shrinkage in time-varying parameter models.
\newblock {\em Journal of Business \& Economic Statistics}, 39(3):669--683.

\bibitem[Huber et~al., 2020]{huber2020bayesian}
Huber, F., Koop, G., and Pfarrhofer, M. (2020).
\newblock Bayesian inference in high-dimensional time-varying parameter models
  using integrated rotated gaussian approximations.
\newblock {\em arXiv preprint arXiv:2002.10274}.

\bibitem[Ito et~al., 2014]{ito2014GLS}
Ito, M., Noda, A., and Wada, T. (2014).
\newblock International stock market efficiency: a non-bayesian time-varying
  model approach.
\newblock {\em Applied Economics}, 46(23):2744--2754.

\bibitem[Ito et~al., 2017]{ito2017alternative}
Ito, M., Noda, A., and Wada, T. (2017).
\newblock An alternative estimation method of a time-varying parameter model.
\newblock Technical report, Working Paper, Faculty of Economics, Keio
  University, Japan.

\bibitem[Jord{\`a}, 2005]{jorda2005}
Jord{\`a}, {\`O}. (2005).
\newblock Estimation and inference of impulse responses by local projections.
\newblock {\em American economic review}, 95(1):161--182.

\bibitem[Kapetanios et~al., 2019]{kapetanios2017large}
Kapetanios, G., Marcellino, M., and Venditti, F. (2019).
\newblock Large time-varying parameter vars: A nonparametric approach.
\newblock {\em Journal of Applied Econometrics}, 34(7):1027--1049.

\bibitem[Kelly et~al., 2022]{kellyvirtue}
Kelly, B.~T., Malamud, S., and Zhou, K. (2022).
\newblock The virtue of complexity in return prediction.
\newblock Technical report, National Bureau of Economic Research.

\bibitem[Kelly et~al., 2017]{kelly2017ipca}
Kelly, B.~T., Pruitt, S., and Su, Y. (2017).
\newblock Instrumented principal component analysis.

\bibitem[Kilian and L{\"u}tkepohl, 2017]{kilian2017SVAR}
Kilian, L. and L{\"u}tkepohl, H. (2017).
\newblock {\em Structural vector autoregressive analysis}.
\newblock Cambridge University Press.

\bibitem[Kimeldorf and Wahba, 1970]{kimeldorf1970}
Kimeldorf, G.~S. and Wahba, G. (1970).
\newblock A correspondence between bayesian estimation on stochastic processes
  and smoothing by splines.
\newblock {\em The Annals of Mathematical Statistics}, 41(2):495--502.

\bibitem[Knaus et~al., 2021]{knaus2021shrinkage}
Knaus, P., Bitto-Nemling, A., Cadonna, A., and Fr{\"u}hwirth-Schnatter, S.
  (2021).
\newblock Shrinkage in the time-varying parameter model framework using the r
  package shrinktvp.
\newblock {\em Journal of Statistical Software}, 100(13).

\bibitem[Koop and Korobilis, 2013]{koop2013large}
Koop, G. and Korobilis, D. (2013).
\newblock Large time-varying parameter vars.
\newblock {\em Journal of Econometrics}, 177(2):185--198.

\bibitem[Koop, 2003]{koop2003bayesian}
Koop, G.~M. (2003).
\newblock {\em Bayesian econometrics}.
\newblock John Wiley \& Sons Inc.

\bibitem[Korobilis, 2014]{korobilis2014data}
Korobilis, D. (2014).
\newblock Data-based priors for vector autoregressions with drifting
  coefficients.

\bibitem[Korobilis, 2019]{korobilis2019high}
Korobilis, D. (2019).
\newblock High-dimensional macroeconomic forecasting using message passing
  algorithms.
\newblock {\em Journal of Business \& Economic Statistics}, pages 1--12.

\bibitem[Kotchoni et~al., 2019]{KLS2019}
Kotchoni, R., Leroux, M., and Stevanovic, D. (2019).
\newblock Macroeconomic forecast accuracy in a data-rich environment.
\newblock {\em Journal of Applied Econometrics}, 34(7):1050--1072.

\bibitem[Lancaster, 2003]{lancaster2003}
Lancaster, T. (2003).
\newblock A note on bootstraps and robustness.
\newblock {\em Available at SSRN 896764}.

\bibitem[Lin and Zhang, 2006]{cosso}
Lin, Y. and Zhang, H.~H. (2006).
\newblock Component selection and smoothing in multivariate nonparametric
  regression.
\newblock {\em The Annals of Statistics}, 34(5):2272--2297.

\bibitem[Liu and Li, 2014]{liuli2014}
Liu, Z. and Li, G. (2014).
\newblock Efficient regularized regression for variable selection with l0
  penalty.
\newblock {\em arXiv preprint arXiv:1407.7508}.

\bibitem[Lusompa, 2020]{lusompa2020local}
Lusompa, A. (2020).
\newblock Local projections, autocorrelation, and efficiency.
\newblock {\em Autocorrelation, and Efficiency (March 29, 2020)}.

\bibitem[MacKinnon, 2006]{mackinnon2006bootstrap}
MacKinnon, J.~G. (2006).
\newblock Bootstrap methods in econometrics.
\newblock {\em Economic Record}, 82:S2--S18.

\bibitem[McCracken and Ng, 2020]{mccracken2020fred}
McCracken, M. and Ng, S. (2020).
\newblock Fred-qd: A quarterly database for macroeconomic research.
\newblock Technical report, National Bureau of Economic Research.

\bibitem[Murphy, 2012]{murphy2012}
Murphy, K.~P. (2012).
\newblock {\em Machine learning: a probabilistic perspective}.
\newblock MIT press.

\bibitem[Newton et~al., 2021]{newton2018weighted}
Newton, M.~A., Polson, N.~G., and Xu, J. (2021).
\newblock Weighted bayesian bootstrap for scalable posterior distributions.
\newblock {\em Canadian Journal of Statistics}, 49(2):421--437.

\bibitem[Newton and Raftery, 1994]{newton1994approximate}
Newton, M.~A. and Raftery, A.~E. (1994).
\newblock Approximate bayesian inference with the weighted likelihood
  bootstrap.
\newblock {\em Journal of the Royal Statistical Society: Series B
  (Methodological)}, 56(1):3--26.

\bibitem[Ng and Newton, 2022]{ng2022random}
Ng, T.~L. and Newton, M.~A. (2022).
\newblock Random weighting in lasso regression.
\newblock {\em Electronic Journal of Statistics}, 16(1):3430--3481.

\bibitem[Nie and Ro{\v{c}}kov{\'a}, 2022]{nie2022bayesian}
Nie, L. and Ro{\v{c}}kov{\'a}, V. (2022).
\newblock Bayesian bootstrap spike-and-slab lasso.
\newblock {\em Journal of the American Statistical Association}, pages 1--16.

\bibitem[Paige and Trindade, 2010]{paige2010ridge}
Paige, R.~L. and Trindade, A.~A. (2010).
\newblock Ridge regression representations of the generalized hodrick-prescott
  filter.
\newblock {\em Journal of the Japan Statistical Society}, 45(2):121--128.

\bibitem[Petrova, 2019]{petrova2016quasi}
Petrova, K. (2019).
\newblock A quasi-bayesian local likelihood approach to time varying parameter
  var models.
\newblock {\em Journal of Econometrics}, 212(1):286--306.

\bibitem[Pettenuzzo and Timmermann, 2017]{pettenuzzo2017forecasting}
Pettenuzzo, D. and Timmermann, A. (2017).
\newblock Forecasting macroeconomic variables under model instability.
\newblock {\em Journal of Business \& Economic Statistics}, 35(2):183--201.

\bibitem[Plagborg-M{\o}ller and Wolf, 2019]{plagborg2019local}
Plagborg-M{\o}ller, M. and Wolf, C.~K. (2019).
\newblock Local projections and vars estimate the same impulse responses.
\newblock {\em Unpublished paper: Department of Economics, Princeton
  University}.

\bibitem[Plagborg-M{\o}ller and Wolf, 2021]{plagborg2021local}
Plagborg-M{\o}ller, M. and Wolf, C.~K. (2021).
\newblock Local projections and vars estimate the same impulse responses.
\newblock {\em Econometrica}, 89(2):955--980.

\bibitem[Primiceri, 2005]{primiceri2005}
Primiceri, G.~E. (2005).
\newblock Time varying structural vector autoregressions and monetary policy.
\newblock {\em The Review of Economic Studies}, 72(3):821--852.

\bibitem[Ramey, 2016]{ramey2016hb}
Ramey, V.~A. (2016).
\newblock Macroeconomic shocks and their propagation.
\newblock In {\em Handbook of macroeconomics}, volume~2, pages 71--162.
  Elsevier.

\bibitem[Romer and Romer, 2004]{romerromer2004}
Romer, C.~D. and Romer, D.~H. (2004).
\newblock A new measure of monetary shocks: Derivation and implications.
\newblock {\em American Economic Review}, 94(4):1055--1084.

\bibitem[Ruisi, 2019]{ruisi2019time}
Ruisi, G. (2019).
\newblock Time-varying local projections.
\newblock Technical report, Working Paper.

\bibitem[Saunders et~al., 1998]{saunders1998ridge}
Saunders, C., Gammerman, A., and Vovk, V. (1998).
\newblock Ridge regression learning algorithm in dual variables.

\bibitem[Schlicht, 2005]{schlicht2005estimating}
Schlicht, E. (2005).
\newblock Estimating the smoothing parameter in the so-called hodrick-prescott
  filter.
\newblock {\em Journal of the Japan Statistical Society}, 35(1):99--119.

\bibitem[Sch{\"o}lkopf et~al., 2001]{scholkopf2001generalized}
Sch{\"o}lkopf, B., Herbrich, R., and Smola, A.~J. (2001).
\newblock {A Generalized Representer Theorem}.
\newblock In {\em International conference on computational learning theory},
  pages 416--426. Springer.

\bibitem[Stevanovic, 2016]{stevanovic2016common}
Stevanovic, D. (2016).
\newblock Common time variation of parameters in reduced-form macroeconomic
  models.
\newblock {\em Studies in Nonlinear Dynamics \& Econometrics}, 20(2):159--183.

\bibitem[Stock and Watson, 1996]{stock1996evidence}
Stock, J.~H. and Watson, M.~W. (1996).
\newblock Evidence on structural instability in macroeconomic time series
  relations.
\newblock {\em Journal of Business \& Economic Statistics}, 14(1):11--30.

\bibitem[Stock and Watson, 1998a]{SW1998comparison}
Stock, J.~H. and Watson, M.~W. (1998a).
\newblock A comparison of linear and nonlinear univariate models for
  forecasting macroeconomic time series.
\newblock Technical report, National Bureau of Economic Research.

\bibitem[Stock and Watson, 1998b]{stock1998median}
Stock, J.~H. and Watson, M.~W. (1998b).
\newblock Median unbiased estimation of coefficient variance in a time-varying
  parameter model.
\newblock {\em Journal of the American Statistical Association},
  93(441):349--358.

\bibitem[Stock and Watson, 2002]{sw2002}
Stock, J.~H. and Watson, M.~W. (2002).
\newblock Macroeconomic forecasting using diffusion indexes.
\newblock {\em Journal of Business \& Economic Statistics}, 20(2):147--162.

\bibitem[Tibshirani et~al., 2005]{fusedlasso2005}
Tibshirani, R., Saunders, M., Rosset, S., Zhu, J., and Knight, K. (2005).
\newblock Sparsity and smoothness via the fused lasso.
\newblock {\em Journal of the Royal Statistical Society: Series B (Statistical
  Methodology)}, 67(1):91--108.

\bibitem[Tibshirani et~al., 2015]{tibshirani2015statistical}
Tibshirani, R., Wainwright, M., and Hastie, T. (2015).
\newblock {\em Statistical learning with sparsity: the lasso and
  generalizations}.
\newblock Chapman and Hall/CRC.

\bibitem[Wahba, 1990]{wahba1990spline}
Wahba, G. (1990).
\newblock {\em Spline models for observational data}, volume~59.
\newblock Siam.

\bibitem[Yamada, 2016]{yamada2016hodrick}
Yamada, H. (2016).
\newblock The hodrick-prescott filter: A special case of penalized spline
  smoothing.
\newblock {\em Electronic Journal of Statistics}, 4:856--874.

\bibitem[Zou, 2006]{zou2006}
Zou, H. (2006).
\newblock The adaptive lasso and its oracle properties.
\newblock {\em Journal of the American statistical association},
  101(476):1418--1429.

\end{thebibliography}

\clearpage

\appendix
%\appendixpage
%\addappheadtotoc
\newcounter{saveeqn}
\setcounter{saveeqn}{\value{section}}
\renewcommand{\theequation}{\mbox{\Alph{saveeqn}.\arabic{equation}}} \setcounter{saveeqn}{1}
\setcounter{equation}{0}
	
\section{Appendix}\label{sec:appendix}
\setstretch{1.25}

%\section{Appendix}

\subsection{Details of MSRR$_{\text{S}}$}\label{gglrrdetails}

To begin with, the penalty part of (\ref{sparsetvp}) in summation notation is
\begin{align*}
\sum_{k=1}^{K}\frac{1}{\sigma_{u_k}^2}{\sum_{t=1}^{T}}u_{k,t}^{2} +\xi{\sum_{k=1}^{K}}| {\sigma_{u_k}}|.
\end{align*}
The E-step of the procedure provides a formula for $\sigma_{u_j}$ in terms of $u$'s. Plugging it in gives 
\begin{align*}
\sum_{k=1}^{K}\frac{1}{\sigma_{u_k}^2}{\sum_{t=1}^{T}} u_{k,t}^{2} +\xi{\sum_{k=1}^{K}}({\sum_{t=1}^{T}}u_{k,t}^{2} )^\frac{1}{2}.
\end{align*}
which is just a Group Lasso penalty with an additional Ridge penalty for each individual coefficients. Hence, classifying parameters into TVP or non-TVP categories is equivalent to group selection of regressors where each $k$ of the $K$ groups is defined as $\{Z_{t,k,\tau}\}_{\tau=1}^{\tau=T}$. If we want a parameter to be constant, we trivially have to drop block-wise its respective basis expansion regressors and only keep $\beta_{0,k}$ in the model. 

This penalty can be obtained by iterating what we already have. \cite{grandvalet1998} shows that the  Lasso solution can be obtained by iterative Adaptive Ridge. \cite{frommletnuel2016}  and \cite{liuli2014} extend his results to obtain $l_0$ regularization without the computational burden associated with this type of regularization. \cite{frommletnuel2016}  also argue in favor of a slightly modified version of \cite{grandvalet1998}'s algorithm which I first review before turning to the final MSRR$_{\text{S}}$ problem.

To implement Lasso by Adaptive Ridge, we have at iteration $i$,
\begin{align*}
\boldsymbol{b}_i = \arg \min_{\boldsymbol{b}}{\sum_{t=1}^{T}}\left(y_{t}-X_{t}\boldsymbol{b}\right)^{2}+\lambda{\sum_{=1}^{J}}w_j{b}_{j}^{2} \\
w_{i+1,j} = \frac{1}{(b_{i,j}+\delta)^2}
\end{align*} 
where $\delta>0$ is small value for numerical stability and we set $w_{j,0}=1 \quad \forall j$.
To get some intuition on why this works, it is worth looking at the penalty part of the problem in the final algorithm step:
 $$\lambda{\sum_{j=1}^{J}}\frac{{b}_{j}^{2}}{|\hat{b}_{j}|+\delta} \approx \lambda{\sum_{j=1}^{J}}|{{b}_{j}}|.$$ 
\cite{liuli2014} show that this qualifies as a proper EM algorithm (each step improves the likelihood). Thus, we can expect it to inherit traditional convergence properties.   Algorithm \ref{gglrralgo} summarizes all the steps.

\begin{algorithm}
\caption{MSRR$_{\text{S}}$ \label{gglrralgo}}
\begin{algorithmic}[1]
  %\footnotesize
  \small
\STATE Initiate the procedure with $\boldsymbol{\hat{\theta}_1}$ or $\boldsymbol{\hat{\theta}_2}$ from Algorithm 1. Keep the sequence of $\sigma_{u_k}^{2,(1)}$'s and the chosen $\lambda_{(1)}$. Set $\tilde{\lambda}=\lambda_{(1)}$. Choose a value for $\alpha$. In applications, it is set to 0.5.
\STATE Iterate the following until convergence of $\lambda_{u_k}$'s. For iteration  $i$:
\vspace{-0.3cm}
\begin{enumerate} \itemsep -0.5em
		\item Use solution (\ref{2sdualsol}) to get $\boldsymbol{\hat{\theta}_3^{(i)}}$.
		\item Obtain ${{\hat{\sigma}_{u,k}^{2,(i)}}}$ the usual way and normalize them to have mean of 1. Generate next step's weights using
$$\lambda_{u_k}^{(i+1)} \leftarrow \tilde{\lambda} \left[ \alpha \frac{1}{\sigma_{u_k}^{2,(1)}} + (1-\alpha) \frac{1}{\sigma_{u_k}^{(i)}} \right] $$ on the diagonal of $\Omega_{u}^{-1,(i)}$. The formula is derived in Appendix \ref{gglrrdetails}.
\item Obtain ${{\hat{\sigma}_{\epsilon,t}^2}}$ by fitting a volatility model to the residuals from step 1. Normalize ${{\hat{\sigma}_{\epsilon,t}^2}}$'s mean to 1 and input it to $\Omega_{\epsilon}^{(i)}$.
\end{enumerate}
\vspace{-0.3cm}

\STATE Use solution (\ref{2sdualsol}) with the converged $\Omega_{\epsilon}$ and $\Omega_{u}$ to get $\boldsymbol{\hat{\theta}_3}$, the final estimator.\footnote{As in Algorithm \ref{2srralgo}, it is possible to run CV again with the converged $\Omega_{u}$ to tune the average level of smoothness.}
\end{algorithmic}
\end{algorithm}

\subsubsection{Building Iterative Weights for MSRR$_{\text{S}}$}

The above methodology can be adapted for a case which is substantially more complicated. The complications are twofold. First, we are doing Group Lasso rather than plain Lasso. Second, the individual ridge penalty must be maintained on top of the Group Lasso penalty. I devise a simple algorithm that will split the original Ridge penalty into two parts, one that we will keep as is and one that will be iterated. The first is the 2SRR part and the second implements Group-Lasso.

Let us first focus on the Group penalty and display why iterating the Ridge solution with updating weights converges to be equivalent to Group Lasso.  In the last step of the algorithm, we have
$$\xi{\sum_{k=1}^{K}} \frac{1}{\hat{\sigma}_{u_k}}{\sum_{t=1}^{T}}{u}_{k,t}^{2} \approx \xi{\sum_{k=1}^{K}}({\sum_{t=1}^{T}}{u}_{k,t}^{2} )^{\frac{1}{2}}$$
where $\hat{\sigma}_{u_k}=({\sum_{t=1}^{T}}{u}_{k,t}^{2})^{\frac{1}{2}}$. The two penalties must be combined in a single penalizing weight that enters the closed-form solution. I split the original penalty into two parts, one that will remain as such and one that will be iterated to generate group selection. A useful observation is the following. For a given iteration $i$, 
\begin{align*}
\lambda \sum_{k=1}^{K}\frac{1}{\sigma_{u_k}^2}{\sum_{t=1}^{T}}{u}_{k,t}^{2} +\xi{\sum_{k=1}^{K}}  \frac{1}{\sigma_{u_k}^{(i)}} {\sum_{t=1}^{T}}{u}_{k,t}^{2} 
\end{align*}
can be re-arranged as
\begin{align*}
\sum_{k=1}^{K}\left[ \frac{\lambda}{\sigma_{u_k}^2} + \frac{\xi}{{\sigma_{u_k}^{(i)}}} \right]{\sum_{t=1}^{T}}{u}_{k,t}^{2} .
\end{align*}
To make this more illuminating, define $\alpha = \frac{\lambda}{\lambda+\xi}$ and $\tilde{\lambda}=(\lambda+\xi)$. We now have
\begin{align*}
\tilde{\lambda}\sum_{k=1}^{K}\left[ \alpha \frac{1}{\sigma_{u_k}^2} + (1-\alpha) \frac{1}{\sigma_{u_k}^{(i)}} \right]{\sum_{t=1}^{T}}{u}_{k,t}^{2} .
\end{align*}
where $\alpha \in (0,1)$ is a tuning parameter controlling how the original ridge penalty is split between smoothness and group-wise sparsity. It is now easy to plug this into the closed-form formula: stack $\lambda_{u_k}^{(i)} = \tilde{\lambda} \left[ \alpha \frac{1}{\sigma_{u_k}^2} + (1-\alpha) \frac{1}{\sigma_{u_k}^{(i-1)}} \right] $ on the diagonal of $\Omega_{u_{i}}^{-1}$ at iteration $i$ in \ref{gensol}. The reader is now referred to the main text (section \ref{cosso}) for the benchmark algorithm that uses these derivations to implement MSRR$_{\text{S}}$.

%Since heteroscedasticity is incorporated in a GLS fashion, credible regions can be obtained by using the formula above with the properly re-weighted data matrix $\boldsymbol{Z^*}$. 

%\hat{\sigma}_{\epsilon}^2.

\subsection{MSRR$_{\text{D}}$ Details}\label{MVGGRRRR}

Noting that 
\begin{align}\label{vectricks}
vec(\Lambda \boldsymbol{F})= (I_T \otimes \Lambda)\boldsymbol{f} \\
vec(\Lambda \boldsymbol{F})= (\boldsymbol{F'} \otimes I_K)l.
\end{align}
it is possible to solve (\ref{densetvp}) with a  maximization-maximization procedure that alternates between two regularized regressions.
\vspace{-0.3cm} 
\begin{enumerate} \itemsep -0.5em
\item Given $\Lambda$, we can solve
\begin{align}\label{densetvpgivenl}
\min_{ \boldsymbol{f},\beta_0}\left(\boldsymbol{y}-\boldsymbol{X}\beta_0-\boldsymbol{Z}^\Lambda\boldsymbol{f}\right)'\Omega_{\epsilon}^{-1} \left(\boldsymbol{y}-\boldsymbol{X}\beta_0-\boldsymbol{Z}^\Lambda\boldsymbol{f}\right)+ \lambda\boldsymbol{f'f}
\end{align}
where $\boldsymbol{Z}^\Lambda = \boldsymbol{Z}(I_T \otimes \Lambda)$.  This is a ridge regression of the likes we have been conducting thus far.
\item  Given $\boldsymbol{F}$, we can get the solution to
\begin{align}\label{densetvpgivenF}
\min_{ l,\beta_0}\left(\boldsymbol{y}-\boldsymbol{X}\beta_0-\boldsymbol{Z^F}l\right)'\Omega_{\epsilon}^{-1} \left(\boldsymbol{y}-\boldsymbol{X}\beta_0-\boldsymbol{Z^F}l\right)+ \xi \Vert l \Vert_1
\end{align}
where $\boldsymbol{Z^F} = \boldsymbol{Z}(\boldsymbol{F'} \otimes I_K)$. This is a Lasso regression of modest size.\footnote{This could also be a RR if we wished to implement dense parameters only. In practice, elastic net with $\alpha=0.5$ is the wiser choice (vs Lasso) given the strong correlation between the generated predictors.}
\end{enumerate} 
\vspace{-0.3cm} 
A first observation is that this problem is biconvex. A second one is that at each step, the objective function is further minimized and the objective is bounded from below. Hence, alternating these steps generate a monotonic sequence that converges to a (local) minima.\footnote{The other legitimate question is whether this algorithm converges to the solution of \ref{densetvp}. It turns out to be a modification of \cite{tibshirani2015statistical} (Chapter 4) alternative algorithm for \cite{cosso}'s COSSO. The additional steps are orthogonalization of factors and loadings as in \cite{bai2017principal}.} In terms of implementation, one must be carefully imposing the identification restriction of the factor model at all times. Algorithm \ref{grrrralgo} summarizes this and other practical aspects.

\begin{algorithm}
\caption{MSRR$_{\text{D}}$ \label{grrrralgo}}
\begin{algorithmic}[1]
  %\footnotesize
  \small
  \STATE Get $\boldsymbol{\hat{\theta}_2}$ from Algorithm 1 or plain RR. Estimate $\boldsymbol{F}^{(1)}$ and $\Lambda^{(1)}$ by fitting a factor model to the $u$'s. Choose $r$ the number of factor using a citerion of choice.\footnote{I use a simple variance reconstruction threshold.}
\STATE Iterate the following until convergence. For iteration  $i>1$:
\vspace{-0.3cm}
\begin{enumerate} \itemsep -0.5em
		\item Run (\ref{densetvpgivenl}) to get $\boldsymbol{F}^{(i)}$ given  $\Lambda^{(i-1)}$. Orthogonalize factors.
		\item Run (\ref{densetvpgivenF}) to get $\Lambda^{(i)}$ given $\boldsymbol{F}^{(i)}$. Orthogonalize loadings. 
		\item Obtain ${{\hat{\sigma}_{\epsilon,t}^2}}$ by fitting a volatility model to (current) residuals. Normalize ${{\hat{\sigma}_{\epsilon,t}^2}}$'s mean to 1 and input it to $\Omega_{\epsilon}^{(i)}$. 
\end{enumerate}
\vspace{-0.3cm} 
%\end{enumerate}

\end{algorithmic}
\end{algorithm}

It is noteworthy that doing Lasso on the loadings $\Lambda$ operates a fusion of sparse and dense TVPs. If a parameter $\beta_k$ does not "load" on any of the factors (because the vector $\Lambda_{k}$ is shrunk perfectly to 0), we effectively get a constant $\beta_k$. In the resulting model, a given parameter can vary or not, and when it does, it shares a common structure with fellow parameters also selected as time-varying.

I now present the multivariate extension to MSRR$_{\text{D}}$ and discuss its connection to \cite{kelly2017ipca}'s Instrumented PCA estimator for asset pricing models.   Dense TVPs as proposed (among others) by \cite{stevanovic2016common} implement a factor structure for parameters of a whole VAR system rather than a single equation. If time-variation is indeed similar for all equations, we can decrease estimation variance significantly by pooling all parameters of the system in a single factor model. First, the factors are better estimated as the number of series increase. Second, the estimated factors are less prone to overfit because they now target $M$ series rather than a single one.\footnote{This is the kind of regularization being used for linear models in \cite{CKM2011}. However, for MV-MSRR$_{\text{D}}$, the reduced-rank matrix is organized differently and the underlying factors have a different interpretation.} The likely case where $r$ is smaller than $M$ (and $P$ not incredibly big) yields a models that will have more observations than parameters, in contrast to everything so far considered in this paper. I briefly describe how to modify Algorithm 3 to obtain Multivariate MSRR$_{\text{D}}$ (MV-MSRR$_{\text{D}}$) estimates.

Starting values for the algorithm below can be obtained from the multivariate RR of section (\ref{MV}). This is done by first re-arranging elements of $\boldsymbol{\hat{\Theta}}$ into $\mathcal{U}=[\boldsymbol{U}_1 \quad \dots \quad \boldsymbol{U}_M]$ and then running PCA on $\mathcal{U}$. Then, the MV-MSRR$_{\text{D}}$ solution can be obtained by alternating the following steps.
\begin{enumerate}
\item Given $\Lambda$, we can solve
\begin{align} %\label{MVdensetvpgivenl}
\min_{ \boldsymbol{f},b_0}\left(vec(\boldsymbol{Y})-(I_M \otimes \boldsymbol{X})b_0-\boldsymbol{Z_M}^\Lambda\boldsymbol{f}\right)'\Omega_{\epsilon_M}^{-1} \left(vec(\boldsymbol{Y})-(I_M \otimes \boldsymbol{X})b_0-\boldsymbol{Z_M}^\Lambda\boldsymbol{f}\right)+ \boldsymbol{f'f}
\end{align}
where $\boldsymbol{Z_M}^\Lambda$  stacks row-wise all the $ \boldsymbol{Z}(I_T \otimes \Lambda_m)$ from $m=1$ to $m=M$. That is, we have the $TM \times Tr$ matrix
$$
\boldsymbol{Z_M}^\Lambda=
\begin{bmatrix}
    \boldsymbol{Z}(I_T \otimes \Lambda_1) \\
    \boldsymbol{Z}(I_T \otimes \Lambda_2)  \\
     \vdots  \\
    \boldsymbol{Z}(I_T \otimes \Lambda_M)  
\end{bmatrix}
$$
as the regressor matrix. $\Lambda_m$ is a sub-matrix of $\Lambda$ that contains the loadings for parameters of equation $m$. Also, $b_0=vec(B_0)$ where $B_0$ is the matrix that corresponds to the multivariate equivalent of $\beta_0$. Unlike a standard multivariate model like a VAR, here, we cannot estimate each equation separately because the $\boldsymbol{f}$ is common across equations. 
\item The loadings updating step is 
\begin{align}\label{MVdensetvpgivenF}
\min_{ l,b_0}\left(vec(\boldsymbol{Y})-(I_M \otimes \boldsymbol{X})b_0-\boldsymbol{Z^F_M}l\right)'\Omega_{\epsilon_M}^{-1} \left(vec(\boldsymbol{Y})-(I_M \otimes \boldsymbol{X})b_0-\boldsymbol{Z^F_M}l\right)+ \xi \Vert l \Vert_1
\end{align}
where $\boldsymbol{Z^F_M} = (I_M \otimes \boldsymbol{Z}(\boldsymbol{F'} \otimes I_K))$. This is just a Lasso regression. The Kronecker structure allows for these Lasso regressions to be ran separately.
\end{enumerate}
As in \cite{bai2017principal} for the estimation of regularized factor models, there is orthogonalization step needed between each of these steps to guarantee identification. 

Note that if $MT > rT + MK$, which is somewhat likely, we have more observations than parameters in step 1. This means standard Ridge regularization is \textit{not} necessary for the inversion of covariance matrix of regressors.\footnote{This also means that it is now computationally more efficient to solve the primal Ridge problem.} Nonetheless, the ridge smoothness prior will still prove useful but can be applied in a much less aggressive way. 

An interesting connection occurs in the MV-MSRR$_{\text{D}}$ case: the time-varying parameter model with a factor structure in parameters can also be seen as a dynamic factor model with deterministically time-varying loadings. By the latter, I mean that loadings change through time because they are interacted with a known set of (random) variables $X_t$. This is a more general version of \cite{kelly2017ipca} Instrumented PCA used to estimate a typical asset-pricing factor model. Formally, this means that the factor TVP model
$$Y_t = X_t\Lambda F_t + \epsilon_t, \quad \quad {F_{t}}={F_{t-1}}+{u}_{t}$$
can be rewritten as 
\begin{align}
Y_t = \Lambda_t F_t + \epsilon_t, \quad \quad {F_{t}}={F_{t-1}}+{u}_{t}, \quad \quad \Lambda_t = X_t\Lambda
\end{align}
which is the so-called Instrumented PCA estimator if we drop the law of motion for $F_t$. An important additional distinction is that \cite{kelly2017ipca} consider cases where the number of instruments is smaller than the size of the cross-section. Here, with the instruments being $X_t$, there is by construction more instruments than the size of the cross-section. Nevertheless, the analogy to the factor model is conceptually useful and can point to further improvements of TVP models inspired by advances in empirical asset pricing research. 

\subsection{Simple MSRR$_{\text{D}}$ Example with $r=1$}\label{caser1}
While Kronecker product operations may seem obscure, they are the generalization of something that much more intuitive: the special case of one factor model ($r=1$). I present here the simpler model when parameters vary according to a single latent source of time-variation. For convenience, I drop evolving volatility and use summation notation. The problem reduces to
\begin{align}
\min_{l, f,\beta_0} \sum_{t=1}^T \left(y_t-X_t \beta_0-\sum_{k=1}^K l_k f_t X_k \right)^2 + \sum_{t=1}^T f_t^2 + \xi \sum_{k=1}^K |l_k| 
\end{align}
which can trivially rewritten as
\begin{align} 
\min_{l, f,\beta_0} \sum_{t=1}^T \left(y_t-X_t \beta_0- f_t \sum_{k=1}^K l_k  X_{k,t} \right)^2 + \sum_{t=1}^T f_t^2 + \xi \sum_{k=1}^K |l_k|.
\end{align}
and this model can be estimated by splitting it two problems. The two steps are 
\begin{enumerate}
\item Given the $l$ vector, we run the TVP regression  
\begin{align*}
\min_{f,\beta_0} \sum_{t=1}^T \left(y_t-X_t \beta_0- \bar{X}_t f_t  \right)^2 + \sum_{t=1}^T f_t^2.
\end{align*}
where $\bar{X}_t \equiv  \sum_{k=1}^K l_k  X_{k,t}$. Hence, the new regressors are just a linear combination of original regressors.
\item Given $f$, the second step is the Lasso regression (or OLS/Ridge if we prefer)
\begin{align*}
\min_{l,\beta_0} \sum_{t=1}^T \left(y_t-X_t \beta_0- \sum_{k=1}^K l_k {X^f}_{k,t} \right)^2 + \xi \sum_{k=1}^K |l_k|.
\end{align*}
where the $K$ new regressors are ${X^f}_{k,t} \equiv  f_t X_{k,t}$. 
\end{enumerate}

\pagebreak
\subsection{Tables}

\begin{table}[!hp]
\caption{Results for Simulation 1 (Cosine) and $T=300$ \label{s1_table}} 
 \begin{threeparttable}
\setlength{\tabcolsep}{0.2em}
\vspace{-0.3cm}
%\hspace*{-0.5cm}
%\begin{center}
\footnotesize 
\begin{tabular}{lcccccccccccccc}
\hline\hline
\multicolumn{1}{l}{\bfseries }&\multicolumn{4}{c}{\bfseries $K=6$}&\multicolumn{1}{c}{\bfseries }&\multicolumn{4}{c}{\bfseries $K=20$}&\multicolumn{1}{c}{\bfseries }&\multicolumn{4}{c}{\bfseries $K=100$}\tabularnewline
\cline{2-5} \cline{7-10} \cline{12-15}
\multicolumn{1}{l}{}&\multicolumn{1}{c}{BVAR}&\multicolumn{1}{c}{2SRR}&\multicolumn{1}{c}{MSRR$_{\text{S}}$}&\multicolumn{1}{c}{MSRR$_{\text{D}}$}&\multicolumn{1}{c}{}&\multicolumn{1}{c}{BVAR}&\multicolumn{1}{c}{2SRR}&\multicolumn{1}{c}{MSRR$_{\text{S}}$}&\multicolumn{1}{c}{MSRR$_{\text{D}}$}&\multicolumn{1}{c}{}&\multicolumn{1}{c}{BVAR}&\multicolumn{1}{c}{2SRR}&\multicolumn{1}{c}{MSRR$_{\text{S}}$}&\multicolumn{1}{c}{MSRR$_{\text{D}}$}\tabularnewline
\hline
{\bfseries $\mathbf{\sfrac{K^*}{K}=0.2}$}&&&&&&&&&&&&&&\tabularnewline
~~$\sigma_{\epsilon} =\text{Low}$& 0.128&{\color{blue}  0.110}&\textbf{ 0.097}& 0.136&&-& 0.115&\textbf{ 0.095}& 0.114&&-& 0.163&\textbf{ 0.160}& 0.200\tabularnewline
~~$\sigma_{\epsilon} =\text{Medium}$&{\color{blue} \textbf{ 0.159}}& 0.165& 0.163& 0.193&&-& 0.165&\textbf{ 0.161}& 0.169&&-& 0.197&\textbf{ 0.192}& 0.314\tabularnewline
~~$\sigma_{\epsilon} =\text{High}$&{\color{blue} \textbf{ 0.228}}& 0.245& 0.244& 0.271&&-& 0.262&\textbf{ 0.262}& 0.269&&-& 0.320&\textbf{ 0.316}& 0.580\tabularnewline
~~$\sigma_{\epsilon,t} = \text{SV Low-Med}$& 0.129&{\color{blue}  0.121}&\textbf{ 0.110}& 0.145&&-& 0.131&\textbf{ 0.120}& 0.132&&-& 0.168&\textbf{ 0.166}& 0.242\tabularnewline
~~$\sigma_{\epsilon,t}  = \text{SV Low-High}$&{\color{blue} \textbf{ 0.143}}& 0.151& 0.152& 0.174&&-& 0.159&\textbf{ 0.158}& 0.175&&-& 0.189&\textbf{ 0.189}& 0.293\tabularnewline
\hline
{\bfseries $\mathbf{\sfrac{K^*}{K}=0.5}$}&&&&&&&&&&&&&&\tabularnewline
~~$\sigma_{\epsilon} =\text{Low}$& 0.169&{\color{blue}  0.130}&\textbf{ 0.120}& 0.144&&-& 0.150& 0.135&\textbf{ 0.129}&&-& 0.256&\textbf{ 0.256}& 0.263\tabularnewline
~~$\sigma_{\epsilon} =\text{Medium}$& 0.224&{\color{blue}  0.207}&\textbf{ 0.206}& 0.247&&-& 0.227& 0.224&\textbf{ 0.221}&&-& 0.283&\textbf{ 0.278}& 0.395\tabularnewline
~~$\sigma_{\epsilon} =\text{High}$&{\color{blue} \textbf{ 0.274}}& 0.291& 0.292& 0.330&&-& 0.314&\textbf{ 0.311}& 0.316&&-& 0.371&\textbf{ 0.365}& 0.700\tabularnewline
~~$\sigma_{\epsilon,t} = \text{SV Low-Med}$& 0.186&{\color{blue}  0.147}&\textbf{ 0.138}& 0.184&&-& 0.171& 0.162&\textbf{ 0.158}&&-&\textbf{ 0.259}& 0.260& 0.303\tabularnewline
~~$\sigma_{\epsilon,t}  = \text{SV Low-High}$& 0.211&{\color{blue} \textbf{ 0.189}}& 0.189& 0.229&&-& 0.216&\textbf{ 0.214}& 0.225&&-&\textbf{ 0.273}& 0.274& 0.361\tabularnewline
\hline
{\bfseries $\mathbf{\sfrac{K^*}{K}=1}$}&&&&&&&&&&&&&&\tabularnewline
~~$\sigma_{\epsilon} =\text{Low}$&{\color{blue} \textbf{ 0.134}}& 0.149& 0.152& 0.157&&-& 0.185& 0.191&\textbf{ 0.141}&&-& 0.367& 0.370&\textbf{ 0.284}\tabularnewline
~~$\sigma_{\epsilon} =\text{Medium}$& 0.302&{\color{blue} \textbf{ 0.242}}& 0.247& 0.282&&-& 0.278& 0.289&\textbf{ 0.252}&&-& 0.389&\textbf{ 0.388}& 0.467\tabularnewline
~~$\sigma_{\epsilon} =\text{High}$&{\color{blue} \textbf{ 0.337}}& 0.355& 0.360& 0.376&&-&\textbf{ 0.388}& 0.389& 0.391&&-& 0.454&\textbf{ 0.451}& 0.766\tabularnewline
~~$\sigma_{\epsilon,t} = \text{SV Low-Med}$& 0.171&{\color{blue} \textbf{ 0.168}}& 0.172& 0.180&&-& 0.208& 0.215&\textbf{ 0.156}&&-& 0.380& 0.381&\textbf{ 0.351}\tabularnewline
~~$\sigma_{\epsilon,t}  = \text{SV Low-High}$& 0.262&{\color{blue} \textbf{ 0.222}}& 0.232& 0.268&&-& 0.262& 0.274&\textbf{ 0.261}&&-& 0.383& 0.382&\textbf{ 0.368}\tabularnewline
\hline\hline
\end{tabular}
%\end{center}
\begin{tablenotes}[para,flushleft]
  \footnotesize Notes: This table reports the average MAE of estimated $\beta_t$'s for various models. The number in bold is the lowest MAE of all models for a given setup. The number in blue is the lowest MAE between BVAR and 2SRR for a given setup.
  \end{tablenotes}
  \end{threeparttable}
\end{table}

\begin{table}[!hp]
\caption{Results for Simulation 2 (Break) and $T=300$ \label{s2_table}} 
 \begin{threeparttable}
\setlength{\tabcolsep}{0.2em}
\vspace{-0.3cm}
%\hspace*{-0.5cm}
%\begin{center}
\footnotesize 
\begin{tabular}{lcccccccccccccc}
\hline\hline
\multicolumn{1}{l}{\bfseries }&\multicolumn{4}{c}{\bfseries $K=6$}&\multicolumn{1}{c}{\bfseries }&\multicolumn{4}{c}{\bfseries $K=20$}&\multicolumn{1}{c}{\bfseries }&\multicolumn{4}{c}{\bfseries $K=100$}\tabularnewline
\cline{2-5} \cline{7-10} \cline{12-15}
\multicolumn{1}{l}{}&\multicolumn{1}{c}{BVAR}&\multicolumn{1}{c}{2SRR}&\multicolumn{1}{c}{MSRR$_{\text{S}}$}&\multicolumn{1}{c}{MSRR$_{\text{D}}$}&\multicolumn{1}{c}{}&\multicolumn{1}{c}{BVAR}&\multicolumn{1}{c}{2SRR}&\multicolumn{1}{c}{MSRR$_{\text{S}}$}&\multicolumn{1}{c}{MSRR$_{\text{D}}$}&\multicolumn{1}{c}{}&\multicolumn{1}{c}{BVAR}&\multicolumn{1}{c}{2SRR}&\multicolumn{1}{c}{MSRR$_{\text{S}}$}&\multicolumn{1}{c}{MSRR$_{\text{D}}$}\tabularnewline
\hline
{\bfseries $\mathbf{\sfrac{K^*}{K}=0.2}$}&&&&&&&&&&&&&&\tabularnewline
~~$\sigma_{\epsilon} =\text{Low}$& 0.154&{\color{blue}  0.113}&\textbf{ 0.098}& 0.146&&-& 0.149&\textbf{ 0.141}& 0.191&&-&\textbf{ 0.331}& 0.337& 0.487\tabularnewline
~~$\sigma_{\epsilon} =\text{Medium}$& 0.216&{\color{blue}  0.176}&\textbf{ 0.165}& 0.296&&-&\textbf{ 0.249}& 0.256& 0.294&&-& 0.587&\textbf{ 0.578}& 1.165\tabularnewline
~~$\sigma_{\epsilon} =\text{High}$& 0.295&{\color{blue} \textbf{ 0.292}}& 0.296& 0.412&&-&\textbf{ 0.473}& 0.480& 0.498&&-& 1.267&\textbf{ 1.236}& 2.484\tabularnewline
~~$\sigma_{\epsilon,t} = \text{SV Low-Med}$& 0.171&{\color{blue}  0.126}&\textbf{ 0.114}& 0.180&&-& 0.175&\textbf{ 0.169}& 0.218&&-&\textbf{ 0.413}& 0.414& 0.708\tabularnewline
~~$\sigma_{\epsilon,t}  = \text{SV Low-High}$& 0.188&{\color{blue}  0.159}&\textbf{ 0.152}& 0.249&&-&\textbf{ 0.232}& 0.248& 0.287&&-&\textbf{ 0.522}& 0.546& 0.984\tabularnewline
\hline
{\bfseries $\mathbf{\sfrac{K^*}{K}=0.5}$}&&&&&&&&&&&&&&\tabularnewline
~~$\sigma_{\epsilon} =\text{Low}$& 0.154&{\color{blue}  0.141}&\textbf{ 0.130}& 0.137&&-& 0.184&\textbf{ 0.180}& 0.232&&-&\textbf{ 0.396}& 0.432& 0.656\tabularnewline
~~$\sigma_{\epsilon} =\text{Medium}$& 0.316&{\color{blue}  0.207}&\textbf{ 0.204}& 0.296&&-&\textbf{ 0.295}& 0.318& 0.362&&-&\textbf{ 0.633}& 0.640& 1.064\tabularnewline
~~$\sigma_{\epsilon} =\text{High}$& 0.370&{\color{blue} \textbf{ 0.335}}& 0.348& 0.465&&-&\textbf{ 0.513}& 0.527& 0.542&&-& 1.291&\textbf{ 1.277}& 2.674\tabularnewline
~~$\sigma_{\epsilon,t} = \text{SV Low-Med}$& 0.197&{\color{blue}  0.156}&\textbf{ 0.148}& 0.156&&-&\textbf{ 0.215}& 0.220& 0.275&&-&\textbf{ 0.469}& 0.487& 0.769\tabularnewline
~~$\sigma_{\epsilon,t}  = \text{SV Low-High}$& 0.242&{\color{blue}  0.193}&\textbf{ 0.190}& 0.298&&-&\textbf{ 0.284}& 0.303& 0.389&&-&\textbf{ 0.587}& 0.620& 1.035\tabularnewline
\hline
{\bfseries $\mathbf{\sfrac{K^*}{K}=1}$}&&&&&&&&&&&&&&\tabularnewline
~~$\sigma_{\epsilon} =\text{Low}$&{\color{blue} \textbf{ 0.143}}& 0.174& 0.183& 0.149&&-& 0.232& 0.430&\textbf{ 0.188}&&-&\textbf{ 0.513}& 0.620& 0.633\tabularnewline
~~$\sigma_{\epsilon} =\text{Medium}$& 0.308&{\color{blue}  0.254}& 0.343&\textbf{ 0.252}&&-& 0.349& 0.493&\textbf{ 0.308}&&-&\textbf{ 0.768}& 0.786& 1.149\tabularnewline
~~$\sigma_{\epsilon} =\text{High}$& 0.506&{\color{blue} \textbf{ 0.414}}& 0.499& 0.447&&-& 0.612&\textbf{ 0.607}& 0.690&&-& 1.437&\textbf{ 1.394}& 2.697\tabularnewline
~~$\sigma_{\epsilon,t} = \text{SV Low-Med}$&{\color{blue}  0.168}& 0.195& 0.205&\textbf{ 0.165}&&-& 0.258& 0.448&\textbf{ 0.220}&&-&\textbf{ 0.571}& 0.663& 0.758\tabularnewline
~~$\sigma_{\epsilon,t}  = \text{SV Low-High}$&{\color{blue} \textbf{ 0.205}}& 0.231& 0.294& 0.231&&-& 0.340& 0.481&\textbf{ 0.334}&&-&\textbf{ 0.695}& 0.726& 1.054\tabularnewline
\hline\hline
\end{tabular}
%\end{center}
\begin{tablenotes}[para,flushleft]
 \footnotesize Notes: see Table \ref{s1_table}.
  \end{tablenotes}
  \end{threeparttable}
\end{table}

\begin{table}[!hp]
\caption{Results for Simulation 3 (Trend and Cosine) and $T=300$ \label{s3_table}} 
 \begin{threeparttable}
\setlength{\tabcolsep}{0.2em}
\vspace{-0.3cm}
%\hspace*{-0.5cm}
%\begin{center}
\footnotesize 
\begin{tabular}{lcccccccccccccc}
\hline\hline
\multicolumn{1}{l}{\bfseries }&\multicolumn{4}{c}{\bfseries $K=6$}&\multicolumn{1}{c}{\bfseries }&\multicolumn{4}{c}{\bfseries $K=20$}&\multicolumn{1}{c}{\bfseries }&\multicolumn{4}{c}{\bfseries $K=100$}\tabularnewline
\cline{2-5} \cline{7-10} \cline{12-15}
\multicolumn{1}{l}{}&\multicolumn{1}{c}{BVAR}&\multicolumn{1}{c}{2SRR}&\multicolumn{1}{c}{MSRR$_{\text{S}}$}&\multicolumn{1}{c}{MSRR$_{\text{D}}$}&\multicolumn{1}{c}{}&\multicolumn{1}{c}{BVAR}&\multicolumn{1}{c}{2SRR}&\multicolumn{1}{c}{MSRR$_{\text{S}}$}&\multicolumn{1}{c}{MSRR$_{\text{D}}$}&\multicolumn{1}{c}{}&\multicolumn{1}{c}{BVAR}&\multicolumn{1}{c}{2SRR}&\multicolumn{1}{c}{MSRR$_{\text{S}}$}&\multicolumn{1}{c}{MSRR$_{\text{D}}$}\tabularnewline
\hline
{\bfseries $\mathbf{\sfrac{K^*}{K}=0.2}$}&&&&&&&&&&&&&&\tabularnewline
~~$\sigma_{\epsilon} =\text{Low}$&{\color{blue} \textbf{ 0.079}}& 0.088& 0.086& 0.098&&-& 0.135& 0.126&\textbf{ 0.114}&&-&\textbf{ 0.258}& 0.259& 0.420\tabularnewline
~~$\sigma_{\epsilon} =\text{Medium}$& 0.179&{\color{blue}  0.142}&\textbf{ 0.135}& 0.191&&-&\textbf{ 0.202}& 0.204& 0.203&&-& 0.348&\textbf{ 0.339}& 0.684\tabularnewline
~~$\sigma_{\epsilon} =\text{High}$&{\color{blue} \textbf{ 0.218}}& 0.222& 0.223& 0.282&&-& 0.290&\textbf{ 0.290}& 0.355&&-& 0.655&\textbf{ 0.638}& 1.313\tabularnewline
~~$\sigma_{\epsilon,t} = \text{SV Low-Med}$&{\color{blue} \textbf{ 0.093}}& 0.103& 0.094& 0.119&&-& 0.156& 0.153&\textbf{ 0.138}&&-& 0.281&\textbf{ 0.278}& 0.498\tabularnewline
~~$\sigma_{\epsilon,t}  = \text{SV Low-High}$& 0.137&{\color{blue}  0.129}&\textbf{ 0.121}& 0.165&&-&\textbf{ 0.196}& 0.199& 0.222&&-& 0.340&\textbf{ 0.339}& 0.602\tabularnewline
\hline
{\bfseries $\mathbf{\sfrac{K^*}{K}=0.5}$}&&&&&&&&&&&&&&\tabularnewline
~~$\sigma_{\epsilon} =\text{Low}$& 0.102&{\color{blue} \textbf{ 0.071}}& 0.082& 0.106&&-&\textbf{ 0.116}& 0.116& 0.126&&-&\textbf{ 0.232}& 0.235& 0.413\tabularnewline
~~$\sigma_{\epsilon} =\text{Medium}$& 0.131&{\color{blue} \textbf{ 0.119}}& 0.122& 0.176&&-&\textbf{ 0.176}& 0.180& 0.193&&-& 0.324&\textbf{ 0.320}& 0.722\tabularnewline
~~$\sigma_{\epsilon} =\text{High}$&{\color{blue} \textbf{ 0.172}}& 0.185& 0.186& 0.234&&-& 0.274&\textbf{ 0.273}& 0.329&&-& 0.646&\textbf{ 0.634}& 1.382\tabularnewline
~~$\sigma_{\epsilon,t} = \text{SV Low-Med}$& 0.115&{\color{blue} \textbf{ 0.085}}& 0.089& 0.116&&-&\textbf{ 0.133}& 0.138& 0.146&&-& 0.264&\textbf{ 0.262}& 0.433\tabularnewline
~~$\sigma_{\epsilon,t}  = \text{SV Low-High}$& 0.120&{\color{blue} \textbf{ 0.105}}& 0.106& 0.143&&-&\textbf{ 0.166}& 0.171& 0.195&&-&\textbf{ 0.316}& 0.319& 0.611\tabularnewline
\hline
{\bfseries $\mathbf{\sfrac{K^*}{K}=1}$}&&&&&&&&&&&&&&\tabularnewline
~~$\sigma_{\epsilon} =\text{Low}$& 0.067&{\color{blue} \textbf{ 0.047}}& 0.050& 0.054&&-& 0.075& 0.091&\textbf{ 0.064}&&-&\textbf{ 0.177}& 0.186& 0.319\tabularnewline
~~$\sigma_{\epsilon} =\text{Medium}$& 0.086&{\color{blue} \textbf{ 0.080}}& 0.087& 0.097&&-& 0.128& 0.139&\textbf{ 0.127}&&-&\textbf{ 0.302}& 0.304& 0.607\tabularnewline
~~$\sigma_{\epsilon} =\text{High}$&{\color{blue} \textbf{ 0.131}}& 0.139& 0.139& 0.179&&-&\textbf{ 0.238}& 0.246& 0.244&&-& 0.638&\textbf{ 0.628}& 1.470\tabularnewline
~~$\sigma_{\epsilon,t} = \text{SV Low-Med}$& 0.067&{\color{blue}  0.056}& 0.059&\textbf{ 0.052}&&-& 0.088& 0.102&\textbf{ 0.081}&&-&\textbf{ 0.211}& 0.221& 0.402\tabularnewline
~~$\sigma_{\epsilon,t}  = \text{SV Low-High}$& 0.071&{\color{blue} \textbf{ 0.066}}& 0.075& 0.085&&-&\textbf{ 0.115}& 0.131& 0.137&&-&\textbf{ 0.281}& 0.289& 0.501\tabularnewline
\hline\hline
\end{tabular}
%\end{center}
\begin{tablenotes}[para,flushleft]
 \footnotesize Notes: see Table \ref{s1_table}.
  \end{tablenotes}
  \end{threeparttable}
\end{table}

\begin{table}[!hp]
\caption{Results for Simulation 4 (Mixture) and $T=300$ \label{s4_table}} 
 \begin{threeparttable}
\setlength{\tabcolsep}{0.2em}
\vspace{-0.3cm}
%\hspace*{-0.5cm}
%\begin{center}
\footnotesize 
\begin{tabular}{lcccccccccccccc}
\hline\hline
\multicolumn{1}{l}{\bfseries }&\multicolumn{4}{c}{\bfseries $K=6$}&\multicolumn{1}{c}{\bfseries }&\multicolumn{4}{c}{\bfseries $K=20$}&\multicolumn{1}{c}{\bfseries }&\multicolumn{4}{c}{\bfseries $K=100$}\tabularnewline
\cline{2-5} \cline{7-10} \cline{12-15}
\multicolumn{1}{l}{}&\multicolumn{1}{c}{BVAR}&\multicolumn{1}{c}{2SRR}&\multicolumn{1}{c}{MSRR$_{\text{S}}$}&\multicolumn{1}{c}{MSRR$_{\text{D}}$}&\multicolumn{1}{c}{}&\multicolumn{1}{c}{BVAR}&\multicolumn{1}{c}{2SRR}&\multicolumn{1}{c}{MSRR$_{\text{S}}$}&\multicolumn{1}{c}{MSRR$_{\text{D}}$}&\multicolumn{1}{c}{}&\multicolumn{1}{c}{BVAR}&\multicolumn{1}{c}{2SRR}&\multicolumn{1}{c}{MSRR$_{\text{S}}$}&\multicolumn{1}{c}{MSRR$_{\text{D}}$}\tabularnewline
\hline
{\bfseries $\mathbf{\sfrac{K^*}{K}=0.2}$}&&&&&&&&&&&&&&\tabularnewline
~~$\sigma_{\epsilon} =\text{Low}$&{\color{blue}  0.054}& 0.054&\textbf{ 0.051}& 0.066&&-& 0.068&\textbf{ 0.066}& 0.073&&-& 0.150&\textbf{ 0.147}& 0.269\tabularnewline
~~$\sigma_{\epsilon} =\text{Medium}$&{\color{blue} \textbf{ 0.079}}& 0.082& 0.080& 0.100&&-& 0.115&\textbf{ 0.113}& 0.122&&-& 0.283&\textbf{ 0.278}& 0.561\tabularnewline
~~$\sigma_{\epsilon} =\text{High}$&{\color{blue} \textbf{ 0.126}}& 0.138& 0.136& 0.160&&-& 0.232&\textbf{ 0.228}& 0.246&&-& 0.628&\textbf{ 0.613}& 1.326\tabularnewline
~~$\sigma_{\epsilon,t} = \text{SV Low-Med}$&{\color{blue} \textbf{ 0.058}}& 0.062& 0.060& 0.074&&-& 0.078&\textbf{ 0.077}& 0.080&&-& 0.185&\textbf{ 0.183}& 0.338\tabularnewline
~~$\sigma_{\epsilon,t}  = \text{SV Low-High}$&{\color{blue} \textbf{ 0.065}}& 0.073& 0.077& 0.097&&-&\textbf{ 0.104}& 0.106& 0.133&&-&\textbf{ 0.258}& 0.263& 0.487\tabularnewline
\hline
{\bfseries $\mathbf{\sfrac{K^*}{K}=0.5}$}&&&&&&&&&&&&&&\tabularnewline
~~$\sigma_{\epsilon} =\text{Low}$& 0.076&{\color{blue}  0.066}&\textbf{ 0.062}& 0.082&&-& 0.086&\textbf{ 0.085}& 0.090&&-& 0.169&\textbf{ 0.167}& 0.289\tabularnewline
~~$\sigma_{\epsilon} =\text{Medium}$&{\color{blue} \textbf{ 0.095}}& 0.097& 0.096& 0.124&&-& 0.130&\textbf{ 0.127}& 0.135&&-& 0.294&\textbf{ 0.290}& 0.571\tabularnewline
~~$\sigma_{\epsilon} =\text{High}$&{\color{blue} \textbf{ 0.138}}& 0.151& 0.149& 0.183&&-& 0.238&\textbf{ 0.234}& 0.254&&-& 0.633&\textbf{ 0.623}& 1.304\tabularnewline
~~$\sigma_{\epsilon,t} = \text{SV Low-Med}$& 0.078&{\color{blue}  0.075}&\textbf{ 0.072}& 0.090&&-& 0.097&\textbf{ 0.096}& 0.099&&-& 0.204&\textbf{ 0.200}& 0.323\tabularnewline
~~$\sigma_{\epsilon,t}  = \text{SV Low-High}$&{\color{blue} \textbf{ 0.085}}& 0.089& 0.091& 0.111&&-&\textbf{ 0.120}& 0.123& 0.151&&-&\textbf{ 0.268}& 0.271& 0.475\tabularnewline
\hline
{\bfseries $\mathbf{\sfrac{K^*}{K}=1}$}&&&&&&&&&&&&&&\tabularnewline
~~$\sigma_{\epsilon} =\text{Low}$& 0.098&{\color{blue} \textbf{ 0.075}}& 0.078& 0.102&&-&\textbf{ 0.110}& 0.114& 0.116&&-& 0.201&\textbf{ 0.198}& 0.358\tabularnewline
~~$\sigma_{\epsilon} =\text{Medium}$& 0.121&{\color{blue} \textbf{ 0.118}}& 0.119& 0.150&&-& 0.155&\textbf{ 0.154}& 0.163&&-& 0.309&\textbf{ 0.306}& 0.629\tabularnewline
~~$\sigma_{\epsilon} =\text{High}$&{\color{blue} \textbf{ 0.161}}& 0.176& 0.177& 0.229&&-& 0.264&\textbf{ 0.257}& 0.288&&-& 0.641&\textbf{ 0.635}& 1.403\tabularnewline
~~$\sigma_{\epsilon,t} = \text{SV Low-Med}$& 0.107&{\color{blue} \textbf{ 0.087}}& 0.087& 0.108&&-&\textbf{ 0.121}& 0.122& 0.126&&-& 0.230&\textbf{ 0.225}& 0.374\tabularnewline
~~$\sigma_{\epsilon,t}  = \text{SV Low-High}$& 0.112&{\color{blue} \textbf{ 0.105}}& 0.109& 0.132&&-&\textbf{ 0.145}& 0.147& 0.172&&-&\textbf{ 0.287}& 0.289& 0.567\tabularnewline
\hline\hline
\end{tabular}
%\end{center}
\begin{tablenotes}[para,flushleft]
 \footnotesize Notes: see Table \ref{s1_table}.
  \end{tablenotes}
  \end{threeparttable}
\end{table}

 %%%%%%%%%%%%%%%%%%%%%%%%%%%%%%%%%%%%%%%%%%%%%%%%%%%%%
 
\begin{table}[!hp]
\caption{Results for Simulation 1 (Cosine) and $T=150$ \label{s1_table_150}} 
 \begin{threeparttable}
\setlength{\tabcolsep}{0.2em}
\vspace{-0.3cm}
%\hspace*{-0.5cm}
%\begin{center}
\footnotesize 
\begin{tabular}{lcccccccccccccc}
\hline\hline
\multicolumn{1}{l}{\bfseries }&\multicolumn{4}{c}{\bfseries $K=6$}&\multicolumn{1}{c}{\bfseries }&\multicolumn{4}{c}{\bfseries $K=20$}&\multicolumn{1}{c}{\bfseries }&\multicolumn{4}{c}{\bfseries $K=100$}\tabularnewline
\cline{2-5} \cline{7-10} \cline{12-15}
\multicolumn{1}{l}{}&\multicolumn{1}{c}{BVAR}&\multicolumn{1}{c}{2SRR}&\multicolumn{1}{c}{MSRR$_{\text{S}}$}&\multicolumn{1}{c}{MSRR$_{\text{D}}$}&\multicolumn{1}{c}{}&\multicolumn{1}{c}{BVAR}&\multicolumn{1}{c}{2SRR}&\multicolumn{1}{c}{MSRR$_{\text{S}}$}&\multicolumn{1}{c}{MSRR$_{\text{D}}$}&\multicolumn{1}{c}{}&\multicolumn{1}{c}{BVAR}&\multicolumn{1}{c}{2SRR}&\multicolumn{1}{c}{MSRR$_{\text{S}}$}&\multicolumn{1}{c}{MSRR$_{\text{D}}$}\tabularnewline
\hline
{\bfseries $\mathbf{\sfrac{K^*}{K}=0.2}$}&&&&&&&&&&&&&&\tabularnewline
~~$\sigma_{\epsilon} =\text{Low}$& 0.136&{\color{blue}  0.119}&\textbf{ 0.109}& 0.154&&-& 0.139&\textbf{ 0.133}& 0.159&&-&\textbf{ 0.244}& 0.252& 0.294\tabularnewline
~~$\sigma_{\epsilon} =\text{Medium}$&{\color{blue} \textbf{ 0.182}}& 0.186& 0.183& 0.214&&-& 0.212&\textbf{ 0.205}& 0.257&&-&\textbf{ 0.321}& 0.328& 0.337\tabularnewline
~~$\sigma_{\epsilon} =\text{High}$&{\color{blue} \textbf{ 0.337}}& 0.354& 0.344& 0.377&&-&\textbf{ 0.393}& 0.396& 0.568&&-&\textbf{ 0.569}& 0.584& 0.617\tabularnewline
~~$\sigma_{\epsilon,t} = \text{SV Low-Med}$& 0.150&{\color{blue} \textbf{ 0.148}}& 0.149& 0.183&&-& 0.162&\textbf{ 0.158}& 0.208&&-&\textbf{ 0.273}& 0.277& 0.297\tabularnewline
~~$\sigma_{\epsilon,t}  = \text{SV Low-High}$&{\color{blue} \textbf{ 0.200}}& 0.212& 0.216& 0.252&&-& 0.244&\textbf{ 0.242}& 0.338&&-&\textbf{ 0.376}& 0.385& 0.385\tabularnewline
\hline
{\bfseries $\mathbf{\sfrac{K^*}{K}=0.5}$}&&&&&&&&&&&&&&\tabularnewline
~~$\sigma_{\epsilon} =\text{Low}$& 0.200&{\color{blue}  0.142}&\textbf{ 0.136}& 0.187&&-& 0.170&\textbf{ 0.170}& 0.215&&-& 0.358& 0.368&\textbf{ 0.347}\tabularnewline
~~$\sigma_{\epsilon} =\text{Medium}$& 0.228&{\color{blue} \textbf{ 0.219}}& 0.222& 0.291&&-&\textbf{ 0.252}& 0.256& 0.305&&-& 0.418& 0.432&\textbf{ 0.406}\tabularnewline
~~$\sigma_{\epsilon} =\text{High}$&{\color{blue} \textbf{ 0.360}}& 0.383& 0.374& 0.396&&-&\textbf{ 0.432}& 0.436& 0.591&&-&\textbf{ 0.624}& 0.633& 0.712\tabularnewline
~~$\sigma_{\epsilon,t} = \text{SV Low-Med}$& 0.203&{\color{blue}  0.175}&\textbf{ 0.173}& 0.227&&-&\textbf{ 0.209}& 0.212& 0.253&&-& 0.387& 0.400&\textbf{ 0.365}\tabularnewline
~~$\sigma_{\epsilon,t}  = \text{SV Low-High}$&{\color{blue} \textbf{ 0.240}}& 0.243& 0.248& 0.285&&-&\textbf{ 0.287}& 0.292& 0.397&&-& 0.465& 0.485&\textbf{ 0.436}\tabularnewline
\hline
{\bfseries $\mathbf{\sfrac{K^*}{K}=1}$}&&&&&&&&&&&&&&\tabularnewline
~~$\sigma_{\epsilon} =\text{Low}$& 0.262&{\color{blue} \textbf{ 0.158}}& 0.161& 0.189&&-&\textbf{ 0.206}& 0.225& 0.231&&-& 0.492& 0.511&\textbf{ 0.429}\tabularnewline
~~$\sigma_{\epsilon} =\text{Medium}$& 0.284&{\color{blue} \textbf{ 0.239}}& 0.247& 0.270&&-&\textbf{ 0.302}& 0.316& 0.352&&-& 0.532& 0.547&\textbf{ 0.500}\tabularnewline
~~$\sigma_{\epsilon} =\text{High}$&{\color{blue} \textbf{ 0.390}}& 0.421& 0.431& 0.433&&-&\textbf{ 0.477}& 0.481& 0.678&&-&\textbf{ 0.717}& 0.737& 0.722\tabularnewline
~~$\sigma_{\epsilon,t} = \text{SV Low-Med}$& 0.270&{\color{blue} \textbf{ 0.191}}& 0.197& 0.241&&-&\textbf{ 0.253}& 0.270& 0.269&&-& 0.508& 0.526&\textbf{ 0.453}\tabularnewline
~~$\sigma_{\epsilon,t}  = \text{SV Low-High}$& 0.295&{\color{blue} \textbf{ 0.270}}& 0.282& 0.324&&-&\textbf{ 0.338}& 0.356& 0.454&&-& 0.572& 0.590&\textbf{ 0.546}\tabularnewline
\hline\hline
\end{tabular}
%\end{center}
\begin{tablenotes}[para,flushleft]
  \footnotesize Notes: see Table \ref{s1_table}.
  \end{tablenotes}
  \end{threeparttable}
\end{table}

\begin{table}[!hp]
\caption{Results for Simulation 2 (Break) and $T=150$ \label{s2_table_150}} 
 \begin{threeparttable}
\setlength{\tabcolsep}{0.2em}
\vspace{-0.3cm}
%\hspace*{-0.5cm}
%\begin{center}
\footnotesize 
\begin{tabular}{lcccccccccccccc}
\hline\hline
\multicolumn{1}{l}{\bfseries }&\multicolumn{4}{c}{\bfseries $K=6$}&\multicolumn{1}{c}{\bfseries }&\multicolumn{4}{c}{\bfseries $K=20$}&\multicolumn{1}{c}{\bfseries }&\multicolumn{4}{c}{\bfseries $K=100$}\tabularnewline
\cline{2-5} \cline{7-10} \cline{12-15}
\multicolumn{1}{l}{}&\multicolumn{1}{c}{BVAR}&\multicolumn{1}{c}{2SRR}&\multicolumn{1}{c}{MSRR$_{\text{S}}$}&\multicolumn{1}{c}{MSRR$_{\text{D}}$}&\multicolumn{1}{c}{}&\multicolumn{1}{c}{BVAR}&\multicolumn{1}{c}{2SRR}&\multicolumn{1}{c}{MSRR$_{\text{S}}$}&\multicolumn{1}{c}{MSRR$_{\text{D}}$}&\multicolumn{1}{c}{}&\multicolumn{1}{c}{BVAR}&\multicolumn{1}{c}{2SRR}&\multicolumn{1}{c}{MSRR$_{\text{S}}$}&\multicolumn{1}{c}{MSRR$_{\text{D}}$}\tabularnewline
\hline
{\bfseries $\mathbf{\sfrac{K^*}{K}=0.2}$}&&&&&&&&&&&&&&\tabularnewline
~~$\sigma_{\epsilon} =\text{Low}$&{\color{blue}  0.083}& 0.085&\textbf{ 0.082}& 0.110&&-& 0.162&\textbf{ 0.155}& 0.175&&-&\textbf{ 0.534}& 0.547& 0.611\tabularnewline
~~$\sigma_{\epsilon} =\text{Medium}$&{\color{blue} \textbf{ 0.154}}& 0.162& 0.160& 0.188&&-& 0.330&\textbf{ 0.312}& 0.449&&-&\textbf{ 1.070}& 1.073& 1.283\tabularnewline
~~$\sigma_{\epsilon} =\text{High}$&{\color{blue} \textbf{ 0.380}}& 0.396& 0.399& 0.472&&-& 0.745&\textbf{ 0.727}& 1.059&&-&\textbf{ 2.364}& 2.430& 2.378\tabularnewline
~~$\sigma_{\epsilon,t} = \text{SV Low-Med}$&{\color{blue} \textbf{ 0.115}}& 0.125& 0.126& 0.165&&-& 0.229&\textbf{ 0.214}& 0.304&&-&\textbf{ 0.766}& 0.784& 0.894\tabularnewline
~~$\sigma_{\epsilon,t}  = \text{SV Low-High}$&{\color{blue} \textbf{ 0.196}}& 0.213& 0.216& 0.268&&-& 0.387&\textbf{ 0.380}& 0.592&&-& 1.341& 1.370&\textbf{ 1.305}\tabularnewline
\hline
{\bfseries $\mathbf{\sfrac{K^*}{K}=0.5}$}&&&&&&&&&&&&&&\tabularnewline
~~$\sigma_{\epsilon} =\text{Low}$& 0.084&{\color{blue}  0.084}&\textbf{ 0.083}& 0.105&&-& 0.164&\textbf{ 0.157}& 0.178&&-&\textbf{ 0.536}& 0.544& 0.652\tabularnewline
~~$\sigma_{\epsilon} =\text{Medium}$&{\color{blue} \textbf{ 0.153}}& 0.162& 0.162& 0.205&&-& 0.336&\textbf{ 0.317}& 0.445&&-&\textbf{ 1.080}& 1.081& 1.180\tabularnewline
~~$\sigma_{\epsilon} =\text{High}$&{\color{blue} \textbf{ 0.382}}& 0.385& 0.385& 0.502&&-& 0.743&\textbf{ 0.731}& 1.021&&-&\textbf{ 2.364}& 2.426& 2.411\tabularnewline
~~$\sigma_{\epsilon,t} = \text{SV Low-Med}$&{\color{blue} \textbf{ 0.113}}& 0.121& 0.120& 0.136&&-& 0.225&\textbf{ 0.213}& 0.305&&-&\textbf{ 0.770}& 0.792& 0.905\tabularnewline
~~$\sigma_{\epsilon,t}  = \text{SV Low-High}$&{\color{blue} \textbf{ 0.198}}& 0.215& 0.220& 0.240&&-& 0.382&\textbf{ 0.376}& 0.580&&-&\textbf{ 1.341}& 1.385& 1.379\tabularnewline
\hline
{\bfseries $\mathbf{\sfrac{K^*}{K}=1}$}&&&&&&&&&&&&&&\tabularnewline
~~$\sigma_{\epsilon} =\text{Low}$&{\color{blue} \textbf{ 0.085}}& 0.089& 0.086& 0.099&&-& 0.163&\textbf{ 0.157}& 0.167&&-& 0.534&\textbf{ 0.532}& 0.717\tabularnewline
~~$\sigma_{\epsilon} =\text{Medium}$&{\color{blue} \textbf{ 0.158}}& 0.173& 0.170& 0.203&&-& 0.330&\textbf{ 0.320}& 0.460&&-&\textbf{ 1.063}& 1.073& 1.264\tabularnewline
~~$\sigma_{\epsilon} =\text{High}$&{\color{blue} \textbf{ 0.383}}& 0.426& 0.419& 0.460&&-& 0.751&\textbf{ 0.736}& 0.953&&-&\textbf{ 2.353}& 2.408& 2.613\tabularnewline
~~$\sigma_{\epsilon,t} = \text{SV Low-Med}$&{\color{blue} \textbf{ 0.114}}& 0.124& 0.126& 0.154&&-& 0.214&\textbf{ 0.212}& 0.280&&-&\textbf{ 0.771}& 0.785& 0.894\tabularnewline
~~$\sigma_{\epsilon,t}  = \text{SV Low-High}$&{\color{blue} \textbf{ 0.198}}& 0.224& 0.225& 0.278&&-& 0.386&\textbf{ 0.373}& 0.538&&-&\textbf{ 1.362}& 1.414& 1.467\tabularnewline
\hline\hline
\end{tabular}
%\end{center}
\begin{tablenotes}[para,flushleft]
 \footnotesize Notes: see Table \ref{s1_table}.
  \end{tablenotes}
  \end{threeparttable}
\end{table}

\begin{table}[!hp]
\caption{Results for Simulation 3 (Trend and Cosine) and $T=150$ \label{s3_table_150}} 
 \begin{threeparttable}
\setlength{\tabcolsep}{0.2em}
\vspace{-0.3cm}
%\hspace*{-0.5cm}
%\begin{center}
\footnotesize 
\begin{tabular}{lcccccccccccccc}
\hline\hline
\multicolumn{1}{l}{\bfseries }&\multicolumn{4}{c}{\bfseries $K=6$}&\multicolumn{1}{c}{\bfseries }&\multicolumn{4}{c}{\bfseries $K=20$}&\multicolumn{1}{c}{\bfseries }&\multicolumn{4}{c}{\bfseries $K=100$}\tabularnewline
\cline{2-5} \cline{7-10} \cline{12-15}
\multicolumn{1}{l}{}&\multicolumn{1}{c}{BVAR}&\multicolumn{1}{c}{2SRR}&\multicolumn{1}{c}{MSRR$_{\text{S}}$}&\multicolumn{1}{c}{MSRR$_{\text{D}}$}&\multicolumn{1}{c}{}&\multicolumn{1}{c}{BVAR}&\multicolumn{1}{c}{2SRR}&\multicolumn{1}{c}{MSRR$_{\text{S}}$}&\multicolumn{1}{c}{MSRR$_{\text{D}}$}&\multicolumn{1}{c}{}&\multicolumn{1}{c}{BVAR}&\multicolumn{1}{c}{2SRR}&\multicolumn{1}{c}{MSRR$_{\text{S}}$}&\multicolumn{1}{c}{MSRR$_{\text{D}}$}\tabularnewline
\hline
{\bfseries $\mathbf{\sfrac{K^*}{K}=0.2}$}&&&&&&&&&&&&&&\tabularnewline
~~$\sigma_{\epsilon} =\text{Low}$& 0.149&{\color{blue}  0.096}&\textbf{ 0.091}& 0.115&&-& 0.159&\textbf{ 0.156}& 0.202&&-& 0.405& 0.415&\textbf{ 0.381}\tabularnewline
~~$\sigma_{\epsilon} =\text{Medium}$& 0.181&{\color{blue}  0.153}&\textbf{ 0.149}& 0.182&&-& 0.244&\textbf{ 0.243}& 0.342&&-&\textbf{ 0.604}& 0.624& 0.628\tabularnewline
~~$\sigma_{\epsilon} =\text{High}$&{\color{blue} \textbf{ 0.253}}& 0.275& 0.276& 0.313&&-&\textbf{ 0.419}& 0.421& 0.680&&-&\textbf{ 1.218}& 1.248& 1.235\tabularnewline
~~$\sigma_{\epsilon,t} = \text{SV Low-Med}$& 0.167&{\color{blue}  0.121}&\textbf{ 0.114}& 0.140&&-&\textbf{ 0.192}& 0.192& 0.246&&-& 0.485& 0.501&\textbf{ 0.480}\tabularnewline
~~$\sigma_{\epsilon,t}  = \text{SV Low-High}$& 0.193&{\color{blue} \textbf{ 0.177}}& 0.178& 0.210&&-&\textbf{ 0.261}& 0.262& 0.420&&-& 0.749& 0.767&\textbf{ 0.609}\tabularnewline
\hline
{\bfseries $\mathbf{\sfrac{K^*}{K}=0.5}$}&&&&&&&&&&&&&&\tabularnewline
~~$\sigma_{\epsilon} =\text{Low}$& 0.104&{\color{blue} \textbf{ 0.077}}& 0.085& 0.118&&-&\textbf{ 0.138}& 0.139& 0.186&&-& 0.365& 0.377&\textbf{ 0.342}\tabularnewline
~~$\sigma_{\epsilon} =\text{Medium}$& 0.128&{\color{blue} \textbf{ 0.127}}& 0.131& 0.161&&-& 0.223&\textbf{ 0.221}& 0.337&&-&\textbf{ 0.585}& 0.599& 0.602\tabularnewline
~~$\sigma_{\epsilon} =\text{High}$&{\color{blue} \textbf{ 0.224}}& 0.249& 0.250& 0.276&&-& 0.397&\textbf{ 0.397}& 0.654&&-&\textbf{ 1.211}& 1.243& 1.262\tabularnewline
~~$\sigma_{\epsilon,t} = \text{SV Low-Med}$& 0.112&{\color{blue} \textbf{ 0.098}}& 0.100& 0.147&&-&\textbf{ 0.173}& 0.175& 0.223&&-& 0.455& 0.474&\textbf{ 0.445}\tabularnewline
~~$\sigma_{\epsilon,t}  = \text{SV Low-High}$&{\color{blue} \textbf{ 0.145}}& 0.151& 0.154& 0.194&&-&\textbf{ 0.241}& 0.241& 0.374&&-& 0.726& 0.743&\textbf{ 0.657}\tabularnewline
\hline
{\bfseries $\mathbf{\sfrac{K^*}{K}=1}$}&&&&&&&&&&&&&&\tabularnewline
~~$\sigma_{\epsilon} =\text{Low}$&{\color{blue} \textbf{ 0.062}}& 0.062& 0.065& 0.071&&-&\textbf{ 0.107}& 0.108& 0.130&&-&\textbf{ 0.299}& 0.311& 0.317\tabularnewline
~~$\sigma_{\epsilon} =\text{Medium}$&{\color{blue} \textbf{ 0.096}}& 0.103& 0.107& 0.116&&-& 0.174&\textbf{ 0.173}& 0.279&&-&\textbf{ 0.547}& 0.559& 0.604\tabularnewline
~~$\sigma_{\epsilon} =\text{High}$&{\color{blue} \textbf{ 0.204}}& 0.221& 0.221& 0.259&&-& 0.384&\textbf{ 0.378}& 0.552&&-&\textbf{ 1.204}& 1.234& 1.280\tabularnewline
~~$\sigma_{\epsilon,t} = \text{SV Low-Med}$&{\color{blue} \textbf{ 0.073}}& 0.077& 0.081& 0.093&&-&\textbf{ 0.126}& 0.128& 0.164&&-& 0.406& 0.423&\textbf{ 0.404}\tabularnewline
~~$\sigma_{\epsilon,t}  = \text{SV Low-High}$&{\color{blue} \textbf{ 0.111}}& 0.121& 0.126& 0.162&&-&\textbf{ 0.202}& 0.202& 0.323&&-& 0.713& 0.734&\textbf{ 0.693}\tabularnewline
\hline\hline
\end{tabular}
%\end{center}
\begin{tablenotes}[para,flushleft]
 \footnotesize Notes: see Table \ref{s1_table}.
  \end{tablenotes}
  \end{threeparttable}
\end{table}

\begin{table}[!hp]
\caption{Results for Simulation 4 (Mixture) and $T=150$ \label{s4_table_150}} 
 \begin{threeparttable}
\setlength{\tabcolsep}{0.2em}
\vspace{-0.3cm}
%\hspace*{-0.5cm}
%\begin{center}
\footnotesize 
\begin{tabular}{lcccccccccccccc}
\hline\hline
\multicolumn{1}{l}{\bfseries }&\multicolumn{4}{c}{\bfseries $K=6$}&\multicolumn{1}{c}{\bfseries }&\multicolumn{4}{c}{\bfseries $K=20$}&\multicolumn{1}{c}{\bfseries }&\multicolumn{4}{c}{\bfseries $K=100$}\tabularnewline
\cline{2-5} \cline{7-10} \cline{12-15}
\multicolumn{1}{l}{}&\multicolumn{1}{c}{BVAR}&\multicolumn{1}{c}{2SRR}&\multicolumn{1}{c}{MSRR$_{\text{S}}$}&\multicolumn{1}{c}{MSRR$_{\text{D}}$}&\multicolumn{1}{c}{}&\multicolumn{1}{c}{BVAR}&\multicolumn{1}{c}{2SRR}&\multicolumn{1}{c}{MSRR$_{\text{S}}$}&\multicolumn{1}{c}{MSRR$_{\text{D}}$}&\multicolumn{1}{c}{}&\multicolumn{1}{c}{BVAR}&\multicolumn{1}{c}{2SRR}&\multicolumn{1}{c}{MSRR$_{\text{S}}$}&\multicolumn{1}{c}{MSRR$_{\text{D}}$}\tabularnewline
\hline
{\bfseries $\mathbf{\sfrac{K^*}{K}=0.2}$}&&&&&&&&&&&&&&\tabularnewline
~~$\sigma_{\epsilon} =\text{Low}$& 0.065&{\color{blue}  0.063}&\textbf{ 0.061}& 0.069&&-& 0.092&\textbf{ 0.091}& 0.094&&-&\textbf{ 0.277}& 0.280& 0.337\tabularnewline
~~$\sigma_{\epsilon} =\text{Medium}$& 0.097&{\color{blue} \textbf{ 0.096}}& 0.097& 0.113&&-& 0.175&\textbf{ 0.172}& 0.201&&-&\textbf{ 0.542}& 0.546& 0.568\tabularnewline
~~$\sigma_{\epsilon} =\text{High}$&{\color{blue} \textbf{ 0.203}}& 0.212& 0.206& 0.227&&-& 0.380&\textbf{ 0.375}& 0.585&&-&\textbf{ 1.187}& 1.214& 1.230\tabularnewline
~~$\sigma_{\epsilon,t} = \text{SV Low-Med}$&{\color{blue} \textbf{ 0.080}}& 0.082& 0.081& 0.089&&-& 0.125&\textbf{ 0.122}& 0.144&&-&\textbf{ 0.386}& 0.392& 0.404\tabularnewline
~~$\sigma_{\epsilon,t}  = \text{SV Low-High}$&{\color{blue} \textbf{ 0.115}}& 0.118& 0.120& 0.145&&-& 0.200&\textbf{ 0.197}& 0.311&&-&\textbf{ 0.696}& 0.720& 0.699\tabularnewline
\hline
{\bfseries $\mathbf{\sfrac{K^*}{K}=0.5}$}&&&&&&&&&&&&&&\tabularnewline
~~$\sigma_{\epsilon} =\text{Low}$& 0.086&{\color{blue}  0.073}&\textbf{ 0.071}& 0.092&&-& 0.106&\textbf{ 0.104}& 0.119&&-&\textbf{ 0.290}& 0.299& 0.349\tabularnewline
~~$\sigma_{\epsilon} =\text{Medium}$& 0.111&{\color{blue} \textbf{ 0.107}}& 0.108& 0.126&&-& 0.184&\textbf{ 0.183}& 0.222&&-&\textbf{ 0.547}& 0.559& 0.594\tabularnewline
~~$\sigma_{\epsilon} =\text{High}$&{\color{blue} \textbf{ 0.208}}& 0.215& 0.215& 0.242&&-& 0.381&\textbf{ 0.378}& 0.547&&-&\textbf{ 1.183}& 1.212& 1.286\tabularnewline
~~$\sigma_{\epsilon,t} = \text{SV Low-Med}$& 0.096&{\color{blue}  0.093}&\textbf{ 0.092}& 0.105&&-& 0.132&\textbf{ 0.131}& 0.154&&-&\textbf{ 0.397}& 0.410& 0.439\tabularnewline
~~$\sigma_{\epsilon,t}  = \text{SV Low-High}$&{\color{blue} \textbf{ 0.127}}& 0.129& 0.130& 0.159&&-& 0.210&\textbf{ 0.206}& 0.326&&-& 0.707& 0.733&\textbf{ 0.678}\tabularnewline
\hline
{\bfseries $\mathbf{\sfrac{K^*}{K}=1}$}&&&&&&&&&&&&&&\tabularnewline
~~$\sigma_{\epsilon} =\text{Low}$& 0.106&{\color{blue} \textbf{ 0.080}}& 0.082& 0.101&&-&\textbf{ 0.124}& 0.125& 0.138&&-&\textbf{ 0.315}& 0.328& 0.334\tabularnewline
~~$\sigma_{\epsilon} =\text{Medium}$& 0.129&{\color{blue}  0.124}&\textbf{ 0.123}& 0.142&&-& 0.198&\textbf{ 0.197}& 0.227&&-&\textbf{ 0.561}& 0.579& 0.640\tabularnewline
~~$\sigma_{\epsilon} =\text{High}$&{\color{blue} \textbf{ 0.218}}& 0.224& 0.225& 0.258&&-& 0.410&\textbf{ 0.409}& 0.583&&-&\textbf{ 1.208}& 1.225& 1.343\tabularnewline
~~$\sigma_{\epsilon,t} = \text{SV Low-Med}$& 0.120&{\color{blue} \textbf{ 0.100}}& 0.105& 0.129&&-&\textbf{ 0.147}& 0.149& 0.170&&-&\textbf{ 0.417}& 0.434& 0.473\tabularnewline
~~$\sigma_{\epsilon,t}  = \text{SV Low-High}$& 0.140&{\color{blue} \textbf{ 0.134}}& 0.139& 0.169&&-& 0.221&\textbf{ 0.218}& 0.348&&-&\textbf{ 0.709}& 0.734& 0.712\tabularnewline
\hline\hline
\end{tabular}
%\end{center}
\begin{tablenotes}[para,flushleft]
 \footnotesize Notes: see Table \ref{s1_table}.
  \end{tablenotes}
  \end{threeparttable}
\end{table}

%%%%%%%%%%%%%%%%%%%%%%%%%%%%%%%%%%%%%%%%%%%%%%%%%%%%%

\begin{table}[!hp]
\caption{Results for Simulation 1 (Cosine) and $T=600$ \label{s1_table_600}} 
 \begin{threeparttable}
\setlength{\tabcolsep}{0.2em}
\vspace{-0.3cm}
%\hspace*{-0.5cm}
%\begin{center}
\footnotesize 
\begin{tabular}{lcccccccccccccc}
\hline\hline
\multicolumn{1}{l}{\bfseries }&\multicolumn{4}{c}{\bfseries $K=6$}&\multicolumn{1}{c}{\bfseries }&\multicolumn{4}{c}{\bfseries $K=20$}&\multicolumn{1}{c}{\bfseries }&\multicolumn{4}{c}{\bfseries $K=100$}\tabularnewline
\cline{2-5} \cline{7-10} \cline{12-15}
\multicolumn{1}{l}{}&\multicolumn{1}{c}{BVAR}&\multicolumn{1}{c}{2SRR}&\multicolumn{1}{c}{MSRR$_{\text{S}}$}&\multicolumn{1}{c}{MSRR$_{\text{D}}$}&\multicolumn{1}{c}{}&\multicolumn{1}{c}{BVAR}&\multicolumn{1}{c}{2SRR}&\multicolumn{1}{c}{MSRR$_{\text{S}}$}&\multicolumn{1}{c}{MSRR$_{\text{D}}$}&\multicolumn{1}{c}{}&\multicolumn{1}{c}{BVAR}&\multicolumn{1}{c}{2SRR}&\multicolumn{1}{c}{MSRR$_{\text{S}}$}&\multicolumn{1}{c}{MSRR$_{\text{D}}$}\tabularnewline
\hline
{\bfseries $\mathbf{\sfrac{K^*}{K}=0.2}$}&&&&&&&&&&&&&&\tabularnewline
~~$\sigma_{\epsilon} =\text{Low}$&{\color{blue}  0.102}& 0.105&\textbf{ 0.078}& 0.102&&-& 0.112& 0.080&\textbf{ 0.077}&&-& 0.129& 0.126&\textbf{ 0.095}\tabularnewline
~~$\sigma_{\epsilon} =\text{Medium}$&{\color{blue} \textbf{ 0.133}}& 0.151& 0.147& 0.164&&-& 0.139& 0.135&\textbf{ 0.134}&&-& 0.149&\textbf{ 0.146}& 0.167\tabularnewline
~~$\sigma_{\epsilon} =\text{High}$&{\color{blue} \textbf{ 0.187}}& 0.205& 0.206& 0.251&&-& 0.204&\textbf{ 0.199}& 0.206&&-& 0.218&\textbf{ 0.212}& 0.241\tabularnewline
~~$\sigma_{\epsilon,t} = \text{SV Low-Med}$&{\color{blue}  0.118}& 0.124&\textbf{ 0.113}& 0.125&&-& 0.127& 0.115&\textbf{ 0.098}&&-& 0.134& 0.131&\textbf{ 0.120}\tabularnewline
~~$\sigma_{\epsilon,t}  = \text{SV Low-High}$&{\color{blue} \textbf{ 0.130}}& 0.149& 0.149& 0.174&&-& 0.136&\textbf{ 0.134}& 0.143&&-& 0.148&\textbf{ 0.146}& 0.182\tabularnewline
\hline
{\bfseries $\mathbf{\sfrac{K^*}{K}=0.5}$}&&&&&&&&&&&&&&\tabularnewline
~~$\sigma_{\epsilon} =\text{Low}$&{\color{blue} \textbf{ 0.087}}& 0.132& 0.115& 0.120&&-& 0.145& 0.124&\textbf{ 0.105}&&-& 0.219& 0.219&\textbf{ 0.109}\tabularnewline
~~$\sigma_{\epsilon} =\text{Medium}$&{\color{blue} \textbf{ 0.204}}& 0.208& 0.206& 0.204&&-& 0.212& 0.209&\textbf{ 0.161}&&-& 0.236& 0.233&\textbf{ 0.195}\tabularnewline
~~$\sigma_{\epsilon} =\text{High}$&{\color{blue} \textbf{ 0.242}}& 0.263& 0.262& 0.308&&-& 0.257&\textbf{ 0.253}& 0.266&&-& 0.279&\textbf{ 0.275}& 0.297\tabularnewline
~~$\sigma_{\epsilon,t} = \text{SV Low-Med}$& 0.154&{\color{blue}  0.152}& 0.142&\textbf{ 0.140}&&-& 0.170& 0.156&\textbf{ 0.135}&&-& 0.222& 0.221&\textbf{ 0.117}\tabularnewline
~~$\sigma_{\epsilon,t}  = \text{SV Low-High}$&{\color{blue} \textbf{ 0.205}}& 0.206& 0.207& 0.225&&-& 0.211& 0.210&\textbf{ 0.188}&&-& 0.234& 0.231&\textbf{ 0.191}\tabularnewline
\hline
{\bfseries $\mathbf{\sfrac{K^*}{K}=1}$}&&&&&&&&&&&&&&\tabularnewline
~~$\sigma_{\epsilon} =\text{Low}$&{\color{blue} \textbf{ 0.087}}& 0.146& 0.148& 0.158&&-& 0.180& 0.183&\textbf{ 0.108}&&-& 0.339& 0.340&\textbf{ 0.128}\tabularnewline
~~$\sigma_{\epsilon} =\text{Medium}$& 0.238&{\color{blue} \textbf{ 0.236}}& 0.239& 0.266&&-& 0.275& 0.278&\textbf{ 0.216}&&-& 0.346& 0.346&\textbf{ 0.178}\tabularnewline
~~$\sigma_{\epsilon} =\text{High}$&{\color{blue} \textbf{ 0.312}}& 0.338& 0.336& 0.350&&-& 0.344&\textbf{ 0.343}& 0.346&&-& 0.374&\textbf{ 0.372}& 0.407\tabularnewline
~~$\sigma_{\epsilon,t} = \text{SV Low-Med}$&{\color{blue} \textbf{ 0.115}}& 0.174& 0.176& 0.174&&-& 0.210& 0.214&\textbf{ 0.162}&&-& 0.340& 0.342&\textbf{ 0.141}\tabularnewline
~~$\sigma_{\epsilon,t}  = \text{SV Low-High}$&{\color{blue} \textbf{ 0.230}}& 0.233& 0.241& 0.258&&-& 0.274& 0.280&\textbf{ 0.213}&&-& 0.348& 0.347&\textbf{ 0.199}\tabularnewline
\hline\hline
\end{tabular}
%\end{center}
\begin{tablenotes}[para,flushleft]
  \footnotesize Notes: see Table \ref{s1_table}.
  \end{tablenotes}
  \end{threeparttable}
\end{table}

\begin{table}[!hp]
\caption{Results for Simulation 2 (Break) and $T=600$ \label{s2_table_600}} 
 \begin{threeparttable}
\setlength{\tabcolsep}{0.2em}
\vspace{-0.3cm}
%\hspace*{-0.5cm}
%\begin{center}
\footnotesize 
\begin{tabular}{lcccccccccccccc}
\hline\hline
\multicolumn{1}{l}{\bfseries }&\multicolumn{4}{c}{\bfseries $K=6$}&\multicolumn{1}{c}{\bfseries }&\multicolumn{4}{c}{\bfseries $K=20$}&\multicolumn{1}{c}{\bfseries }&\multicolumn{4}{c}{\bfseries $K=100$}\tabularnewline
\cline{2-5} \cline{7-10} \cline{12-15}
\multicolumn{1}{l}{}&\multicolumn{1}{c}{BVAR}&\multicolumn{1}{c}{2SRR}&\multicolumn{1}{c}{MSRR$_{\text{S}}$}&\multicolumn{1}{c}{MSRR$_{\text{D}}$}&\multicolumn{1}{c}{}&\multicolumn{1}{c}{BVAR}&\multicolumn{1}{c}{2SRR}&\multicolumn{1}{c}{MSRR$_{\text{S}}$}&\multicolumn{1}{c}{MSRR$_{\text{D}}$}&\multicolumn{1}{c}{}&\multicolumn{1}{c}{BVAR}&\multicolumn{1}{c}{2SRR}&\multicolumn{1}{c}{MSRR$_{\text{S}}$}&\multicolumn{1}{c}{MSRR$_{\text{D}}$}\tabularnewline
\hline
{\bfseries $\mathbf{\sfrac{K^*}{K}=0.2}$}&&&&&&&&&&&&&&\tabularnewline
~~$\sigma_{\epsilon} =\text{Low}$&{\color{blue}  0.082}& 0.084&\textbf{ 0.068}& 0.086&&-& 0.109&\textbf{ 0.094}& 0.120&&-&\textbf{ 0.219}& 0.222& 0.236\tabularnewline
~~$\sigma_{\epsilon} =\text{Medium}$& 0.140&{\color{blue}  0.136}&\textbf{ 0.128}& 0.177&&-& 0.185&\textbf{ 0.182}& 0.210&&-& 0.372&\textbf{ 0.365}& 0.418\tabularnewline
~~$\sigma_{\epsilon} =\text{High}$&{\color{blue} \textbf{ 0.213}}& 0.237& 0.241& 0.322&&-& 0.348&\textbf{ 0.347}& 0.358&&-& 0.763&\textbf{ 0.743}& 0.948\tabularnewline
~~$\sigma_{\epsilon,t} = \text{SV Low-Med}$& 0.113&{\color{blue}  0.103}&\textbf{ 0.090}& 0.109&&-& 0.133&\textbf{ 0.123}& 0.155&&-&\textbf{ 0.265}& 0.266& 0.291\tabularnewline
~~$\sigma_{\epsilon,t}  = \text{SV Low-High}$& 0.138&{\color{blue}  0.132}&\textbf{ 0.130}& 0.229&&-&\textbf{ 0.175}& 0.179& 0.233&&-&\textbf{ 0.376}& 0.385& 0.514\tabularnewline
\hline
{\bfseries $\mathbf{\sfrac{K^*}{K}=0.5}$}&&&&&&&&&&&&&&\tabularnewline
~~$\sigma_{\epsilon} =\text{Low}$&{\color{blue}  0.085}& 0.109& 0.096&\textbf{ 0.081}&&-& 0.140& 0.127&\textbf{ 0.124}&&-&\textbf{ 0.275}& 0.298& 0.316\tabularnewline
~~$\sigma_{\epsilon} =\text{Medium}$& 0.182&{\color{blue}  0.163}&\textbf{ 0.152}& 0.169&&-&\textbf{ 0.221}& 0.230& 0.251&&-&\textbf{ 0.417}& 0.431& 0.515\tabularnewline
~~$\sigma_{\epsilon} =\text{High}$&{\color{blue} \textbf{ 0.270}}& 0.274& 0.280& 0.414&&-&\textbf{ 0.384}& 0.397& 0.411&&-& 0.788&\textbf{ 0.778}& 0.908\tabularnewline
~~$\sigma_{\epsilon,t} = \text{SV Low-Med}$&{\color{blue}  0.121}& 0.130& 0.121&\textbf{ 0.108}&&-& 0.165&\textbf{ 0.161}& 0.162&&-&\textbf{ 0.315}& 0.341& 0.355\tabularnewline
~~$\sigma_{\epsilon,t}  = \text{SV Low-High}$& 0.185&{\color{blue} \textbf{ 0.162}}& 0.163& 0.209&&-&\textbf{ 0.211}& 0.230& 0.274&&-&\textbf{ 0.421}& 0.446& 0.539\tabularnewline
\hline
{\bfseries $\mathbf{\sfrac{K^*}{K}=1}$}&&&&&&&&&&&&&&\tabularnewline
~~$\sigma_{\epsilon} =\text{Low}$&{\color{blue} \textbf{ 0.089}}& 0.133& 0.134& 0.092&&-& 0.177& 0.341&\textbf{ 0.121}&&-& 0.355& 0.466&\textbf{ 0.316}\tabularnewline
~~$\sigma_{\epsilon} =\text{Medium}$&{\color{blue} \textbf{ 0.151}}& 0.205& 0.248& 0.163&&-& 0.267& 0.404&\textbf{ 0.218}&&-&\textbf{ 0.505}& 0.524& 0.587\tabularnewline
~~$\sigma_{\epsilon} =\text{High}$&{\color{blue} \textbf{ 0.285}}& 0.338& 0.395& 0.316&&-& 0.455& 0.479&\textbf{ 0.449}&&-& 0.898&\textbf{ 0.817}& 1.252\tabularnewline
~~$\sigma_{\epsilon,t} = \text{SV Low-Med}$&{\color{blue} \textbf{ 0.113}}& 0.154& 0.164& 0.120&&-& 0.203& 0.375&\textbf{ 0.146}&&-& 0.402& 0.483&\textbf{ 0.396}\tabularnewline
~~$\sigma_{\epsilon,t}  = \text{SV Low-High}$&{\color{blue} \textbf{ 0.147}}& 0.198& 0.255& 0.185&&-& 0.255& 0.409&\textbf{ 0.246}&&-&\textbf{ 0.498}& 0.534& 0.789\tabularnewline
\hline\hline
\end{tabular}
%\end{center}
\begin{tablenotes}[para,flushleft]
 \footnotesize Notes: see Table \ref{s1_table}.
  \end{tablenotes}
  \end{threeparttable}
\end{table}

\begin{table}[!hp]
\caption{Results for Simulation 3 (Trend and Cosine) and $T=600$ \label{s3_table_600}} 
 \begin{threeparttable}
\setlength{\tabcolsep}{0.2em}
\vspace{-0.3cm}
%\hspace*{-0.5cm}
%\begin{center}
\footnotesize 
\begin{tabular}{lcccccccccccccc}
\hline\hline
\multicolumn{1}{l}{\bfseries }&\multicolumn{4}{c}{\bfseries $K=6$}&\multicolumn{1}{c}{\bfseries }&\multicolumn{4}{c}{\bfseries $K=20$}&\multicolumn{1}{c}{\bfseries }&\multicolumn{4}{c}{\bfseries $K=100$}\tabularnewline
\cline{2-5} \cline{7-10} \cline{12-15}
\multicolumn{1}{l}{}&\multicolumn{1}{c}{BVAR}&\multicolumn{1}{c}{2SRR}&\multicolumn{1}{c}{MSRR$_{\text{S}}$}&\multicolumn{1}{c}{MSRR$_{\text{D}}$}&\multicolumn{1}{c}{}&\multicolumn{1}{c}{BVAR}&\multicolumn{1}{c}{2SRR}&\multicolumn{1}{c}{MSRR$_{\text{S}}$}&\multicolumn{1}{c}{MSRR$_{\text{D}}$}&\multicolumn{1}{c}{}&\multicolumn{1}{c}{BVAR}&\multicolumn{1}{c}{2SRR}&\multicolumn{1}{c}{MSRR$_{\text{S}}$}&\multicolumn{1}{c}{MSRR$_{\text{D}}$}\tabularnewline
\hline
{\bfseries $\mathbf{\sfrac{K^*}{K}=0.2}$}&&&&&&&&&&&&&&\tabularnewline
~~$\sigma_{\epsilon} =\text{Low}$&{\color{blue} \textbf{ 0.064}}& 0.085& 0.089& 0.093&&-& 0.131& 0.123&\textbf{ 0.103}&&-&\textbf{ 0.217}& 0.218& 0.239\tabularnewline
~~$\sigma_{\epsilon} =\text{Medium}$&{\color{blue} \textbf{ 0.120}}& 0.143& 0.133& 0.157&&-& 0.188& 0.188&\textbf{ 0.169}&&-& 0.264&\textbf{ 0.263}& 0.380\tabularnewline
~~$\sigma_{\epsilon} =\text{High}$&{\color{blue} \textbf{ 0.200}}& 0.216& 0.217& 0.262&&-&\textbf{ 0.238}& 0.239& 0.272&&-& 0.433&\textbf{ 0.419}& 0.799\tabularnewline
~~$\sigma_{\epsilon,t} = \text{SV Low-Med}$&{\color{blue} \textbf{ 0.080}}& 0.104& 0.097& 0.107&&-& 0.156& 0.148&\textbf{ 0.127}&&-&\textbf{ 0.229}& 0.231& 0.279\tabularnewline
~~$\sigma_{\epsilon,t}  = \text{SV Low-High}$&{\color{blue} \textbf{ 0.114}}& 0.139& 0.134& 0.163&&-& 0.183& 0.187&\textbf{ 0.181}&&-& 0.273&\textbf{ 0.272}& 0.480\tabularnewline
\hline
{\bfseries $\mathbf{\sfrac{K^*}{K}=0.5}$}&&&&&&&&&&&&&&\tabularnewline
~~$\sigma_{\epsilon} =\text{Low}$&{\color{blue} \textbf{ 0.059}}& 0.071& 0.099& 0.102&&-& 0.113& 0.123&\textbf{ 0.111}&&-&\textbf{ 0.189}& 0.198& 0.219\tabularnewline
~~$\sigma_{\epsilon} =\text{Medium}$& 0.124&{\color{blue} \textbf{ 0.117}}& 0.119& 0.150&&-& 0.156& 0.163&\textbf{ 0.148}&&-&\textbf{ 0.244}& 0.247& 0.321\tabularnewline
~~$\sigma_{\epsilon} =\text{High}$&{\color{blue} \textbf{ 0.156}}& 0.180& 0.180& 0.201&&-&\textbf{ 0.221}& 0.225& 0.237&&-& 0.418&\textbf{ 0.408}& 0.698\tabularnewline
~~$\sigma_{\epsilon,t} = \text{SV Low-Med}$&{\color{blue} \textbf{ 0.084}}& 0.086& 0.108& 0.116&&-& 0.133& 0.135&\textbf{ 0.119}&&-&\textbf{ 0.202}& 0.210& 0.282\tabularnewline
~~$\sigma_{\epsilon,t}  = \text{SV Low-High}$& 0.118&{\color{blue} \textbf{ 0.112}}& 0.121& 0.145&&-&\textbf{ 0.152}& 0.163& 0.163&&-&\textbf{ 0.249}& 0.258& 0.484\tabularnewline
\hline
{\bfseries $\mathbf{\sfrac{K^*}{K}=1}$}&&&&&&&&&&&&&&\tabularnewline
~~$\sigma_{\epsilon} =\text{Low}$&{\color{blue} \textbf{ 0.034}}& 0.039& 0.041& 0.040&&-& 0.060& 0.072&\textbf{ 0.046}&&-&\textbf{ 0.131}& 0.167& 0.212\tabularnewline
~~$\sigma_{\epsilon} =\text{Medium}$& 0.073&{\color{blue} \textbf{ 0.067}}& 0.073& 0.072&&-& 0.101& 0.125&\textbf{ 0.080}&&-&\textbf{ 0.206}& 0.220& 0.252\tabularnewline
~~$\sigma_{\epsilon} =\text{High}$&{\color{blue} \textbf{ 0.116}}& 0.120& 0.128& 0.130&&-& 0.187& 0.203&\textbf{ 0.179}&&-& 0.400&\textbf{ 0.395}& 0.493\tabularnewline
~~$\sigma_{\epsilon,t} = \text{SV Low-Med}$&{\color{blue} \textbf{ 0.044}}& 0.048& 0.049& 0.050&&-& 0.072& 0.088&\textbf{ 0.057}&&-&\textbf{ 0.152}& 0.181& 0.210\tabularnewline
~~$\sigma_{\epsilon,t}  = \text{SV Low-High}$& 0.065&{\color{blue} \textbf{ 0.064}}& 0.072& 0.068&&-&\textbf{ 0.096}& 0.121& 0.099&&-&\textbf{ 0.217}& 0.236& 0.369\tabularnewline
\hline\hline
\end{tabular}
%\end{center}
\begin{tablenotes}[para,flushleft]
 \footnotesize Notes: see Table \ref{s1_table}.
  \end{tablenotes}
  \end{threeparttable}
\end{table}

\begin{table}[!hp]
\caption{Results for Simulation 4 (Mixture) and $T=600$ \label{s4_table_600}} 
 \begin{threeparttable}
\setlength{\tabcolsep}{0.2em}
\vspace{-0.3cm}
%\hspace*{-0.5cm}
%\begin{center}
\footnotesize 
\begin{tabular}{lcccccccccccccc}
\hline\hline
\multicolumn{1}{l}{\bfseries }&\multicolumn{4}{c}{\bfseries $K=6$}&\multicolumn{1}{c}{\bfseries }&\multicolumn{4}{c}{\bfseries $K=20$}&\multicolumn{1}{c}{\bfseries }&\multicolumn{4}{c}{\bfseries $K=100$}\tabularnewline
\cline{2-5} \cline{7-10} \cline{12-15}
\multicolumn{1}{l}{}&\multicolumn{1}{c}{BVAR}&\multicolumn{1}{c}{2SRR}&\multicolumn{1}{c}{MSRR$_{\text{S}}$}&\multicolumn{1}{c}{MSRR$_{\text{D}}$}&\multicolumn{1}{c}{}&\multicolumn{1}{c}{BVAR}&\multicolumn{1}{c}{2SRR}&\multicolumn{1}{c}{MSRR$_{\text{S}}$}&\multicolumn{1}{c}{MSRR$_{\text{D}}$}&\multicolumn{1}{c}{}&\multicolumn{1}{c}{BVAR}&\multicolumn{1}{c}{2SRR}&\multicolumn{1}{c}{MSRR$_{\text{S}}$}&\multicolumn{1}{c}{MSRR$_{\text{D}}$}\tabularnewline
\hline
{\bfseries $\mathbf{\sfrac{K^*}{K}=0.2}$}&&&&&&&&&&&&&&\tabularnewline
~~$\sigma_{\epsilon} =\text{Low}$&{\color{blue}  0.049}& 0.050&\textbf{ 0.038}& 0.053&&-& 0.057&\textbf{ 0.052}& 0.062&&-& 0.109&\textbf{ 0.105}& 0.108\tabularnewline
~~$\sigma_{\epsilon} =\text{Medium}$&{\color{blue}  0.071}& 0.076&\textbf{ 0.070}& 0.096&&-& 0.092&\textbf{ 0.091}& 0.095&&-& 0.182&\textbf{ 0.178}& 0.206\tabularnewline
~~$\sigma_{\epsilon} =\text{High}$&{\color{blue} \textbf{ 0.107}}& 0.117& 0.114& 0.144&&-& 0.171&\textbf{ 0.169}& 0.170&&-& 0.383&\textbf{ 0.371}& 0.642\tabularnewline
~~$\sigma_{\epsilon,t} = \text{SV Low-Med}$&{\color{blue}  0.057}& 0.059&\textbf{ 0.051}& 0.069&&-& 0.069&\textbf{ 0.066}& 0.075&&-& 0.131&\textbf{ 0.127}& 0.144\tabularnewline
~~$\sigma_{\epsilon,t}  = \text{SV Low-High}$&{\color{blue} \textbf{ 0.066}}& 0.069& 0.068& 0.087&&-&\textbf{ 0.088}& 0.090& 0.107&&-& 0.192&\textbf{ 0.192}& 0.292\tabularnewline
\hline
{\bfseries $\mathbf{\sfrac{K^*}{K}=0.5}$}&&&&&&&&&&&&&&\tabularnewline
~~$\sigma_{\epsilon} =\text{Low}$& 0.062&{\color{blue}  0.061}&\textbf{ 0.052}& 0.066&&-& 0.078&\textbf{ 0.073}& 0.082&&-&\textbf{ 0.134}& 0.135& 0.141\tabularnewline
~~$\sigma_{\epsilon} =\text{Medium}$& 0.091&{\color{blue}  0.089}&\textbf{ 0.085}& 0.109&&-&\textbf{ 0.113}& 0.114& 0.117&&-& 0.205&\textbf{ 0.199}& 0.221\tabularnewline
~~$\sigma_{\epsilon} =\text{High}$&{\color{blue} \textbf{ 0.127}}& 0.134& 0.136& 0.179&&-& 0.189&\textbf{ 0.187}& 0.190&&-& 0.393&\textbf{ 0.382}& 0.532\tabularnewline
~~$\sigma_{\epsilon,t} = \text{SV Low-Med}$& 0.074&{\color{blue}  0.068}&\textbf{ 0.062}& 0.075&&-& 0.089&\textbf{ 0.087}& 0.097&&-& 0.156&\textbf{ 0.154}& 0.163\tabularnewline
~~$\sigma_{\epsilon,t}  = \text{SV Low-High}$& 0.089&{\color{blue} \textbf{ 0.084}}& 0.085& 0.111&&-&\textbf{ 0.110}& 0.114& 0.132&&-&\textbf{ 0.206}& 0.209& 0.333\tabularnewline
\hline
{\bfseries $\mathbf{\sfrac{K^*}{K}=1}$}&&&&&&&&&&&&&&\tabularnewline
~~$\sigma_{\epsilon} =\text{Low}$& 0.076&{\color{blue} \textbf{ 0.070}}& 0.073& 0.090&&-&\textbf{ 0.101}& 0.114& 0.113&&-&\textbf{ 0.179}& 0.188& 0.213\tabularnewline
~~$\sigma_{\epsilon} =\text{Medium}$& 0.121&{\color{blue} \textbf{ 0.108}}& 0.109& 0.131&&-&\textbf{ 0.141}& 0.150& 0.145&&-& 0.244&\textbf{ 0.238}& 0.268\tabularnewline
~~$\sigma_{\epsilon} =\text{High}$&{\color{blue} \textbf{ 0.158}}& 0.162& 0.163& 0.199&&-& 0.224& 0.219&\textbf{ 0.218}&&-& 0.431&\textbf{ 0.411}& 0.686\tabularnewline
~~$\sigma_{\epsilon,t} = \text{SV Low-Med}$& 0.096&{\color{blue} \textbf{ 0.083}}& 0.084& 0.107&&-&\textbf{ 0.114}& 0.128& 0.124&&-&\textbf{ 0.200}& 0.201& 0.244\tabularnewline
~~$\sigma_{\epsilon,t}  = \text{SV Low-High}$& 0.120&{\color{blue} \textbf{ 0.105}}& 0.110& 0.146&&-&\textbf{ 0.138}& 0.148& 0.160&&-& 0.248&\textbf{ 0.248}& 0.381\tabularnewline
\hline\hline
\end{tabular}
%\end{center}
\begin{tablenotes}[para,flushleft]
 \footnotesize Notes: see Table \ref{s1_table}.
  \end{tablenotes}
  \end{threeparttable}
\end{table}

\begin{table}[ht]
    \centering
    \begin{threeparttable}
    \caption{2SRR Computational Time (in seconds) for Various Combinations of $K$ and $T$ for Mixture DGP and  ${\sfrac{K^*}{K}=0.5}$}
    \begin{tabular*}{0.75\textwidth}{@{\extracolsep{\fill}}r| S[table-format=3.2] S[table-format=3.2] S[table-format=3.2]}
        \toprule
        {$K \phantom{.}\backslash \phantom{.} T$} & {\phantom{---.}150}  & {\phantom{---.}300}  & {\phantom{---.}600}    \\
        \midrule
                \rowcolor{gray!15} 
        \multicolumn{4}{l}{\textbf{Estimating a Single Equation}} \\
        6   & 0.33 & 1.68 & 11.77 \\
        20  & 0.68 & 4.03 & 31.36 \\
        50  & 1.61 & 10.38 & 77.94 \\
        100 & 3.24 & 21.55 & 153.07 \\
        \midrule
                \rowcolor{gray!15} 
        \multicolumn{4}{l}{\textbf{Estimating $K$ Equations}} \\
        6   & 0.40 & 1.87 & 12.65 \\
        20  & 1.91 & 11.46 & 88.67 \\
        50  & 9.78 & 59.88 & 473.03 \\
        100 & 37.23 & 232.22 & 1805.77 \\
        \bottomrule
    \end{tabular*}
    \begin{tablenotes}[para,flushleft]
	\scriptsize 
		\textit{Notes}: This table reports the computational time for 2SRR with different values of $K$ and $T$. The times are recorded using a 2020 M1 Macbook Air and are averaged over 5 simulations.	
		\end{tablenotes}
    \label{tab:2SRR_computational_time}
    \end{threeparttable}
\end{table}

\begin{table}[t!]
	\footnotesize
	\centering
	%\rowcolors{2}{white}{gray!15}
\caption{\normalsize {Comparison of Uncertainty Quantification for Mixture DGP,  $T=300$,  and $\sfrac{K^*}{K}=0.5$} \vspace*{-0.3cm}} \label{tab:uq}

\begin{threeparttable}
		\setlength{\tabcolsep}{0.60em}%{0.61em}
		  \setlength\extrarowheight{2.5pt}
 \begin{tabular}{l| ccccccc | ccccccc} 
\toprule \toprule
\addlinespace[2pt]
& & \multicolumn{5}{c}{$K=6$} & & & \multicolumn{5}{c}{$K=20$} \\
\cmidrule(lr){3-7} \cmidrule(lr){10-14} \addlinespace[2pt]
& &  2SRR &  &   \texttt{ShTVP}-R &  &\texttt{ShTVP}-3G &  & &
	  2SRR  &  & \texttt{ShTVP}-R &  &\texttt{ShTVP}-3G & \\
\midrule
\addlinespace[5pt] 
\rowcolor{gray!15} 
 \multicolumn{14}{l}{\textbf{68\% Nominal Coverage}} &\cellcolor{gray!15} \\ \addlinespace[2pt] 
$\sigma_{\epsilon} =\text{Low}$ & & 65.0 & & 73.2  & &  63.0  && &  61.9 & & 75.7 & & 56.4   &  \\ \addlinespace[2pt]  
$\sigma_{\epsilon} =\text{Medium}$ & & 78.2 & & 73.5  & &  57.1  && &  76.4  & & 77.8 & & 54.3  &  \\ \addlinespace[2pt]  
$\sigma_{\epsilon} =\text{High}$ & & 82.0 & & 79.6  & &  59.7  && &  82.0 & & 80.0 & & 40.8  &  \\ \addlinespace[2pt]  
$\sigma_{\epsilon,t} = \text{SV Low-Med}$ & & 66.5 & & 73.4  & &  62.9  && & 68.4 & & 73.6 & & 55.1 &  \\ \addlinespace[2pt]  
$\sigma_{\epsilon,t} = \text{SV Low-High}$ & & 77.9 & & 76.5  & &  59.9  && &  73.8 & & 75.6 & & 50.6  &  \\ \addlinespace[2pt]

\midrule
\addlinespace[5pt] 
\rowcolor{gray!15} 
 \multicolumn{14}{l}{\textbf{95\% Nominal Coverage}} &\cellcolor{gray!15} \\ \addlinespace[2pt] 
$\sigma_{\epsilon} =\text{Low}$ & & 94.0  & & 96.9  & &  89.3  && &  89.1 & & 97.8 & & 88.0  &  \\ \addlinespace[2pt]  
$\sigma_{\epsilon} =\text{Medium}$ & & 97.4 & & 98.0  & &  89.5  && &  96.5 & & 98.8 & & 88.9  &  \\ \addlinespace[2pt]  
$\sigma_{\epsilon} =\text{High}$ & & 99.3  & & 98.7  & &  92.6  && &  98.2 & & 99.3 & & 88.1  &  \\ \addlinespace[2pt]  
$\sigma_{\epsilon,t} = \text{SV Low-Med}$ & & 93.3 & & 96.5  & & 91.3  && &  93.7  & & 97.9 & & 87.6  &  \\ \addlinespace[2pt]  
$\sigma_{\epsilon,t} = \text{SV Low-High}$ & & 97.1 & & 96.9 & &  88.9  && &  96.3 & & 98.4 & & 87.9  &  \\ \addlinespace[2pt]

\midrule
\addlinespace[5pt] 
\rowcolor{gray!15} 
 \multicolumn{14}{l}{\textbf{\texttt{Corr}(IQR, $\sigma_{\epsilon,t}$)}} &\cellcolor{gray!15} \\ \addlinespace[2pt] 

$\sigma_{\epsilon,t} = \text{SV Low-Med}$ & & 0.71 & & 0.28  & &  0.21  && &  0.65 & & 0.23 & & 0.19  &  \\ \addlinespace[2pt]  
$\sigma_{\epsilon,t} = \text{SV Low-High}$ & & 0.80 & & 0.62  & &  0.42  && &  0.87  & & 0. 46 & & 0.20  &  \\ \addlinespace[2pt]

\bottomrule \bottomrule
	\end{tabular}
		\begin{tablenotes}[para,flushleft]
	\scriptsize %\item[] \hspace*{-0.5cm}
		\textit{Notes}: Results are averaged over 20 simulations and the whole simulation path.    The IQR is for the 12\%-84\% range.   %The number in bold is the best statistic (lowest MAE or computing time) of all models for a given setup. 
	\end{tablenotes}
\end{threeparttable}
\end{table}

%%%%%%%%%%%%%%%%%%%%%%%%%%%%%%%%%%%%%%%%%%%%%%%%%%%%%%%%%%%%

\begin{landscape}\begin{table}[!tbp]
\caption{Forecasting Results \label{tab_fcst_main}} 
 \begin{threeparttable}
 %\begin{center}
\setlength{\tabcolsep}{0.2em}
%  \rowcolors{2}{white}{gray!15}
\begin{tabular}{lllllcllllcllllcllll}
\hline\hline
\multicolumn{1}{l}{\bfseries }&\multicolumn{4}{c}{\bfseries AR}&\multicolumn{1}{c}{\bfseries }&\multicolumn{4}{c}{\bfseries ARDI}&\multicolumn{1}{c}{\bfseries }&\multicolumn{4}{c}{\bfseries VAR5}&\multicolumn{1}{c}{\bfseries }&\multicolumn{4}{c}{\bfseries VAR20}\tabularnewline
\cline{2-5} \cline{7-10} \cline{12-15} \cline{17-20}
\multicolumn{1}{l}{}&\multicolumn{1}{c}{Plain}&\multicolumn{1}{c}{2SRR}&\multicolumn{1}{c}{MSRR$_{\text{S}}$}&\multicolumn{1}{c}{MSRR$_{\text{D}}$}&\multicolumn{1}{c}{}&\multicolumn{1}{c}{Plain}&\multicolumn{1}{c}{2SRR}&\multicolumn{1}{c}{MSRR$_{\text{S}}$}&\multicolumn{1}{c}{MSRR$_{\text{D}}$}&\multicolumn{1}{c}{}&\multicolumn{1}{c}{Plain}&\multicolumn{1}{c}{2SRR}&\multicolumn{1}{c}{MSRR$_{\text{S}}$}&\multicolumn{1}{c}{MSRR$_{\text{D}}$}&\multicolumn{1}{c}{}&\multicolumn{1}{c}{Plain}&\multicolumn{1}{c}{2SRR}&\multicolumn{1}{c}{MSRR$_{\text{S}}$}&\multicolumn{1}{c}{MSRR$_{\text{D}}$}\tabularnewline
\hline
{\bfseries GDP}&&&&&&&&&&&&&&&&&&&\tabularnewline
~~$h=1$&1.00&0.98&0.99&\textbf{0.98}&&1.03&1.11&1.10&1.04&&1.04&1.06&1.05&1.01&&1.24&2.07&1.61**&1.45*\tabularnewline
~~$h=2$&\textbf{1.00}&1.10&1.00&1.00&&1.05&1.77&1.06&1.08&&1.08&1.39&1.16*&1.08**&&1.27&1.32**&1.34**&1.70*\tabularnewline
~~$h=4$&\textbf{1.00}&1.23&1.06&1.06&&1.07&1.41&1.06&1.08&&1.06&1.41&1.14&1.15&&1.10&1.27&1.23**&1.12\tabularnewline
\hline
{\bfseries UR}&&&&&&&&&&&&&&&&&&&\tabularnewline
~~$h=1$&1.00&1.11&1.15&1.15&&\textbf{0.99}&1.02&1.05&1.17&&1.10*&1.13&1.20&1.10&&1.63&1.45&1.76*&1.37\tabularnewline
~~$h=2$&1.00&1.46&1.29&1.19&&\textbf{1.00}&1.80&1.48&1.03&&1.11&1.44&1.44&1.39&&1.40&2.11&1.76&2.21\tabularnewline
~~$h=4$&\textbf{1.00}&1.59&1.34&1.13**&&1.00&1.47&1.32&1.19&&1.09&1.40&1.30&1.40&&1.07&1.49&1.33*&1.33*\tabularnewline
\hline
{\bfseries INF}&&&&&&&&&&&&&&&&&&&\tabularnewline
~~$h=1$&1.00&\textbf{0.93}&0.96&0.95&&1.01&0.99&1.09&0.95&&1.00&0.94&0.95&1.22&&1.79&1.79&1.81&1.85\tabularnewline
~~$h=2$&\textbf{1.00}&1.14&1.15&1.00&&1.06&1.35&1.14&1.49&&1.03&1.39&1.17&1.26&&1.15&1.77&1.53&1.09*\tabularnewline
~~$h=4$&\textbf{1.00}&1.09&1.10&1.12&&1.06*&1.20&1.11&1.15&&1.00&1.11&1.09&1.02&&1.38&1.64&1.12&1.10\tabularnewline
\hline
{\bfseries IR}&&&&&&&&&&&&&&&&&&&\tabularnewline
~~$h=1$&1.00&0.64***&0.71***&0.79***&&0.94***&\textbf{0.64}***&0.80***&1.05&&1.07&0.72***&0.82**&1.09**&&1.46*&2.03&2.28&0.71***\tabularnewline
~~$h=2$&1.00&0.72***&\textbf{0.66}***&0.72***&&0.94**&0.86&0.74**&0.99&&0.97&0.95&0.98&0.93&&1.37&2.46&0.79**&1.34\tabularnewline
~~$h=4$&1.00&1.03&1.06&1.12&&0.93&1.04&0.91&1.13&&0.97&1.12&0.96&0.93&&1.08&1.24&0.93&\textbf{0.81}*\tabularnewline
\hline
{\bfseries SPREAD}&&&&&&&&&&&&&&&&&&&\tabularnewline
~~$h=1$&1.00&0.86***&0.86***&0.86**&&0.90**&0.86**&0.85**&0.84***&&0.96&\textbf{0.82}***&0.87**&0.93&&2.13&2.70*&2.48*&1.94\tabularnewline
~~$h=2$&1.00&1.02&1.00&0.96&&0.91*&1.09&1.06&0.95&&\textbf{0.88}**&0.96&0.91&0.92&&2.03&2.31&2.30&2.01\tabularnewline
~~$h=4$&1.00&1.58&1.14&1.32&&0.90**&1.35&1.27&0.95&&\textbf{0.88}**&1.34&1.20&1.01&&0.89&1.50&1.25&1.16\tabularnewline
\hline\hline
\end{tabular}
%\end{center}
\begin{tablenotes}[para,flushleft]
  Notes: This table reports $RMSPE_{v,h,m}/RMSPE_{v,h,\text{Plain AR(2)}}$ for 5 variables, 3 horizons and 16 models considered in the pseudo-out-of-sample experiment. Numbers in bold identifies the best predictive performance of the row. Diebold-Mariano tests are performed to evaluate whether the difference in predictive performance between a model and the AR(2) benchmark is statistically significant. '*', '**' and '***' respectively refer to p-values below 10\%, 5\% and 1\%.
  \end{tablenotes}
  \end{threeparttable}
\end{table}\end{landscape}

\begin{landscape}\begin{table}[!tbp]
\caption{Forecasting Results, \textit{Half \& Half} \label{tab_fcst_modavg}} 
 \begin{threeparttable}
 %\begin{center}
\setlength{\tabcolsep}{0.2em}
\begin{tabular}{lllllcllllcllllcllll}
\hline\hline
\multicolumn{1}{l}{\bfseries }&\multicolumn{4}{c}{\bfseries AR}&\multicolumn{1}{c}{\bfseries }&\multicolumn{4}{c}{\bfseries ARDI}&\multicolumn{1}{c}{\bfseries }&\multicolumn{4}{c}{\bfseries VAR5}&\multicolumn{1}{c}{\bfseries }&\multicolumn{4}{c}{\bfseries VAR20}\tabularnewline
\cline{2-5} \cline{7-10} \cline{12-15} \cline{17-20}
\multicolumn{1}{l}{}&\multicolumn{1}{c}{Plain}&\multicolumn{1}{c}{2SRR}&\multicolumn{1}{c}{MSRR$_{\text{S}}$}&\multicolumn{1}{c}{MSRR$_{\text{D}}$}&\multicolumn{1}{c}{}&\multicolumn{1}{c}{Plain}&\multicolumn{1}{c}{2SRR}&\multicolumn{1}{c}{MSRR$_{\text{S}}$}&\multicolumn{1}{c}{MSRR$_{\text{D}}$}&\multicolumn{1}{c}{}&\multicolumn{1}{c}{Plain}&\multicolumn{1}{c}{2SRR}&\multicolumn{1}{c}{MSRR$_{\text{S}}$}&\multicolumn{1}{c}{MSRR$_{\text{D}}$}&\multicolumn{1}{c}{}&\multicolumn{1}{c}{Plain}&\multicolumn{1}{c}{2SRR}&\multicolumn{1}{c}{MSRR$_{\text{S}}$}&\multicolumn{1}{c}{MSRR$_{\text{D}}$}\tabularnewline
\hline
{\bfseries GDP}&&&&&&&&&&&&&&&&&&&\tabularnewline
~~$h=1$&1.00&\textbf{0.99}&0.99&0.99&&1.03&1.01&1.00&1.04&&1.04&1.02&1.03&1.02&&1.24&1.59&1.34&1.24\tabularnewline
~~$h=2$&1.00&1.00&0.97&1.00&&1.05&1.21&\textbf{0.97}&1.05&&1.08&1.10&1.06&1.07*&&1.27&1.19&1.20&1.33\tabularnewline
~~$h=4$&1.00&1.01&0.96&0.95&&1.07&1.06&\textbf{0.94}&0.97&&1.06&1.09&1.01&0.97&&1.10&1.05&1.05&0.97\tabularnewline
\hline
{\bfseries UR}&&&&&&&&&&&&&&&&&&&\tabularnewline
~~$h=1$&1.00&1.02&1.04&1.02&&0.99&\textbf{0.97}&0.98&1.04&&1.10*&1.09&1.12&1.06&&1.63&1.47&1.59&1.27*\tabularnewline
~~$h=2$&1.00&1.08&1.08&1.03&&\textbf{1.00}&1.16&1.12&1.01&&1.11&1.15&1.17&1.17&&1.40&1.67&1.50&1.70\tabularnewline
~~$h=4$&\textbf{1.00}&1.15&1.10&1.05*&&1.00&1.08&1.04&1.02&&1.09&1.16&1.14*&1.19*&&1.07&1.20&1.15&1.11\tabularnewline
\hline
{\bfseries INF}&&&&&&&&&&&&&&&&&&&\tabularnewline
~~$h=1$&1.00&\textbf{0.94}&0.95&0.96&&1.01&0.98&1.01&0.97&&1.00&0.96&0.96&1.06&&1.79&1.78&1.78&1.74\tabularnewline
~~$h=2$&1.00&0.91&0.92&0.94&&1.06&1.12&\textbf{0.91}*&1.15&&1.03&1.07&0.93&1.07&&1.15&1.30**&1.18&1.07\tabularnewline
~~$h=4$&1.00&0.85*&0.85*&0.86*&&1.06*&0.88&0.84&0.91&&1.00&0.93&\textbf{0.84}&0.87&&1.38&1.45&1.17&1.13\tabularnewline
\hline
{\bfseries IR}&&&&&&&&&&&&&&&&&&&\tabularnewline
~~$h=1$&1.00&0.80***&0.84***&0.88***&&0.94***&\textbf{0.76}***&0.85***&0.96&&1.07&0.87**&0.93&1.04&&1.46*&1.67&1.77&0.98\tabularnewline
~~$h=2$&1.00&0.83***&0.76***&0.82***&&0.94**&0.85**&\textbf{0.76}***&0.93&&0.97&0.88*&0.95&0.92**&&1.37&1.84&0.98&1.26\tabularnewline
~~$h=4$&1.00&0.99&1.01&1.02&&0.93&0.95&0.90&0.99&&0.97&0.99&0.92&0.91&&1.08&1.07&0.97&\textbf{0.88}\tabularnewline
\hline
{\bfseries SPREAD}&&&&&&&&&&&&&&&&&&&\tabularnewline
~~$h=1$&1.00&0.90***&0.92***&0.89***&&0.90**&\textbf{0.86}**&0.87***&0.86***&&0.96&0.87***&0.90**&0.92*&&2.13&2.30&2.21&1.98\tabularnewline
~~$h=2$&1.00&0.97&0.98&0.97&&0.91*&0.94&0.95&0.92&&0.88**&\textbf{0.88}&0.88&0.88**&&2.03&2.12&2.12&1.90\tabularnewline
~~$h=4$&1.00&1.21&1.03&1.12&&0.90**&1.03&1.03&0.91*&&\textbf{0.88}**&1.03&0.98&0.92&&0.89&1.08&0.95&0.98\tabularnewline
\hline\hline
\end{tabular}
%\end{center}
\begin{tablenotes}[para,flushleft]
  Notes: This table reports $RMSPE_{v,h,m}/RMSPE_{v,h,\text{Plain AR(2)}}$ for 5 variables, 3 horizons and 16 models considered in the pseudo-out-of-sample experiment. TVPs of each model are shrunk to constant parameters via model averaging with equal weights for both the TVP model and its constant coefficients counterpart. Numbers in bold identifies the best predictive performance of the row. Diebold-Mariano tests are performed to evaluate whether the difference in predictive performance between a model and the AR(2) benchmark is statistically significant. '*', '**' and '***' respectively refer to p-values below 10\%, 5\% and 1\%.
  \end{tablenotes}
  \end{threeparttable}
\end{table}\end{landscape}

\begin{landscape}\begin{table}[!tbp]
\caption{Forecasting Results,  Blocked Cross-Validation \label{tab_fcst_main_bb}} 
 \begin{threeparttable}
 %\begin{center}
\setlength{\tabcolsep}{0.2em}
%  \rowcolors{2}{white}{gray!15}
\begin{tabular}{lllllcllllcllllcllll}
\hline\hline
\multicolumn{1}{l}{\bfseries }&\multicolumn{4}{c}{\bfseries AR}&\multicolumn{1}{c}{\bfseries }&\multicolumn{4}{c}{\bfseries ARDI}&\multicolumn{1}{c}{\bfseries }&\multicolumn{4}{c}{\bfseries VAR5}&\multicolumn{1}{c}{\bfseries }&\multicolumn{4}{c}{\bfseries VAR20}\tabularnewline
\cline{2-5} \cline{7-10} \cline{12-15} \cline{17-20}
\multicolumn{1}{l}{}&\multicolumn{1}{c}{Plain}&\multicolumn{1}{c}{2SRR}&\multicolumn{1}{c}{MSRR$_{\text{S}}$}&\multicolumn{1}{c}{MSRR$_{\text{D}}$}&\multicolumn{1}{c}{}&\multicolumn{1}{c}{Plain}&\multicolumn{1}{c}{2SRR}&\multicolumn{1}{c}{MSRR$_{\text{S}}$}&\multicolumn{1}{c}{MSRR$_{\text{D}}$}&\multicolumn{1}{c}{}&\multicolumn{1}{c}{Plain}&\multicolumn{1}{c}{2SRR}&\multicolumn{1}{c}{MSRR$_{\text{S}}$}&\multicolumn{1}{c}{MSRR$_{\text{D}}$}&\multicolumn{1}{c}{}&\multicolumn{1}{c}{Plain}&\multicolumn{1}{c}{2SRR}&\multicolumn{1}{c}{MSRR$_{\text{S}}$}&\multicolumn{1}{c}{MSRR$_{\text{D}}$}\tabularnewline
\hline
{\bfseries GDP}&&&&&&&&&&&&&&&&&&&\tabularnewline
~~$h=1$&1.00&\textbf{1.00}&1.00&1.00&&1.06**&1.04&1.07&1.05*&&1.04&1.09*&1.06&1.02&&1.72&1.74&1.80&1.74\tabularnewline
~~$h=2$&1.00&\textbf{0.99}&1.00&1.00&&1.08&1.21&1.19&1.06&&1.13**&1.14&1.11*&1.12&&1.89&1.90&1.79&1.17\tabularnewline
~~$h=4$&1.00&0.98&0.95&1.02&&1.10&1.11&\textbf{0.91}&1.20&&1.13&1.09&1.10&1.08&&1.12&1.45*&1.13&0.99\tabularnewline
\hline
{\bfseries UR}&&&&&&&&&&&&&&&&&&&\tabularnewline
~~$h=1$&1.00&1.01&1.04&1.05&&1.00&\textbf{0.94}&0.99&1.04&&1.12*&1.10&1.17&1.23*&&1.64&1.46&1.52&1.79*\tabularnewline
~~$h=2$&1.00&1.04&1.02&1.39&&\textbf{1.00}&1.00&1.02&1.34&&1.13*&1.12&1.29&1.29&&1.50&1.72&1.64&1.73\tabularnewline
~~$h=4$&1.00&1.05&1.11&1.30&&1.02&1.24&1.23&\textbf{0.97}&&1.14&1.18&1.34&1.13&&1.08&1.61&1.56**&1.69\tabularnewline
\hline
{\bfseries INF}&&&&&&&&&&&&&&&&&&&\tabularnewline
~~$h=1$&1.00&\textbf{0.92}&0.95&0.93&&1.02&0.97&1.08&1.08&&1.00&0.94&1.02&1.00&&1.65&1.65&1.72&1.73\tabularnewline
~~$h=2$&1.00&0.84**&\textbf{0.84}**&0.92**&&1.09&0.95&0.90&0.94&&1.05&0.92&0.87*&0.87**&&1.15&1.65&1.82&1.27**\tabularnewline
~~$h=4$&1.00&0.68&0.68&0.95&&1.09&0.70&0.70&0.93&&1.02*&\textbf{0.68}&0.68&1.25&&1.28&1.71&1.34&0.88\tabularnewline
\hline
{\bfseries IR}&&&&&&&&&&&&&&&&&&&\tabularnewline
~~$h=1$&1.00&\textbf{0.96}**&1.02&1.16***&&0.97&0.96&1.07**&1.26***&&1.17**&1.09&1.23***&1.50***&&2.21**&3.03&2.89&1.14**\tabularnewline
~~$h=2$&1.00&0.98&1.01&1.13**&&0.95&\textbf{0.94}&1.00&1.18***&&1.01&0.97&1.12*&1.10**&&1.62&3.30&2.94&1.47*\tabularnewline
~~$h=4$&1.00&1.05&1.12&1.17*&&0.95&1.03&1.13&1.04&&0.98&1.04&1.19&1.10*&&1.14&1.56&1.37&\textbf{0.91}\tabularnewline
\hline
{\bfseries SPREAD}&&&&&&&&&&&&&&&&&&&\tabularnewline
~~$h=1$&1.00&0.96***&0.99&1.02&&\textbf{0.94}&0.98&0.97&0.96&&1.00&0.96&0.96&0.97&&2.41&3.02*&3.05*&1.70*\tabularnewline
~~$h=2$&1.00&0.99&0.99&0.96&&0.93&0.98&0.98&0.99&&0.91&\textbf{0.88}*&0.94&0.95&&2.01&2.33&2.31&1.76\tabularnewline
~~$h=4$&1.00&1.00&1.02&1.48&&0.90**&0.96&0.96&1.02&&\textbf{0.88}**&0.89*&0.93&0.91*&&0.88&1.10&1.05&1.14\tabularnewline
\hline\hline
\end{tabular}
%\end{center}
\begin{tablenotes}[para,flushleft]
  Notes: This table reports $RMSPE_{v,h,m}/RMSPE_{v,h,\text{Plain AR(2)}}$ for 5 variables, 3 horizons and 16 models considered in the pseudo-out-of-sample experiment. Numbers in bold identifies the best predictive performance of the row. Diebold-Mariano tests are performed to evaluate whether the difference in predictive performance between a model and the AR(2) benchmark is statistically significant. '*', '**' and '***' respectively refer to p-values below 10\%, 5\% and 1\%.
  \end{tablenotes}
  \end{threeparttable}
\end{table}\end{landscape}

\subsection{Additional Graphs}

\begin{figure}[!hp]
			\centering
\includegraphics[scale=.27]{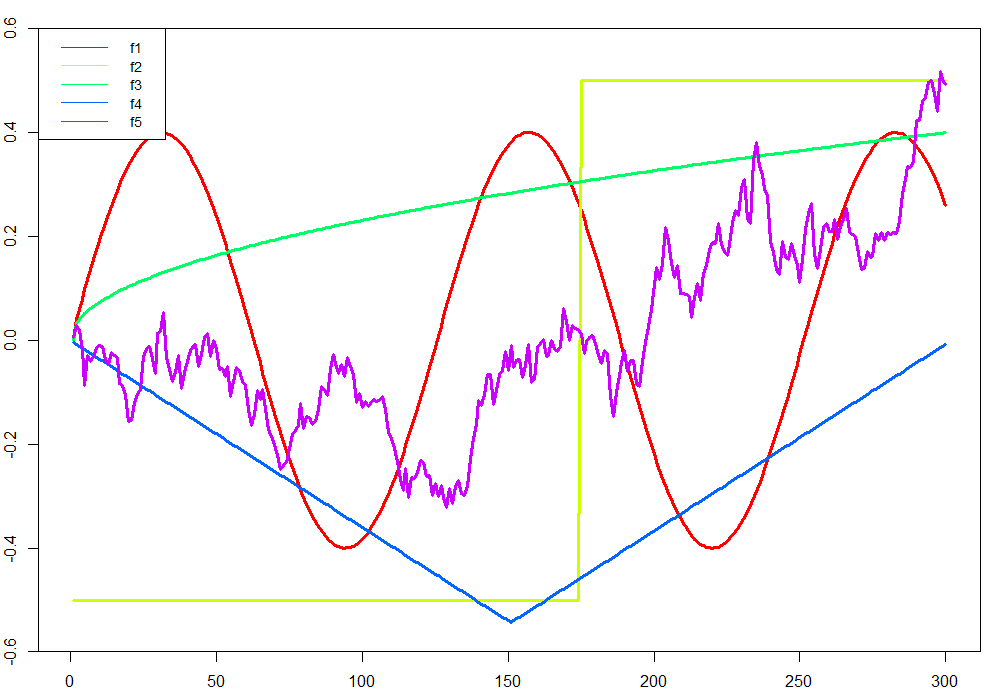}  
		\caption{\footnotesize{This graph displays the 5 paths out of which the true $\beta_{k,t}$'s will be constructed for simulations.}}
		\label{fig:paths}
		\end{figure}
\vspace{-0.5cm}		
\begin{figure}[!h]
  \begin{subfigure}[b]{0.5\textwidth}
   \includegraphics[width=\textwidth]{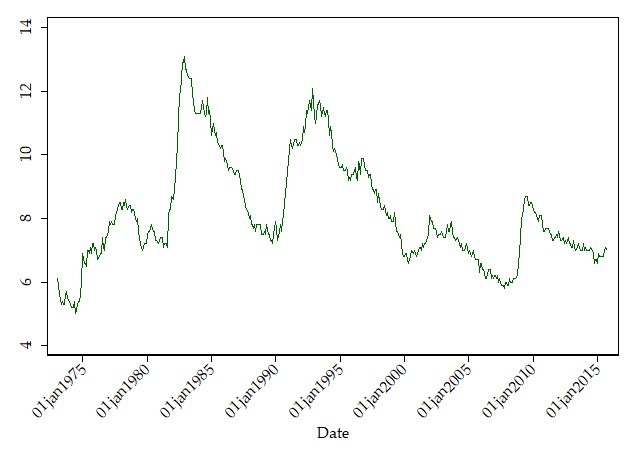}
    \caption{Unemployment Rate}
  \end{subfigure}
  %\hspace{0.2em}
  \begin{subfigure}[b]{0.5\textwidth}
   \includegraphics[width=\textwidth]{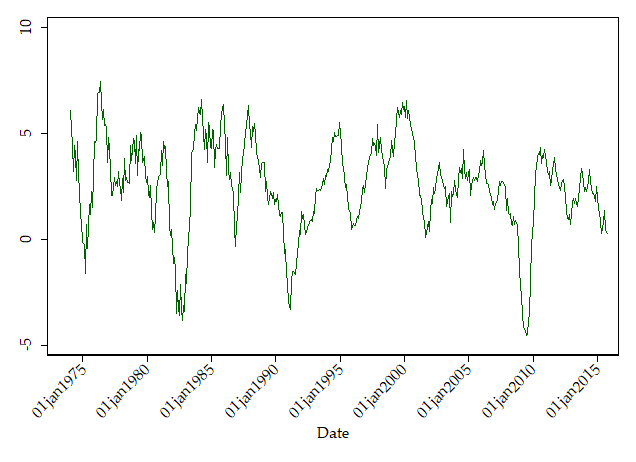}
    \caption{Month over Month GDP growth}
  \end{subfigure} %\\[1ex]
   % \hspace{0.2em}
  \begin{subfigure}[b]{0.5\textwidth}
  \centering
   \includegraphics[width=\textwidth]{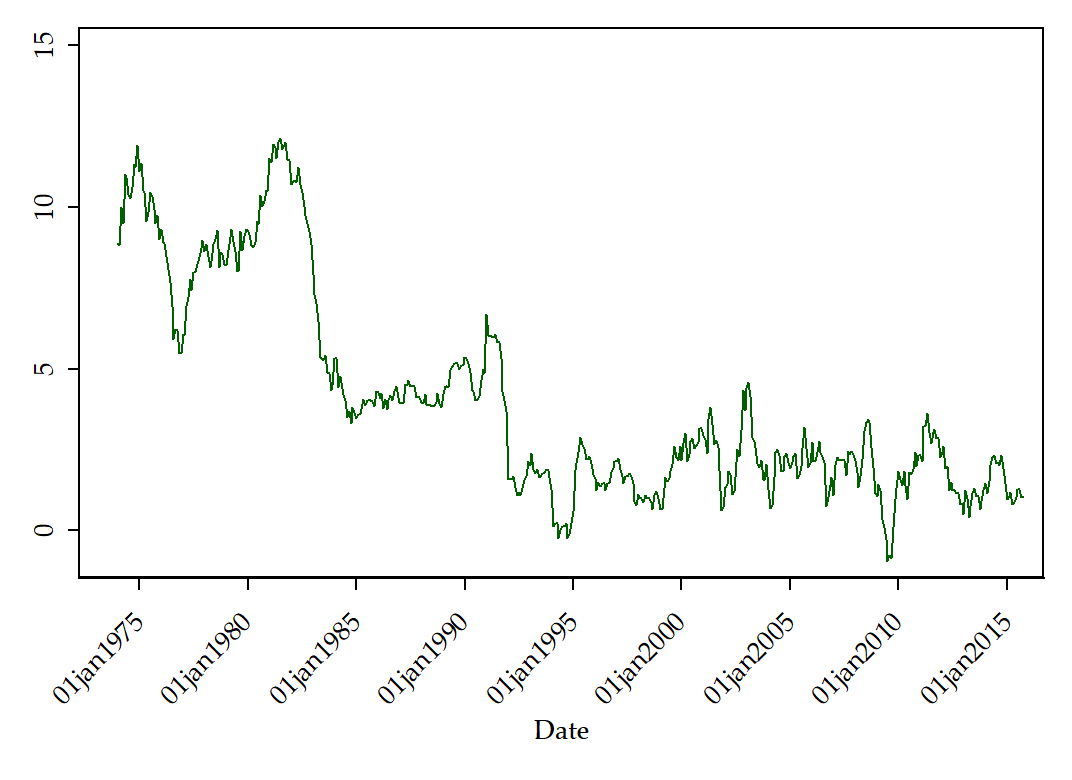}
    \caption{Month over Month Inflation Rate}
      \label{can_inf}
  \end{subfigure}
    \begin{subfigure}[b]{0.5\textwidth}
  \centering
   \includegraphics[width=\textwidth]{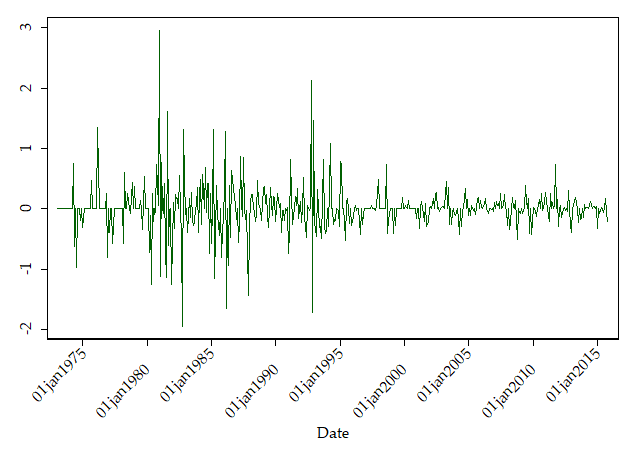}
    \caption{Monetary Policy Shocks}
  \end{subfigure}
  \caption{Four Main Canadian Time series}
  \label{can_ts}
  \end{figure}

\begin{figure}[!hp]
			\centering
\hspace*{-0.35cm}\includegraphics[scale=.6]{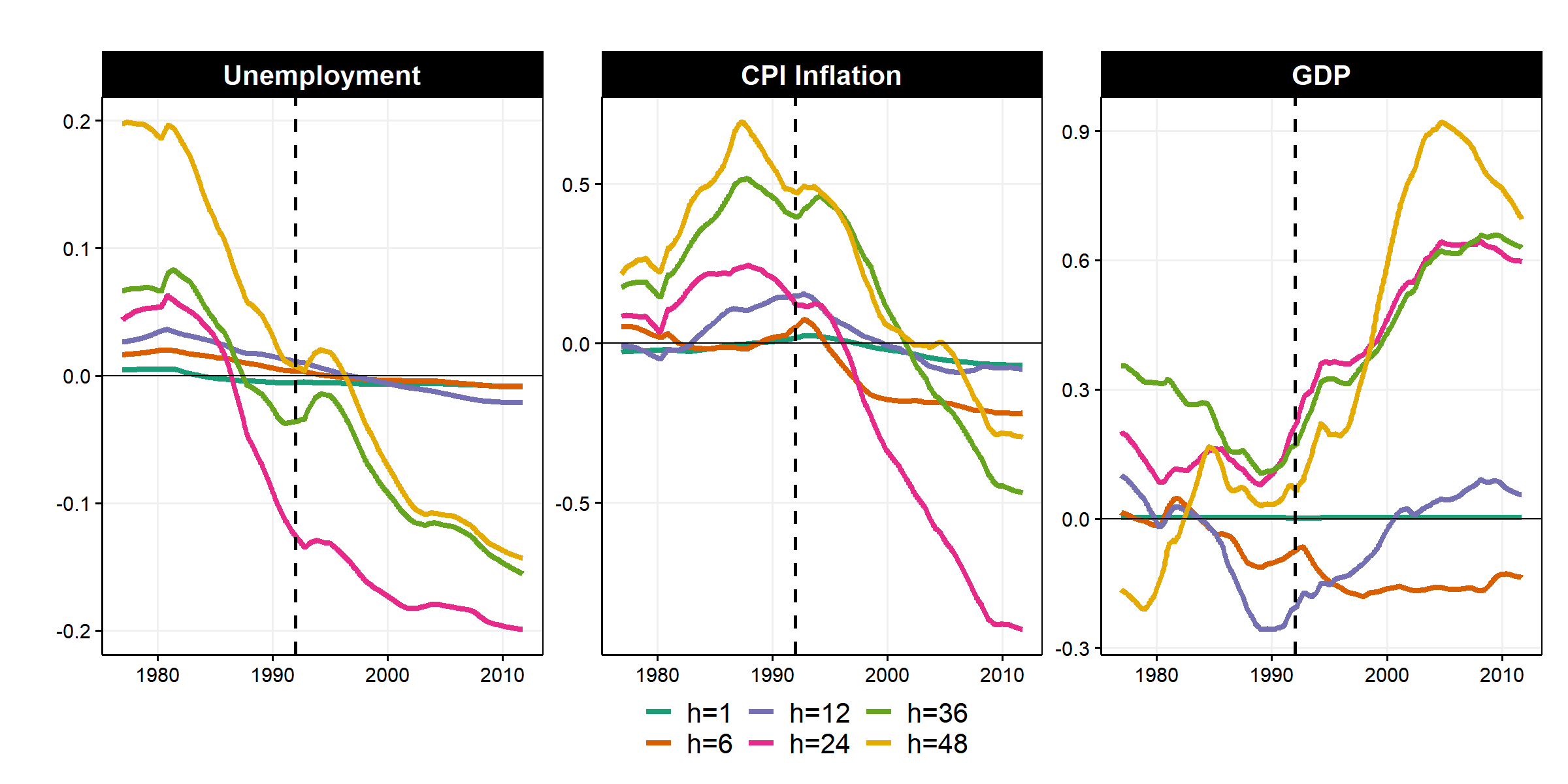}  
		\caption{$\beta_t^{2SRR}-\beta^{OLS}$ for the cumulative effect of MP shocks on variables of interest. Note that for better visibility, GDP and CPI Inflation units are now percentages. Dashed black line is the onset of inflation targeting.}
		\label{TVLPwrtOLS}
		\end{figure}

\end{document}